\documentclass[manuscript]{aastex62}
\usepackage{xspace}
\usepackage{color}
\usepackage{url}
\usepackage{graphicx}

\shorttitle{GMIMS: Faraday Spectral Moments}
\shortauthors{Dickey, J.M.  et al.}

\begin{document}
\title{The Galactic Magneto-Ionic Medium Survey:  Moments of the Faraday Spectra}
\author[0000-0002-6300-7459]{John M. Dickey}
\affiliation{School of Natural Sciences, Private Bag 37, University of Tasmania, Hobart, TAS, 7001, Australia}
\author{T.L. Landecker}%\altaffiliation{2,1,5}
\affiliation{National Research Council Canada, Dominion Radio Astrophysical Observatory, P.O. Box 218, Penticton, British Columbia, V2A 6J9, Canada}
\author[0000-0001-9472-041X]{Alec J.M. Thomson }%\altaffiliation{1}
\affiliation{Research School of Astronomy and Astrophysics, Australian National University, Canberra, ACT 2611, Australia}
\author{M. Wolleben}%\altaffiliation{1}
\affiliation{Skaha Remote Sensing Ltd., 3165 Juniper Drive, Naramata, British Columbia V0H 1N0, Canada}
\author{X. Sun}%\altaffiliation{9}}
\affiliation{Department of Astronomy, Yunnan University, and Key Laboratory of Astroparticle Physics of Yunnan Province, Kunming, 650091, People's Republic of China}
\author[0000-0002-3973-8403]{E. Carretti}%\altaffiliation{9}}
\affiliation{INAF - Instituto de Radioastronomia, Via P. Gobetti 101, I-40129, Bologna, Italy}
%Via della Scienza 5, I-09047 Selargius (CA), Italy}
\author{K. Douglas}%\altaffiliation{9}}
\affiliation{Physics and Astronomy Department, Okanagan College, 1000 KLO Road, Kelowna, British Columbia V1Y 4X8, Canada}
\author[0000-0003-1741-1714]{A. Fletcher}%\altaffiliation{9}}
\affiliation{School of Mathematics, Statistics and Physics, Newcastle University, Newcastle-upon-Tyne, NE13 7RU, UK}
\author[0000-0002-3382-9558]{B.M. Gaensler}
\affiliation{Dunlap Institute for Astronomy and Astrophysics, University of Toronto, Toronto, ON M5S 3H4, Canada}
\author{A. Gray}%\altaffiliation{9}}
\affiliation{National Research Council Canada, Dominion Radio Astrophysical Observatory, P.O. Box 218, Penticton, British Columbia, V2A 6J9, Canada}
%\author{J.L. Han}%\altaffiliation{9}}
%\affiliation{National Astronomical Observatories, Chinese Academy of Sciences, Beijing 100012, China}
\author[0000-0002-5288-312X]{M. Haverkorn}
\affiliation{Department of Astrophysics/IMAPP, Radboud University, PO Box 9010, NL-6500 GL Nijmegen, the Netherlands}
\author[0000-0001-7301-5666]{A.S. Hill}
\affiliation{National Research Council Canada, Dominion Radio Astrophysical Observatory, P.O. Box 218, Penticton, British Columbia, V2A 6J9, Canada}
\affiliation{Department of Physics and Astronomy, University of British Columbia, Vancouver, BC V6T 1Z1, Canada}
\affiliation{Space Science Institute, Boulder, CO, 80301}
\author{S. A. Mao}
\affiliation{Max Planck Institute for Radio Astronomy, Auf dem H\"{u}gel 69, D-53121 Bonn, Germany}
\author[0000-0003-2730-957X]{N. M. McClure-Griffiths}
\affiliation{Research School of Astronomy and Astrophysics, Australian National University, Canberra, ACT 2611, Australia}
\correspondingauthor{John Dickey}
\email{john.dickey@utas.edu.au}
%\author{More...}
%\author{Others, All-the}%\altaffiliation{7}}
%\affiliation{All Over the Place}

\begin{abstract}

Faraday rotation occurs along every line of sight in the Galaxy; Rotation Measure (RM) synthesis allows
a three-dimensional representation of the interstellar magnetic field.  This study uses
data from the Global Magneto-Ionic Medium Survey, a combination of single-antenna spectro-polarimetric
studies, including northern sky data from the DRAO 26-m Telescope (1270-1750 MHz) and southern sky
data from the Parkes 64-m Telescope (300-480 MHz). From the synthesized Faraday spectral cubes 
we compute the zeroth, first, and second moments to find the total polarized emission, 
mean and RM-width of the polarized emission. From DRAO first moments we find a
weak vertical field directed from Galactic North to South, but Parkes data reveal fields directed
towards the Sun at high latitudes in both hemispheres: the two surveys clearly sample different
volumes. DRAO second moments show feature widths in Faraday spectra increasing with decreasing
positive latitudes, implying that longer lines of sight encounter more Faraday rotating medium, but this 
is not seen at negative latitudes. Parkes data show the opposite: at positive latitudes the second
moment decreases with decreasing latitude, but not at negative latitudes. Comparing first moments 
with RMs of pulsars and extragalactic sources and a study of depolarization together confirm
that the DRAO survey samples to larger distances than the Parkes data. Emission regions
in the DRAO survey are typically 700 to 1000 pc away, slightly beyond the scale-height of the
magneto-ionic medium; emission detected in the Parkes survey is entirely within the
magneto-ionic disk, less than 500 pc away.
\end{abstract}


\section{Introduction}

\subsection{Galactic Diffuse Polarized Emission}

  The magnetic field of the Milky Way can be traced qualitatively and measured
quantitatively through various observations, many of them involving polarization
%(Han 2017, Ferri\`ere 2015, Mao et al. 2015a, Planck Collaboration, 2018).
\citep{Han_2017, Ferriere_2015, Mao_etal_2015a, Planck_2018}.
Starlight polarization that shows large scale patterns, and generally increases with 
the distance of the star, was the first evidence for a coherent magnetic field
on a large scale in the Galactic interstellar medium 
%(Hiltner, 1949, Hall, 1949, Mathewson and Ford 1970), 
\citep{Hiltner_1949, Hall_1949, Mathewson_Ford_1970}
and it remains
a valuable tracer of the magnetic field configuration on various scales 
%(Heiles 2000).
\citep{Heiles_2000}.
The same large scale alignment of spinning, aspherical dust grains that causes
the starlight polarization causes polarized far-infrared emission
\citep{Houde_etal_2011}.  In the near infra-red, starlight polarization
%Houde, M., Rao, R., Vaillancourt, J.E., and Hildebrand, R.H. 2011, Ap. J. 733, issue 2, article id. 109.
allows the field configuration to be traced further into dark interstellar clouds 
%(Jones 2003, Clemens, Pavel, and Cashman 2012). 
\citep{Jones_2003, Clemens_etal_2012}.
Quantitative measurement
of the line-of-sight component of the magnetic field is possible with Zeeman
splitting observations of various spectral lines; the 21-cm line of atomic
hydrogen is the most widespread, and it provides opportunities to measure
the splitting either in absorption or in emission 
\citep[e.g.][]{Crutcher_etal_2010}.  One of the most
widespread tracers of the Galactic magnetic field is radio synchrotron
emission, which is linearly polarized due to the motion of the relativistic
electrons around the magnetic field lines.  The fairly strong and consistent
linear polarization of the Galactic diffuse emission at radio frequencies
was one of the first and most convincing arguments in favor of the synchrotron
emission process \citep{AlfvenHerlofsen1950}, reviewed by 
\citet{Ginzburg_Syrovatskii_1965}.
%Ginzburg, V.L. and Syrovatskii, S.I., 1965, Ann. Rev. Astr. Astroph. 3, 297.

At radio frequencies, linearly polarized emission propagating through an
ionized medium with a magnetic field that has a component along the line
of sight will show a rotation of the plane of polarization due to Faraday rotation 
(e.g. \citealt{Harwit_1973} chapter 6,  \citealt{Jokipii_Lerche_1969}).  
The position angle, $\chi$, of the polarization is defined in terms
of the Stokes parameters $Q$ and $U$, as
\[ \chi \ = \ \frac{1}{2} \ \arctan{(\frac{U}{Q})} \]
where the signs of both $U$ and $Q$ are used to determine $\chi$ over
the full $\pm\pi$ phase range.  The position angle changes with wavelength,
$\lambda$; for any value of $\lambda^2$ we can measure the derivative,
\[ RM \ = \ \frac{\mathrm{d} \chi}{\mathrm{d}( \lambda^2)} \]
in units of radians per meter squared.  This empirical definition allows
many different values of $RM$ to be present in a single complex spectrum 
of $Q + i U$ as a function of $\lambda^2$.

Many compact polarized sources show a single value of $RM$, that can be interpreted
as the effect of magnetised plasma along the line of sight from the source
at distance $d$ to the observer (at distance zero):
\begin{equation} \label{eq:FD_def}
RM \ = \ 0.81 \ \int_d^0 \ n_e \ \ \vec{B} \ \cdot \ d\vec{s} 
\end{equation}
If $n_e$, the electron density,  is in units of cm$^{-3}$, $B$, the
magnetic field,  is in $\mu$G, and $d$ is in parsecs, then
$RM$ is given by equation \ref{eq:FD_def} in  rad m$^{-2}$.  
The convention that $d\vec{s}$ points along the
line of sight {\it from the source to the observer} in
equation \ref{eq:FD_def} sets the convention that $RM$
is positive for $\vec{B}$ field pointing toward the observer.
%In more extended extragalactic sources, mapping the different Faraday components
%provides a way to determine the juxtaposition of the synchrotron emission and
%the magnetised thermal plasma (O'Sullivan et al. 2017, Kaczmarek et al. 2018).

Over the last decade, surveys of RMs of larger and larger samples of extragalactic 
continuum sources have been made, some concentrating
on low Galactic latitudes \citep[][figure 8]{Han_2017} 
and others covering all the sky available to the telescope
%(Stil, Taylor, \& Sunstrom 2011).
\citep{Stil_etal_2011}.
These have been combined by 
\citet{Oppermann_etal_2012, Oppermann_etal_2015}
into a grid of the best estimates for the Galactic
contribution to the RM in each cell on the sky.  Since the individual sources
have intrinsic RMs as well as the Galactic RM, the precision of the
estimate of the Galactic foreground depends on the density of
point sources.  Future surveys such as POSSUM \citep{Gaensler_2009} will greatly
improve the precision of maps like those of Oppermann et al.
Surveys of the RM of extragalactic radio sources show
large scale patterns at high 
\citep{Mao_etal_2010, Mao_etal_2012, Mao_etal_2018, Taylor_etal_2009} and
low latitudes \citep{Ordog_etal_2017}, 
somewhat similar to those seen in the starlight polarization.  
Rotation measure surveys of pulsars are particularly valuable, because 
the rotation measure divided by the
dispersion measure, DM = $\int_0^d \ n_e \ ds$, provides a measure
of the $\overline{B_{||}} = \frac{RM}{DM}$ averaged along the line of sight
%(Han et al., 2018a, Yao, Manchester \& Wang 2017).
\citep{Han_etal_2018a, Yao_etal_2017}.

\subsection{The Faraday Depth ($\phi$) Axis}

In contrast to the Rotation Measure, the Faraday Depth, $\phi$, is an
{\bf independent variable} with units of rad m$^{-2}$ over which we compute the distribution of
% Copy Editor:  please leave the \bf in the above line
linearly polarized brightness as the Faraday spectrum, 
the polarized intensity $F$, as a function of $\phi$,  
 \begin{equation}
\label{eq:Fourier}
F(\phi) \ = \ \frac{1}{\pi} \ \int_{-\infty}^{+\infty}\  P(\lambda^2) \ 
e^{-2i \phi \lambda^2} \ d(\lambda^2) 
\end{equation}
%(Burn 1966 eq. 11).
\citep[][eq. 11]{Burn_1966}.
%The goal of the GMIMS survey is to produce a Faraday cube. This
A broadband polarization survey of $Q$ and $U$ over a wide range of $\lambda$
can be transformed into a Faraday depth cube. This is
analogous to a spectral line cube for which the axes are two
sky coordinates and Doppler velocity (measured as frequency or
wavelength).  For linear polarization surveys, the third axis is not velocity but
Faraday depth, $\phi$. 
$F(\phi)$ is the Fourier conjugate function to $P(\lambda^2)$;
it is also complex, with real part Stokes $Q(\phi)$ and imaginary
part Stokes $U(\phi)$. 
The Faraday spectrum may be represented as polarized brightness temperature:
\[ T(\phi) \ = \ |F| \ = \ \sqrt{Q(\phi)^2 \ + \ U(\phi)^2} \]
where $Q$, $U$, $F$, $P$, and $T$ all have units of K since the diffuse emission is
calibrated as brightness temperature using the Rayleigh-Jeans approximation.  
The symbol $T_p$ is often used for the linearly polarized brightness temperature,
to distinguish it from the unpolarized emission; in this paper we do not discuss
the Stokes I or V parameters at all, so we can abbreviate $T_p$ by simply $T$.
As functions of $\phi$, or Faraday spectra, the true distributions of these
quantities are distorted by the resolving function or rotation measure spread
%14sep18 mh function, RMTF, that is determined by the limited range
function, RMSF, that is determined by the limited range
of wavelength-squared in the observations.  This distortion can be partially corrected by
deconvolution with the \verb|RM-CLEAN| algorithm 
\citep{Heald_2009} that changes the resolving function from
a messy dirty beam to a smoother clean beam that is chosen to be a Gaussian.
%(Brentjens and de Bruyn 2005, Heald, 2009, Purcell et al. 2018).  
The polarized brightness temperature 
in the cleaned spectrum then has units K per beam, where the beam is the
%14sep18 mh clean RMTF used in the deconvolution process.  For brevity we will use simply
clean RMSF used in the deconvolution process.  For brevity we will use simply
K units for $T(\phi)$.

It is only since the mid-2000s that the necessary parameters of a survey of
diffuse polarization have been 
understood.  This is because the requirements of bandwidth and 
resolution imposed by the Fourier relationship between
$F(\phi)$ and $P(\lambda^2)$, 
derived originally by \citet{Burn_1966}, were not widely appreciated until the
seminal paper by \citet{Brentjens_deBruyn_2005}.  
An ambitious international
collaboration to use large, single-dish radio telescopes with broad-band
spectro-polarimeters to determine the structure of the Galactic magneto-ionic medium
was begun in 2008, called GMIMS (the Galactic Magneto-Ionic Medium Survey, 
\citealt{Wolleben_etal_2009}).
GMIMS uses the variation of the strength of the Stokes $Q$ and $U$ components with
$\lambda^2$ through the Fourier transform to determine the distribution
of the polarized emission as a continuous function of $\phi$ \citep{deBruyn_Brentjens_2005}.  

Two of the GMIMS surveys have been
completed and the data are fully reduced and calibrated: the Dominion Radio Astrophysical 
Observatory (DRAO) Survey of the Northern sky (87$^o\ > \ \delta \ > \ -30^o$) at 
frequencies 1270 to 1750 MHz and the Parkes Survey of the Southern sky 
($-90^o\ < \ \delta \ < \ +20^o$) at frequencies 300 to 480 MHz \citep{Wolleben_etal_2018}.
The corresponding 
wavelength and rotation measure coverage are summarised on table \ref{tab:surveys}.  
The numbers on table \ref{tab:surveys} are computed using the full bandwidth used to construct the
Faraday cube.  In some directions some spectral channels were flagged due to interference.  This
flagging causes variation in the parameters on table 1 from place to place in the two Faraday cubes.  
The spectral cubes of the DRAO survey data used for this analysis were smoothed to 2$^o$ resolution.

\begin{deluxetable}{lrrcrr}[h]
\tablecolumns{7}
\tablecaption{Survey Details \label{tab:surveys}}
\tablehead{\colhead{survey} &
\multicolumn{2}{c}{Parkes} & &
\multicolumn{2}{c}{DRAO}\\
& \colhead{min} &  \colhead{max} & 
& \colhead{min} &  \colhead{max} \\
}
\startdata
declination range $\ \ \ \ $& -90$^o$ & +20$^o$ & $\ \ \ \ \ $ & -30$^o$ & +87$^o$ \\
angular resolution & 83.6\arcmin & 79.4\arcmin & & 40\arcmin & 30.5\arcmin \\
frequency range & 300.25 MHz   & 479.75 MHz & & 1270 MHz &   1750 MHz \\
$\lambda^2$ range & 0.391 m$^2$ & 1.0 m$^2$ & & 0.029 m$^2$ & 0.056 m$^2$  \\ 
$\Delta \lambda^2$ & 0.608 m$^2$ &&& 0.026 m$^2$ \\
%$\delta \lambda^2$ & 1.52$\cdot 10^{-3}$ m$^2$ &&& 6.2$\cdot 10^{-5}$ m$^2$ \\
$\delta \lambda^2$ & 3.32$\cdot 10^{-3}$ m$^2$ &&& 6.2$\cdot 10^{-5}$ m$^2$ \\
%RM resolution $\delta \phi$ & 5.7 rad m$^{-2}$ && &1.3$\cdot 10^2$ rad m$^{-2}$\\
RM resolution $\delta \phi$ & 6.2 rad m$^{-2}$ && &1.4$\cdot 10^2$ rad m$^{-2}$\\
%RM range $\phi_{max}$& 1.1$\cdot 10^3$ rad m$^{-2}$ &  &  & 2.8$\cdot 10^4$ rad m$^{-2}$\\
RM range $\phi_{max}$& 1.3$\cdot 10^3$ rad m$^{-2}$ &  &  & 3.1$\cdot 10^4$ rad m$^{-2}$\\
RM feature width $\phi_{max-scale}$ & 8.0 rad m$^{-2}$ & & & 1.1$\cdot$10$^2$ rad m$^{-2}$\\ 
Cleaned $\phi$ spectral range & -100 rad m$^{-2}$ & +100 rad m$^{-2}$ & & -400 rad m$^{-2}$ & +400 rad m$^{-2}$ \\ 
Faraday spectrum channel width & 0.5 rad m$^{-2}$ && &5 rad m$^{-2}$  \\
\enddata
\end{deluxetable}

\subsection{The Rotation Measure Spread Function \label{subsec:rmsf}}

If the spectrometer provides a bandwidth and channel separation translated
to wavelength squared that has some sensitivity function,
$W(\lambda^2)$ in the notation of \citet{Brentjens_deBruyn_2005}, 
then the resolving function in the RM dimension is 
the Fourier transform of $W$. This is the RMSF,
$R(\phi)$.  For a simple $W(\lambda^2)$ that is a 
top-hat (boxcar) function centred on $\lambda_c^2$ with width
$\Delta\lambda^2= \lambda_2^2 \ - \ \lambda_1^2$
then the corresponding $R$ is a sinc function with a phase wind:
\begin{equation} R(\phi) \ = \ e^{i(\phi \lambda_c^2)} 
\ \ \frac{\sin{(\phi \ \Delta\lambda^2)}}{\phi \ \Delta\lambda^2} 
\label{eq:sinc} \end{equation}
(illustrated in appendix \ref{app2}).
Note that $\Delta\lambda^2$ indicates $\Delta(\lambda^2)$ rather than
$(\Delta\lambda)^2$.
The width of a sinc($\theta$) function measured between half-power points
is $ \delta \theta \ = \ 3.79$, so
the resolution in $\phi$ of the survey is roughly the width of the 
main lobe of the function $R(\phi)$, which for the simple form
of equation (\ref{eq:sinc}) has full width to half maximum: 
\[\delta \phi = \frac{3.79}{\Delta\lambda^2}\]
Similarly, the {\bf maximum} rotation measure that can be detected is one that would give
% Copy Editor:  please leave the \bf in the above line
a drop of a factor of one half over a single step $\delta \lambda^2$ in 
the spectrum:
\[ \phi_{max} \ = \ \frac{1.9}{\delta\lambda^2} \]
As \citet{Schnitzeler_Lee_2015} explain, the upper limit $\phi$ is somewhat lower
than this depending on the computational approach taken to compute the Faraday
spectrum, i.e. the discrete form of equation \ref{eq:Fourier}.  Their equation
14 gives slightly lower values of $\phi_{max}$ of 9.84$\times10^2$ and 
2.92$\times10^4$ rad m$^{-2}$ for the Parkes and DRAO Surveys, respectively.

Since the $\phi$ axis of a Faraday spectrum is the Fourier conjugate of the $\lambda^2$ spectrum
derived from the spectrometer output, the relationship between the spectrometer sensitivity,
in $\lambda^2$ space, and the RMSF in $\phi$ space is similar to the relationship 
in aperture synthesis between the extent or coverage of observed baselines in $u,v$ space
and the beam, or resolving function in two dimensions on the plane of the sky.  
If there is a broad emission feature in Faraday space, the absence of the ``zero-spacing'' or
infinite frequency measurement means that the observed Faraday spectrum is high-pass filtered, 
so that the edges of the broad feature are enhanced, but the rest is attenuated nearly to
zero.  The broadest feature that is not attenuated in this way has width $\phi_{max-scale}$,
given by 
\[ \phi_{max-scale} \ \simeq \ \frac{\pi}{\lambda_1^2} \]
An additional complication is the spectral index of the synchrotron emission,
which generally has a power law with polarized brightness temperature 
$T(\nu) \propto \nu^{-\beta}$.  This can lead to enhanced sidelobes in the
un-cleaned Faraday spectrum \citep[][figure 1]{Schnitzeler_2018}.  

The aim of the GMIMS surveys is to make $\delta\phi$ {\bf less} than $\phi_{max-scale}$
% Copy Editor:  please leave the \bf in the above line
{\bf for the first time} {\bf at frequencies above 250 MHz in the Milky Way}.  The weakness
of polarization surveys taken with narrow-band
% Copy Editor:  please leave the \bf in the above line
receivers is that the RMSF function is broader than the maximum detectable scale
in $\phi$.  This happens whenever the bandwidth, $\Delta\lambda^2$ is less than the 
minimum wavelength squared, $\lambda_1^2$.
The result is that even a relatively simple $\phi$ spectrum is converted into a messy
function, see examples in appendix \ref{app2} and other examples in appendix 2 of  
\citet{Brentjens_deBruyn_2005}.
As table \ref{tab:surveys} shows, the Parkes survey has $\frac{\Delta\lambda^2}{\lambda_1^2}$
about 1.6, which is quite safe.  For the DRAO survey the value is $\sim0.9$, so the RMSF
is marginally affected by missing large scale Fourier components.  Features in the Faraday
spectra that are much wider than $\phi_{max-scale}$ will still be hollowed-out, i.e.
edge-filtered by the RMSF (see appendix \ref{app2}).   
% Alex' version:
Surveys with the LOFAR and MWA telescopes at low frequencies ($\nu < 25$ MHz) have achieved $\delta \phi < \phi_{max-scale}$ over several degree-square areas 
%with relatively small phi_max-scale (REFS) 
providing a rich set of resolved features in the Faraday spectrum
\citep{Iacobelli_etal_2013, Jelic_etal_2014, Jelic_etal_2015, Lenc_etal_2016, vanEck_etal_2017}.

%Surveys with the LOFAR and MWA telescopes at low frequencies ($\nu <$250 MHz) have achieved $\delta\phi < \phi_{max-scale}$ 
%\citep{Iacobelli_etal_2013, Jelic_etal_2014, Jelic_etal_2015, Lenc_etal_2016, vanEck_etal_2017}
%providing resolved features in the Faraday spectrum similar to those in the Parkes survey discussed below.}

In this paper we study the Faraday cubes of the two GMIMS surveys by computing the
moments of the emission spectra and comparing them with other RM tracers. 
This is the first application of spectral moment techniques to the
study of the diffuse polarized emission from the Galaxy.  
Note that a different set of parameters, also called Faraday moments,
is proposed by \citet{Farnes_etal_2018} as statistical parameters to develop an optimal
detection strategy for finding sources of polarized emission in the
presence of radiometer noise.  These are computed directly from $Q(\lambda)$ and
$U(\lambda)$ for efficiency in searching large survey data sets.

Representative spectra from the Faraday cubes of the two surveys
are presented in section \ref{sec:data}, the method of calculating
the moments is discussed, and
the zeroth, first, and second moments are shown for the full areas of the two surveys.
These are two-dimensional representations of the survey data that can be easily
compared with other RM data in section \ref{sec:comparison}.  In particular,
comparison with RMs of nearby pulsars with known distances provides a distance estimate for
the polarized emission in the DRAO survey, but not for the 
Parkes data, as discussed in section \ref{sec:distances}.  The very different skies seen in
the two surveys can be explained as the result of
the polarization horizon, i.e. 
the limit to the distance from which polarized emission can reach us,
determined by depolarization processes
%the furthest distance that we can see due to depolarization
\citep{Uyaniker_etal_2003}, with the result that they sample quite different
volumes, as discussed in section \ref{sec:depol}.

\section{The Survey Data \label{sec:data} }

\subsection{All Sky Averages}

\begin{figure}[h]
\hspace{.5in}\includegraphics[width=5.5in]{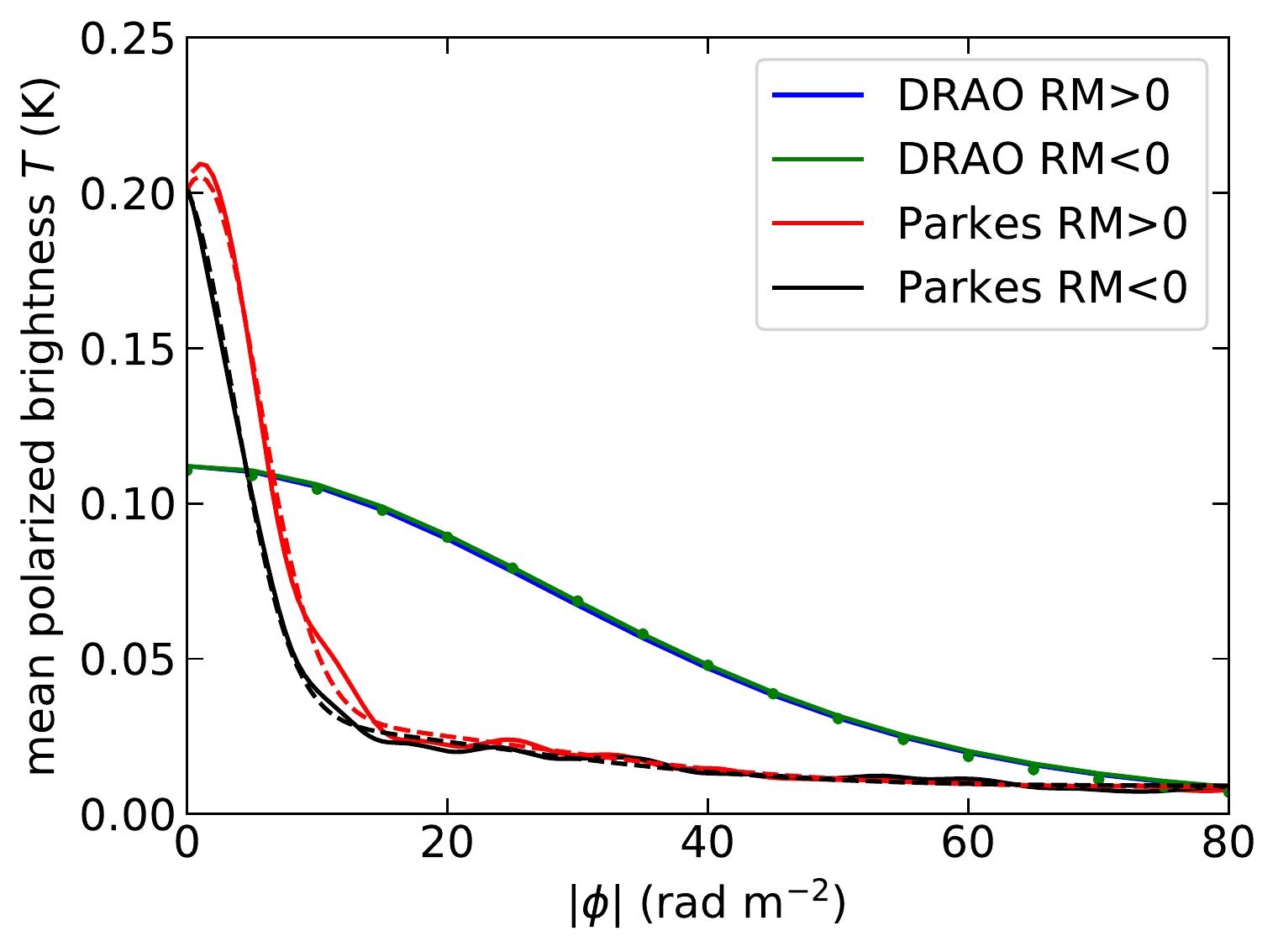}
\caption{The average linearly polarized brightness, $T(\phi)$, over the area of each survey, computed separately
for each plane of the Faraday cube. The fitted Gaussians are indicated by the green dots (DRAO)
and the red and black dashes (Parkes) using the parameters on table \ref{tab:Gaussians}.  The
DRAO fit is a single Gaussian, the Parkes fit is the sum of two Gaussians, each of the form
in equation \ref{eq:Gaussian_eq}.
The error in $T(\phi)$ is dominated by fluctuations introduced by the limited wavelength coverage,
and at low latitudes by leakage of Stokes I into Stokes Q and U. 
In this figure, the residuals about the best fit Gaussian in the Parkes data are $\sim 4\cdot 10^{-3}$ K,
in the DRAO data they are $\sim 1 \cdot 10^{-3}$ K.
\label{fig:skyav} }
\end{figure}

The Parkes and DRAO surveys are very complementary in several ways.  The DRAO telescopes in
British Columbia 
can observe the entire northern sky, and in the South down to $\delta \simeq -30^o$, and
the Parkes telescope in New South Wales can observe the entire southern sky, and in the North
as high as $\delta \simeq +20^o$, thus there is an overlap band of width about $50^o$. 
Since the ranges of $\lambda^2$ are so different, the
RMSFs of the two surveys are very different also
(Table \ref{tab:surveys}).  Most important, the synchrotron emission has 
spectral index $\beta \sim -2.75$, 
%which applies to linear polarization as well as total brightness, 
so the much lower frequencies of the Parkes
survey see brighter emission.  That emission is spread over a much narrower
range of $\phi$ than for the higher frequencies of the DRAO survey.  This is shown
in figure \ref{fig:skyav}, which plots the mean brightness temperatures of
the polarized intensity of the two surveys as functions of $\phi$,
averaged over the entire survey areas.  

The x-axis of figure \ref{fig:skyav} is $|\phi|$, to make the symmetry between
the positive and negative values of $\phi$ clear, although the fitting
was done for the full range.   The y-axis plots the mean of
$T(\phi)$ over the full area of each survey.  The DRAO survey does not resolve the 
structure of the emission in Faraday depth when averaged over the full area, but
when individual Faraday spectra are measured, or Faraday cubes for small regions, then
structure appears, as shown in section \ref{subsec:spectra} below.
The DRAO survey average profile is very well fit by a Gaussian 
as:
\begin{equation} \label{eq:Gaussian_eq}
{\Large 
T(\phi)\ = \ T_o \ e^{\left(-\ \frac{(\phi - \phi_o)^2}{2 \sigma_{\phi}^2}\right)} }
\noindent
\end{equation}
Least-squares fitted values of the Gaussian parameters are given on table \ref{tab:Gaussians}.
The width of the DRAO Faraday spectrum is artificially made smaller than the 
nominal resolution of the survey (table \ref{tab:surveys}) because in the Faraday cleaning
step of the data reduction the ``clean beam'' or restoring function was set as a 
Gaussian of width 60 rad m$^{-2}$.  The RMSF of the Parkes survey is much narrower, and
it allows resolution of two Gaussian components in the survey average Faraday spectrum, one with half-width
$\sigma_{\phi}$ = 4.5 rad m$^{-2}$ and the second fainter but much broader with
$\sigma_{\phi}$ = 23.5 rad m$^{-2}$ (Table \ref{tab:Gaussians}).
The polarized brightness measured in these two surveys has not been de-biased to reduce the
contribution of noise to $T(\phi)$; there is a non-zero baseline that is fitted along
with the Gaussian parameters (fifth column, table \ref{tab:Gaussians}).

\begin{deluxetable}{lrrrr}
\tablecolumns{5}
\tablecaption{ Survey Mean $T(\phi)$ Gaussian Fits \label{tab:Gaussians}}
\tablehead{\colhead{survey} & \colhead{$\phi_o$} &  \colhead{$\sigma_{\phi}$} & \colhead{$T_o$} & \colhead{baseline} }
\startdata
DRAO & -0.3 rad m$^{-2}$ & 30.3 rad m$^{-2}$ & 0.11 K & 0.004 K\\
Parkes 1 & +1.0 rad m$^{-2}$ & 4.5 rad m$^{-2}$ & 0.17 K & 0.008 K\\
Parkes 2 & +1.7 rad m$^{-2}$ & 23.5 rad m$^{-2}$ & 0.02 K & \\
\enddata
\end{deluxetable}
%# Maik's Cube: Parkes & +1.4 rad m$^{-2}$ & 4.6 rad m$^{-2}$ & 0.22 K & 0.018 \\

\subsection{Sample Faraday Spectra \label{subsec:spectra}}

\begin{figure}[h]
\hspace{.1in}\includegraphics[width=3.5in]{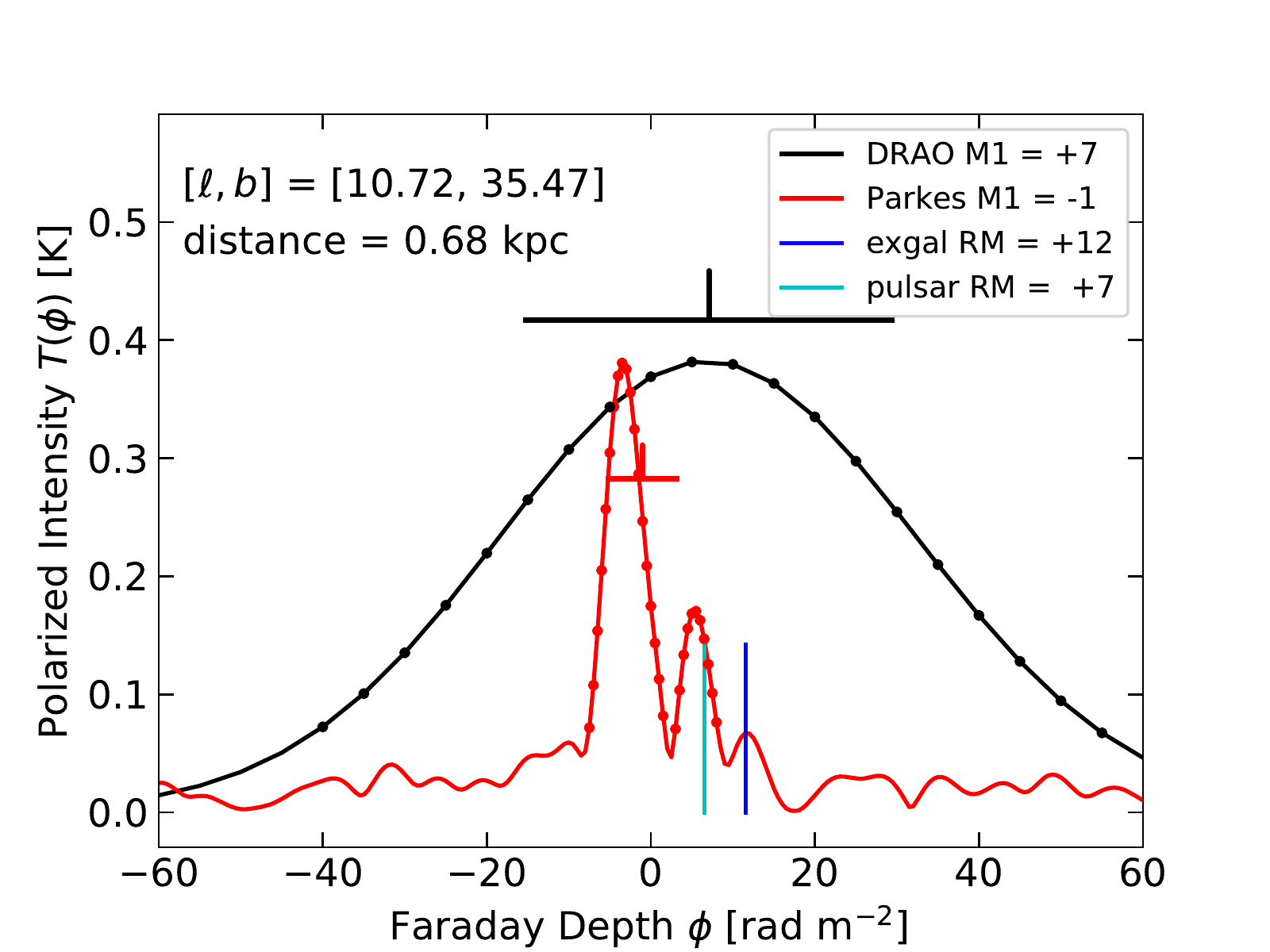}
\hspace{.1in}\includegraphics[width=3.5in]{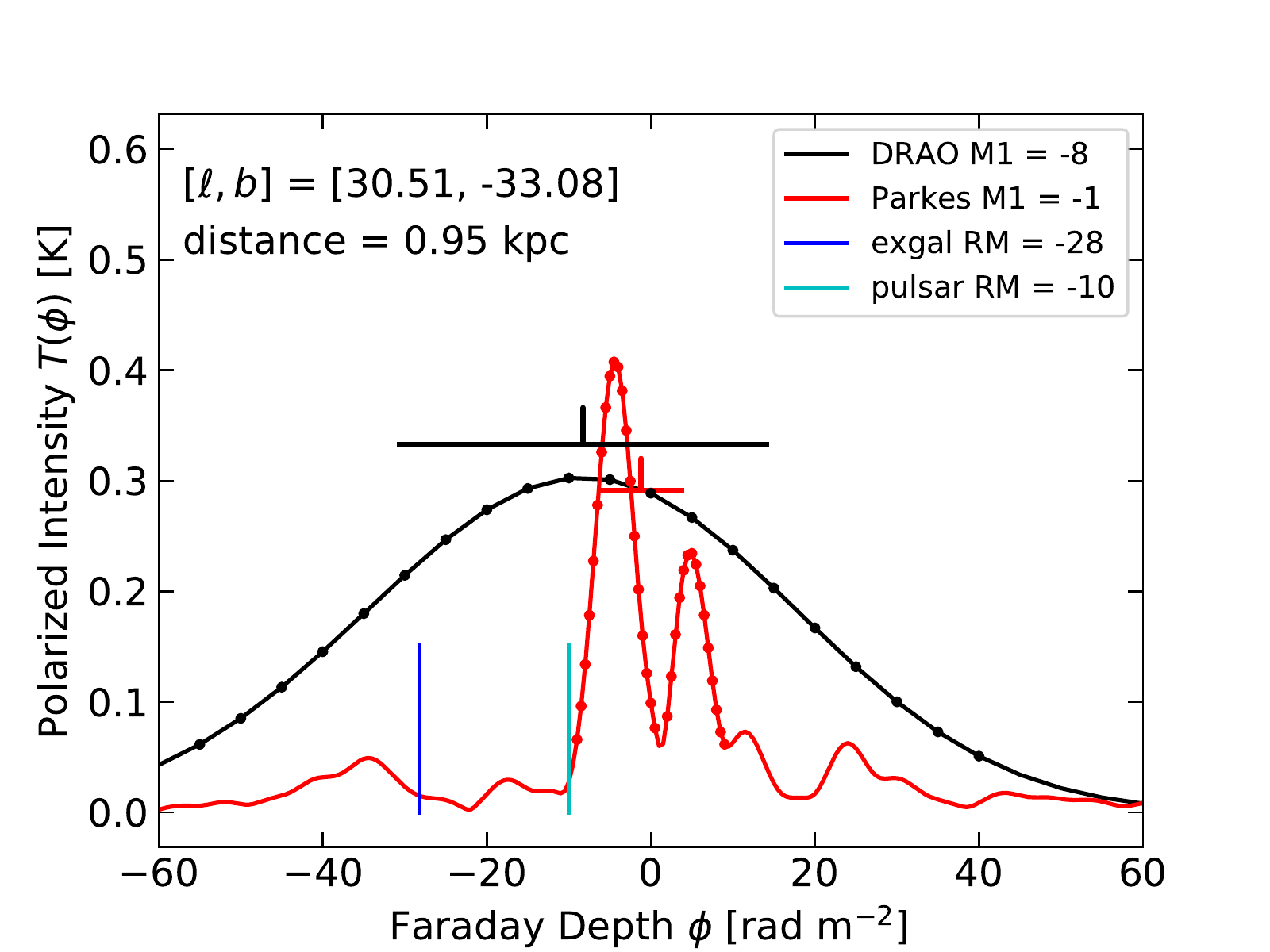}

\caption{Two pairs of Faraday spectra in the directions of pulsars
J1607-0032 (left) and J2048-1617 (right).  The x-axis shows Faraday depth, $\phi$,
and the y-axis shows polarized intensity, T, in K.  The black curves show 
the DRAO spectra, the red curves show the Parkes spectra.  The black and
red horizontal bars show the first and second moments of the spectra 
(section \ref{sec:moments} below).  The blue and cyan markers indicate
the rotation measures of the extragalactic foreground and the 
pulsar, respectively (section \ref{sec:comparison} below).
The dots on the spectra indicate channels
above a threshold set at the greater of 15\% of the peak value, or a minumum of 0.04 K 
%(for the DRAO survey)  or 0.04 K for the Parkes survey 
(see section \ref{sec:thresh}).  
The DRAO spectra generally show only one spectral feature, while the Parkes spectra often show two
or more features.
%possibly as a result of the missing short wavelengths in the Parkes survey (discussed in appendix \ref{app2}).
}

\label{fig:spectra_1}
\end{figure}

\begin{figure}[h]
\hspace{.1in}\includegraphics[width=3.5in]{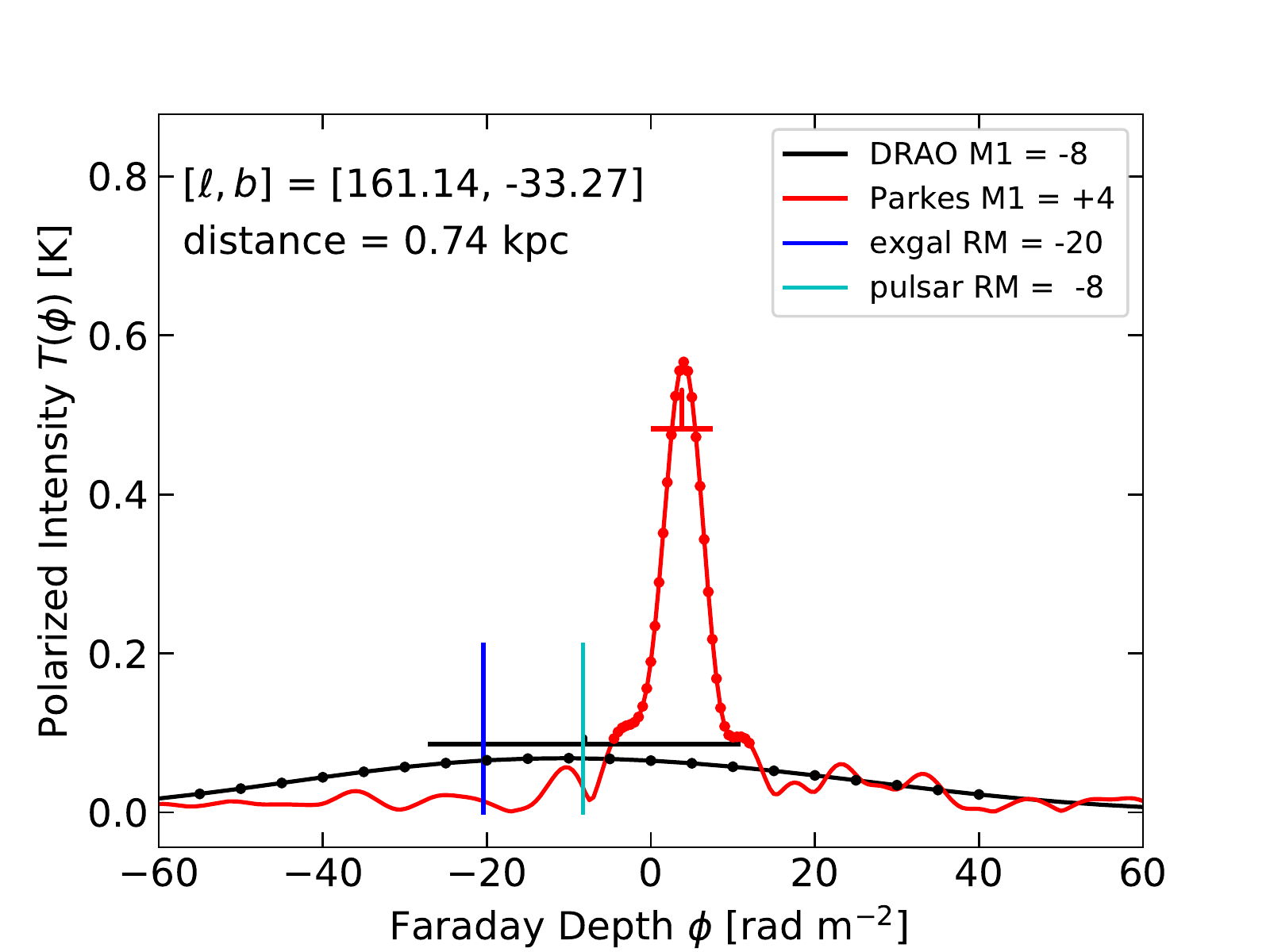}
\hspace{.1in}\includegraphics[width=3.5in]{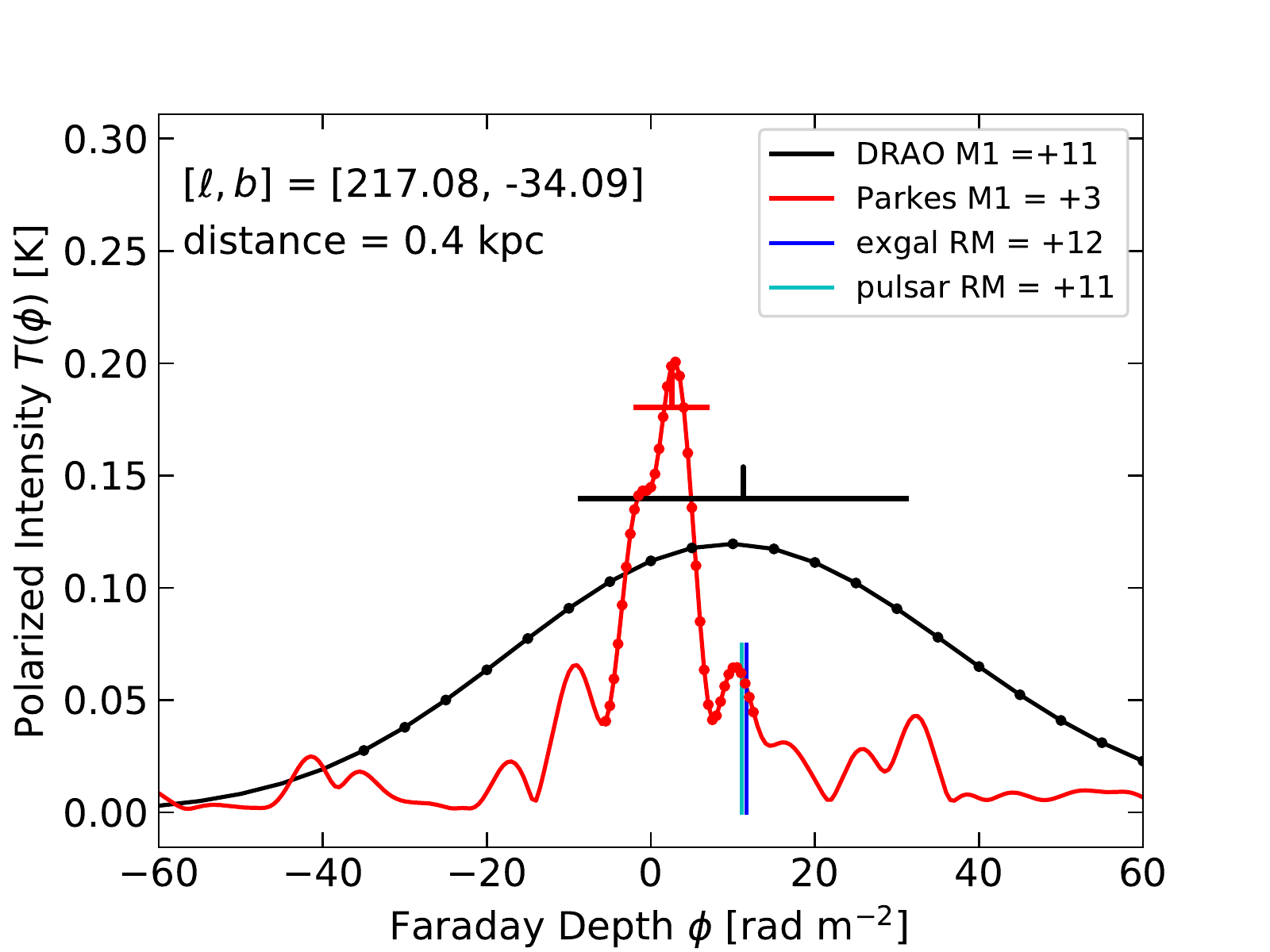}

\caption{
Faraday spectra in the directions of pulsars J0304+1932 (left) and
J0452-1759 (right).  
The colors and symbols are the same as in figure \ref{fig:spectra_1}.
}
\label{fig:spectra_2}
\end{figure}

\begin{figure}[h]
\hspace{.1in}\includegraphics[width=3.5in]{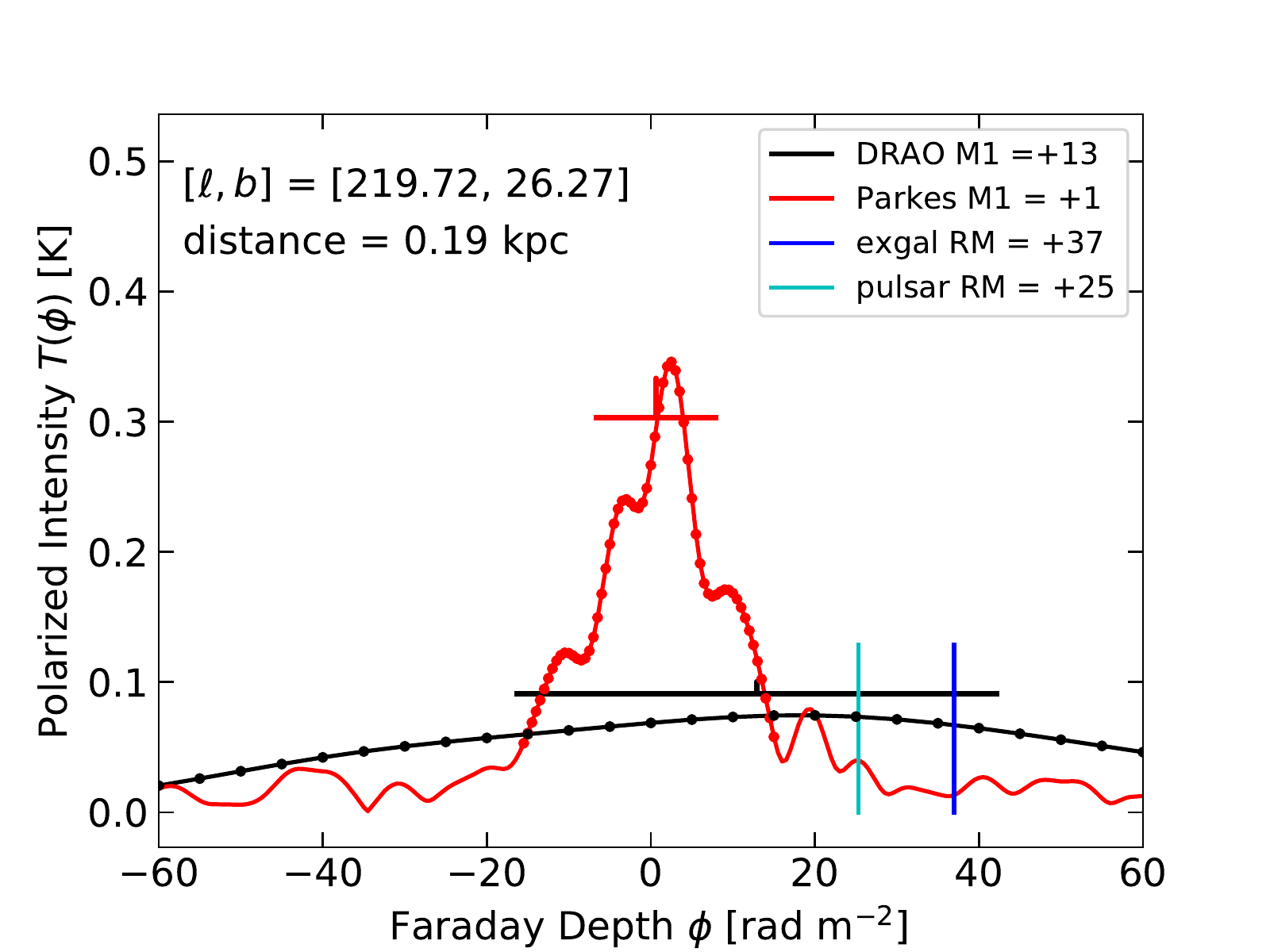}
\hspace{.1in}\includegraphics[width=3.5in]{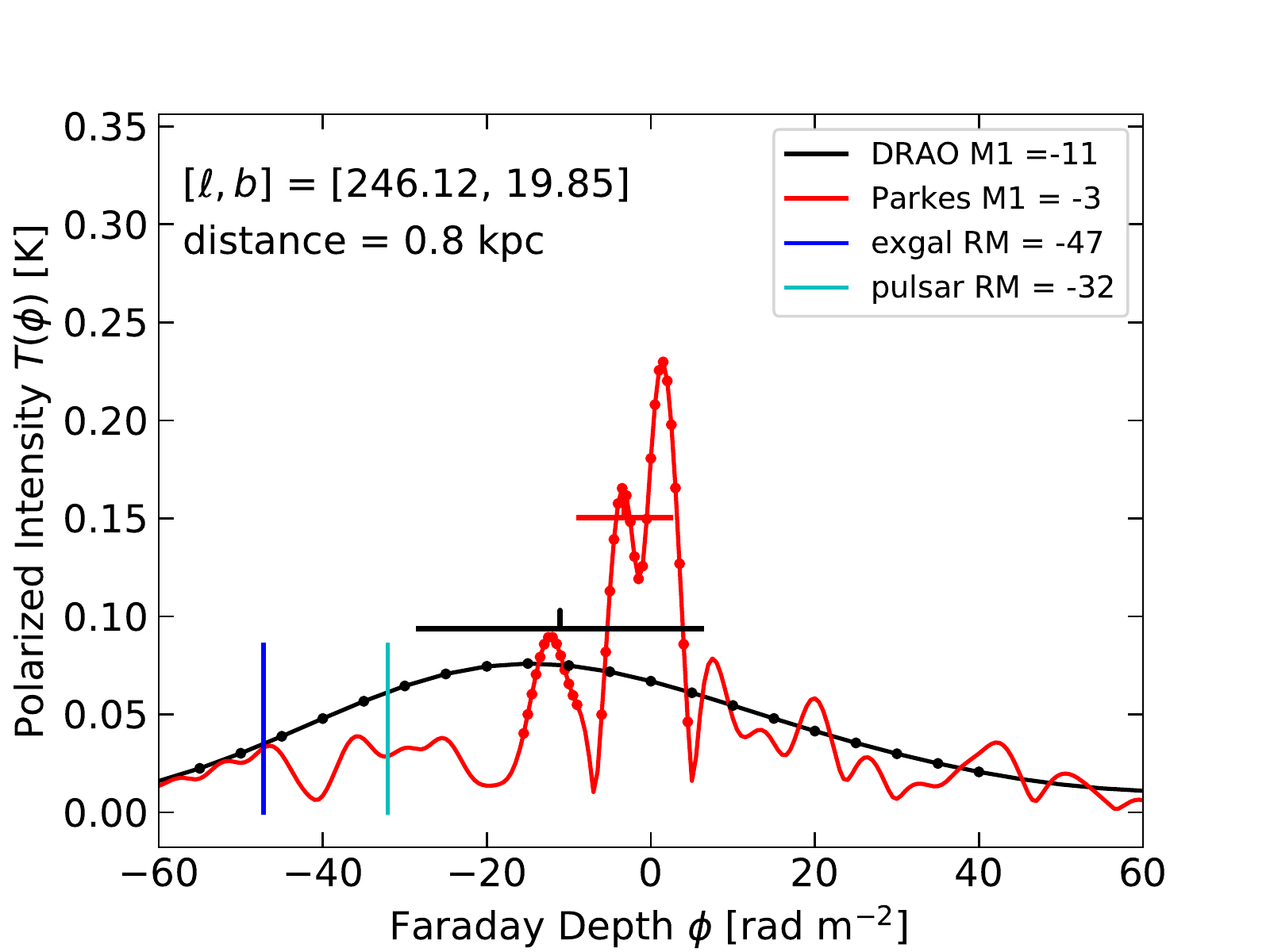}

\caption{
Faraday survey spectra in the direction of pulsars J0837+0610 (left)
and J0908-1739 (right).
The colors and symbols are the same as in figure \ref{fig:spectra_1}.
}
\label{fig:spectra_3}
\end{figure}

Figures \ref{fig:spectra_1} - \ref{fig:spectra_3} show six example rotation measure spectra.  The first two (fig. \ref{fig:spectra_1}) are in the first quadrant, at longitudes $\ell \sim 11^o$ and 31$^o$, the rest 
are in the outer galaxy.  All are at intermediate latitudes (here meaning roughly $15^o < |b| < 40^o$), the first
four at $|b| \sim $ 33$^o$ to 35$^o$, the last two at $|b|\sim 26^o$ and
$\sim20^o$.  These directions are all in the overlap region covered by both
the Parkes and DRAO surveys.  They are in the directions of pulsars
with distances less than one kiloparsec,
and with measured values of $RM$ as discussed below in
section \ref{sec:pulsar}.  For comparison, Faraday spectra
at lower latitudes ($b=10.6^o$) have been studied in detail by \citet[][figure 6]{vanEck_etal_2017} 
with better spatial resolution and excellent RMSF cleaning.

The effect of smoothing in $\phi$ in the DRAO spectra is clear on figures
\ref{fig:spectra_1} - \ref{fig:spectra_3}.  The
Parkes spectra have much higher resolution in $\phi$.  But the two
spectra are not consistent with each other even after accounting for the
different resolutions.  This is because of the very different wavelength ranges;
the path lengths sampled by the two spectra are therefore very different, with the
shorter wavelengths sensitive to much greater distances due to depolarization,
discussed in section \ref{sec:depol} below.
In some cases, such as those shown on figures \ref{fig:spectra_1} and \ref{fig:spectra_2}, the peak
of the DRAO feature corresponds well with the pulsar RM.  This is not always the case, as discussed
in section \ref{sec:pulsar} below.

%The dots on the spectra in figures \ref{fig:spectra_1} - \ref{fig:spectra_3} indicate channels
%above a threshold set at the greater of 15\% of the peak value, or a minumum of 0.03 K 
%(for the DRAO survey)  or 0.08 K for the Parkes survey (see below, section \ref{sec:thresh}).  
%The DRAO spectra generally show only one spectral feature, while the Parkes spectra often show two features
%well above the threshold.  For the calculation of the spectral moments (section \ref{sec:moments}
%below, we use only spectral channels contiguous with the first or second strongest feature, down to the threshold. 
Some of the weaker features in the Parkes spectra are very likely real, but determining
the dynamic range of the Faraday spectrum, i.e. the ratio of the brightest spurious feature to the
peak of the brightest feature, will require more careful analysis of both Faraday cubes (Thomson et al.
2018 in preparation, Ordog et al. 2018 in preparation).

Figures \ref{fig:spectra_1} - \ref{fig:spectra_3} show only the middle channels ($-60 < \phi < +60$ rad m$^{-2}$)
of the Faraday cube.  The Gaussian features in the DRAO spectra extend to at least $\pm 100$ rad m$^{-2}$, and
the Parkes spectra show some features outside this rotation measure range as well.   The full $\phi$ ranges of the cleaned 
Faraday cubes are $\pm 100$ and $\pm 400$ rad m$^{-2}$ for the Parkes and DRAO surveys,
respectively (table \ref{tab:surveys}).

\subsection{Faraday Moments \label{sec:moments}}

\begin{figure}[h]
\hspace{.5in}\includegraphics[width=5.5in]{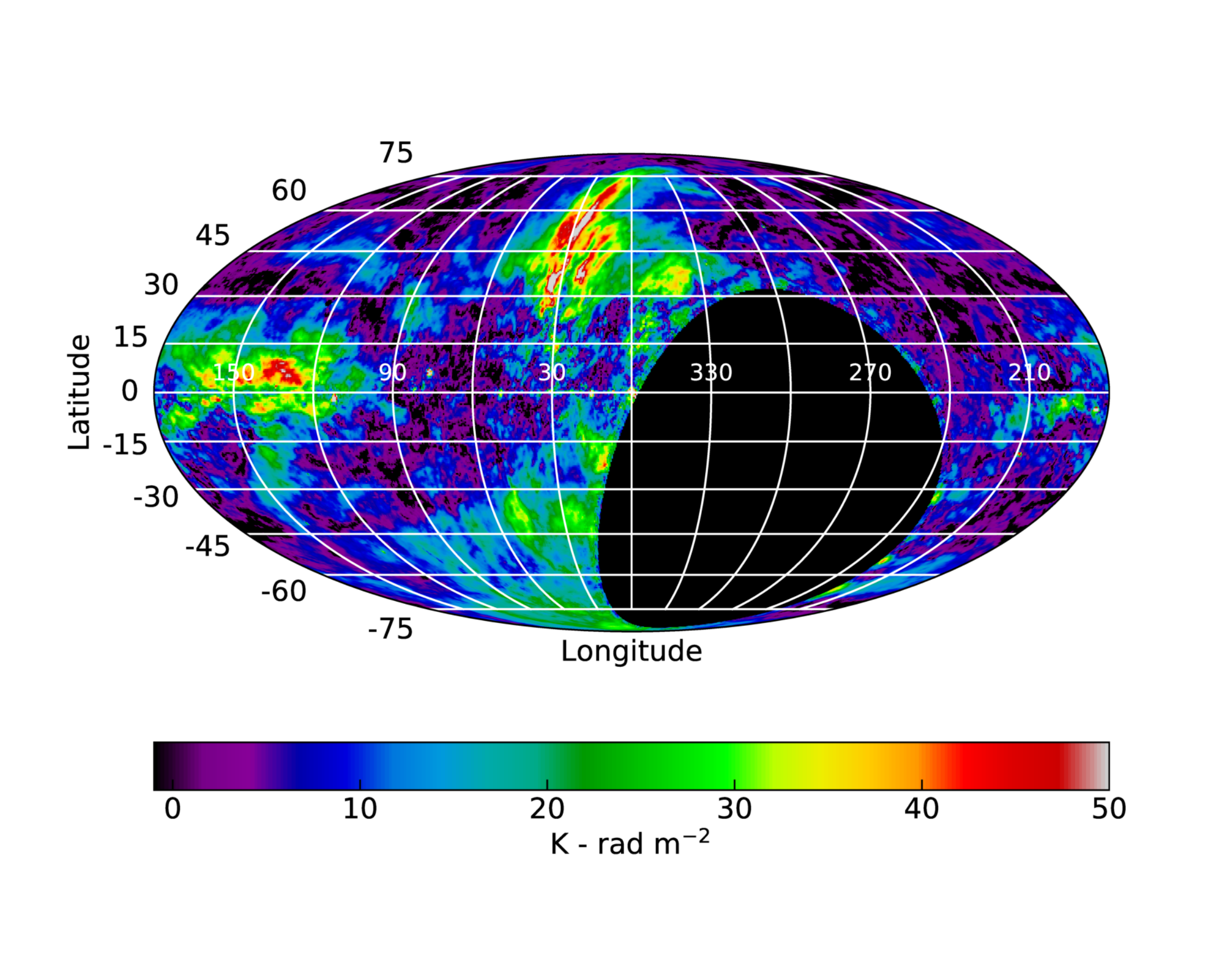}
%\vspace*{2.5in}
\vspace{-.35in}

%\vspace*{2.5in}
\hspace{.5in}\includegraphics[width=5.5in]{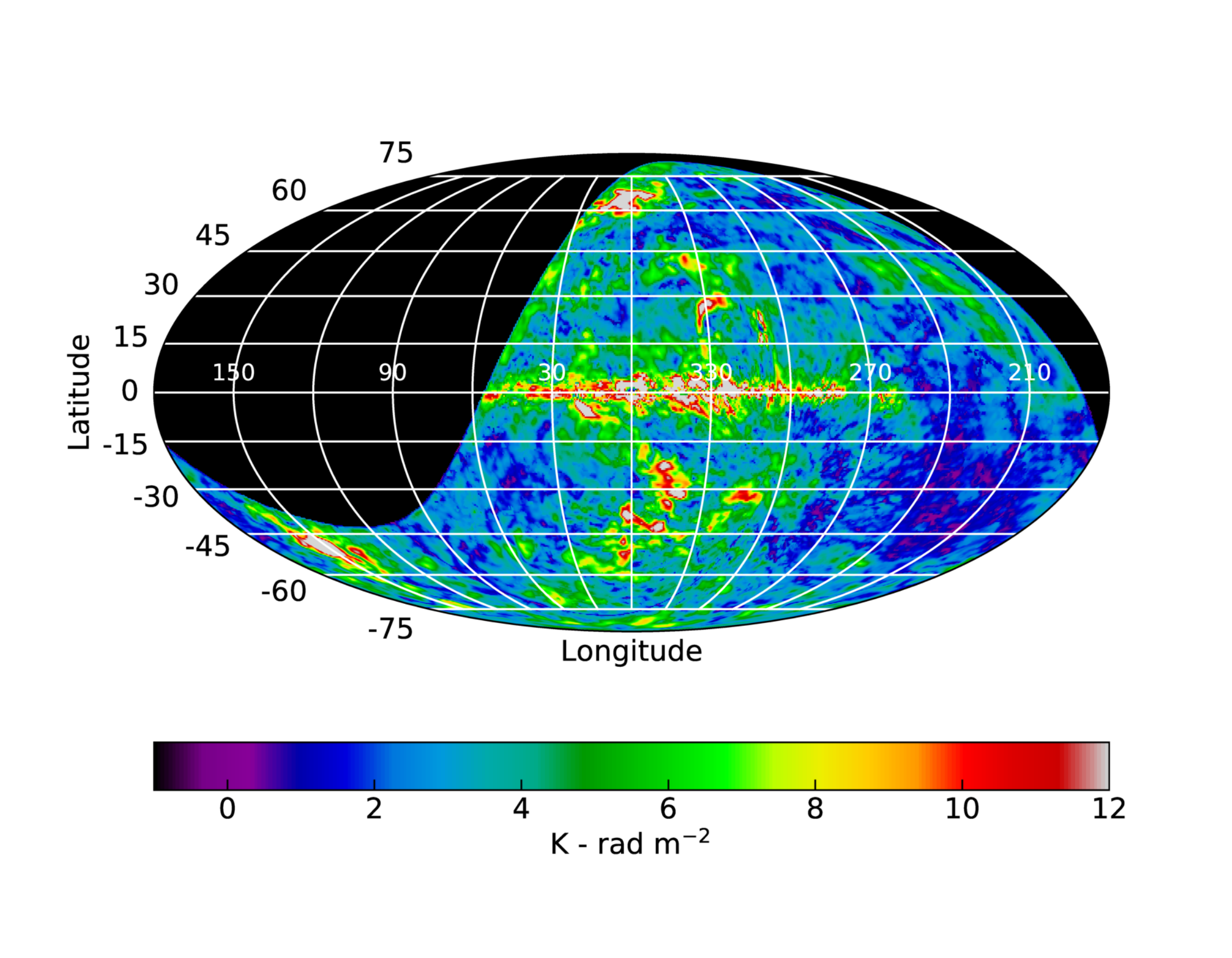}
\caption{The zero moments ($M_0$) of the Faraday cubes of the DRAO (upper) and Parkes (lower) surveys,
shown in Galactic coordinates and a Mollweide projection.  The black areas are
either outside the declination limits of the surveys,
or positions where the Faraday spectra do not show any features above the minimum threshold of 0.04 K.
%0.03 K (DRAO survey) or 0.08 K (Parkes survey).
The units are K rad m$^{-2}$ from equation \ref{eq:mom0}.
\label{fig:mom0} }
\end{figure}
%  xxx rechecked through here 11 oct 18

The distribution of polarized brightness on the sky shows interesting structures on a range of
angular scales.  
Distinct structures in the GMIMS surveys have been studied individually 
%(Wolleben et al. 2010, Sun et al. 2015, Hill et al. 2017, Thomson et al. 2018b),
\citep{Wolleben_etal_2010, Sun_etal_2015, Hill_etal_2017, Thomson_etal_2018b}
but the purpose of
this paper is to study the properties of the 
entire sky in polarized emission, rather than individual objects. 
To study the properties of the Faraday cube over a large area, the spectral moments are useful tools.
%For this purpose, a useful representation is the moments of the Faraday cube.
These are analogous to moments in velocity space for a spectral line cube.  The zero moment, $M_0$ is defined as:
\begin{equation} \label{eq:mom0}
M_0 \ \equiv \ \sum\limits_{i=1}^{n} \ T_i \ d \phi
\end{equation}
with units K rad m$^{-2}$, where $d \phi$ is the width of each of the $n$ channels of
the Faraday spectrum contributing to the sum.  The first moment, $M_1$, is defined as
\begin{equation} \label{eq:mom1}
M_1 \ \equiv \  \frac{\sum\limits_{i=1}^n \ T_i \cdot \phi_i}{\sum\limits_{i=1}^n  T_i}
\end{equation}
with units  rad m$^{-2}$.  The second moment, $M_2$, is defined as:
\begin{equation} \label{eq:mom2}
M_2 \ \equiv \ \frac{\sum\limits_{i=1}^n \ T_i \cdot (\phi_i - M_1)^2}{\sum\limits_{i=1}^n  T_i}
\end{equation}
with units (rad m$^{-2})^2$.
The sums are taken over the channels of the $\phi$ spectrum, or selected ranges of
channels where the signal is well above the noise, and $T_i$
is the polarized intensity, $T(\phi_i)$, in brightness temperature units.  For a 
continuous distribution, $T(\phi)$, the moments are integrals, $M_0 = \int\limits_{-\infty}^{\infty} T(\phi) \ d\phi$, $M_1 \ = \ \frac{\int\limits_{-\infty}^{\infty} T(\phi) \cdot \phi \ d\phi}{M_0} $, and
$M_2 \ = \ \frac{\int\limits_{-\infty}^{\infty} T(\phi) \cdot (\phi - M_1)^2 \ d\phi}{M_0}$.  For a single Gaussian
spectral feature with no noise, the moments correspond to $M_0=\sqrt{2 \pi}\ T_o \ \sigma_{\phi}$,
$M_1= \phi_o$, and $M_2 = \sigma_{\phi}^2$.  To simplify
comparison between the 
moments, we compute the square root of the second moment,
$m_2 = \sqrt{M_2}$; all plots involving second moments in this paper use $m_2$ for the second moment,
with dimension rad m$^{-2}$.  Note that $M_2$ is the second {\bf central} moment, because it is taken about the mean, $M_1$.
% Copy Editor:  please leave the \bf in the above line
The effect of taking the moments is to reduce the Faraday cube to a series of images, having just the two 
angular dimensions of the survey, but with the images representing the distribution of brightness over the 
third dimension, $\phi$.  Simpler alternatives to the spectral moments are discussed in appendix \ref{app1}.

For an intuitive understanding, the zero moment is the total polarized brightness integrated over the full range
of $\phi$, the first moment is the intensity weighted mean of $\phi$, and the square root of the second
moment, $m_2$ is the half-width of the brightness distribution along the $\phi$ axis.  Neither the peak $T(\phi)$ 
nor the value of $\phi$ at the peak are measured by the moments, although $T_o$ can be estimated
assuming a Gaussian
or other functional form for the line shape.  The red and black bars on figures \ref{fig:spectra_1} - \ref{fig:spectra_3}
are placed at the height of an equivalent Gaussian profile with the same $M_0$ and $m_2$ values, which is
$T_{peak} = \frac{M_0}{\sqrt{2 \pi} \  m_2}$.

\subsection{Thresholding \label{sec:thresh}}

   Because of the weighting by $\phi_i$ and $(\phi_i - M_1)^2$ in equations \ref{eq:mom1} and
\ref{eq:mom2}, the first and second moments are strongly affected by noise or spurious features in
the spectra at high positive and negative values of $\phi$.  Since $T_i$ is positive definite,
this is an even worse problem for computing the moments of Faraday spectra than it is for more
familiar velocity spectra, that are usually dominated by Gaussian noise.  In most directions in
both of the surveys considered here, the noise is primarily from residuals left by the Faraday deconvolution
process.  To mitigate the effect of spurious emission at high positive and negative values of $\phi$,
we use a threshold to restrict the range of channels contributing to the sums in equations 
\ref{eq:mom0}, \ref{eq:mom1}, and \ref{eq:mom2}.  

For each pixel in the cube, the thresholds are set at the larger of 
either 15\% of the peak of the emission spectrum in that pixel, or a minimum set at 0.04 K.
%0.03 K for the DRAO survey and 0.08 K for the Parkes survey.
%{\bf or} $T_p \times f$, where $T_p$ is the 
%highest value of the emission in the spectrum at that pixel and $f = 0.25$.
% and $f=0.25$ for the DRAO cube, and $f=0.25$ for the Parkes cube.
Reducing the 15\% threshold causes little change in the zero and first moment maps, but the 
second moment map becomes less smooth and has small scale structure that does not seem to be
real based on the spectra themselves.
%Galactic plane stands out more clearly in the second moment.
%This suggests that leakage of Stokes I is beginning to affect the Faraday spectra for $f < 0.25$ at low latitudes.
%The statistical analysis below gives nearly identical results using 15\% vs. 25\%.
Similarly, reducing the minimum thresholds below 0.04 
%and 0.08 K 
appears to introduce noise in the second moment results in areas of low $M_0$.

Channels on either side of the peak are included in the moment calculation until the spectrum drops
below the threshold.  For the DRAO data, only those channels are used.  In some directions, the Parkes
spectra show two separate features well above threshold, so we extend the range of channels by 
fitting a Gaussian to the first feature, then subtracting it from the data and finding the next peak.
If the height of that peak is more than two times the threshold, then we find the range of channels
for which $T_i$ is above the threshold again.  These supplement the channels already selected 
(from the first peak), and they together make up the channel ranges $i = 1 \dots n$ in equations 
\ref{eq:mom0}, \ref{eq:mom1}, and \ref{eq:mom2}.  Although this thresholding clearly biases the
resulting moments against emission in faint features well separated from the dominant peaks,
the moments that result are very consistent with the values of the integral, center, and width
of the best fit Gaussians to each spectrum (see appendix \ref{app1}).  Removing the
threshold entirely gives very similar results
for $M_0$, but the results for $M_1$ and $M_2$ jump discontinuously from one pixel to another in
some areas.

\subsection{Moment Maps}

The zero moment maps for the Parkes and DRAO surveys are shown in figure \ref{fig:mom0}. 
Features in $M_0$ for the DRAO survey have good correspondence with known 
structures, particularly the North Polar Spur (NPS) that reaches from latitude $b \sim 25^o$ at
longitude $\ell \sim 45^o$ to near the north Galactic pole at $b \sim 75^o$ where it
arches over to $\ell \sim 320^o$, see \citet{Sun_etal_2015} and references therein and \citet{Wolleben_etal_2010}.
Another bright structure in the DRAO $M_0$ map
is the Fan region near the Galactic plane ($b \sim 0^o$) at longitudes $110^o < \ell < 160^o$
\citep{Hill_etal_2017}.  The angular scale of the brightness variations is 
larger (smoother) at high latitudes and smaller near the Galactic plane.
There are also some residual effects of the survey scanning pattern that surround the
empty region south of the DRAO declination limit ($\delta = -30^o$) in the lower right.

In the Parkes $M_0$ map much of the NPS and all of the Fan Region are north of the declination
limit ($\delta = +20^o$), although there is a hint of a feature aligned with the NPS
near $\ell = 0^o$ and $+60^o < b < +75^o$.  In general there is very little correspondence
between bright regions in the two zeroth moment maps.  The Galactic Plane stands out on 
both, but differently.   The plane appears bright in the Parkes map due to leakage of Stokes I into
the Stokes $Q$ and $U$ beams.  It is dark on the DRAO map, in part because the leakage has been estimated
and subtracted using the low latitudes for calibration ($|b|< 2^o$).
 In the Parkes map there is less of a change in angular scale
between high, intermediate, and low latitudes.
The lack of correspondence between structures even in the region of overlap between the
two surveys (-30$^o < \delta < +20^o$)
suggests that they are sampling different physical volumes.

\begin{figure}[h]
%\vspace*{2.5in}
\hspace{.5in}\includegraphics[width=5.5in]{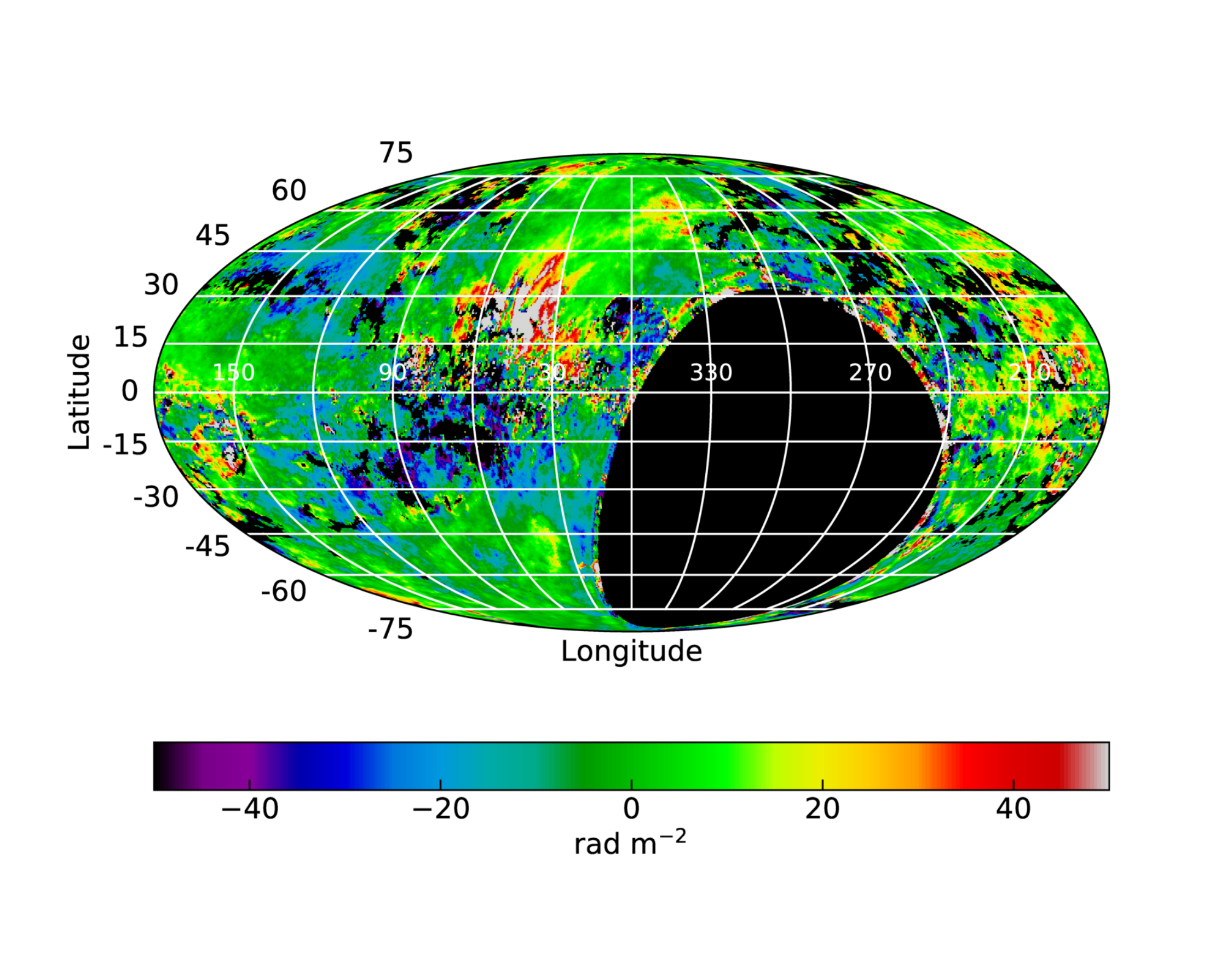}
\vspace{-.35in}

%\vspace*{2.5in}
\hspace{.5in}\includegraphics[width=5.5in]{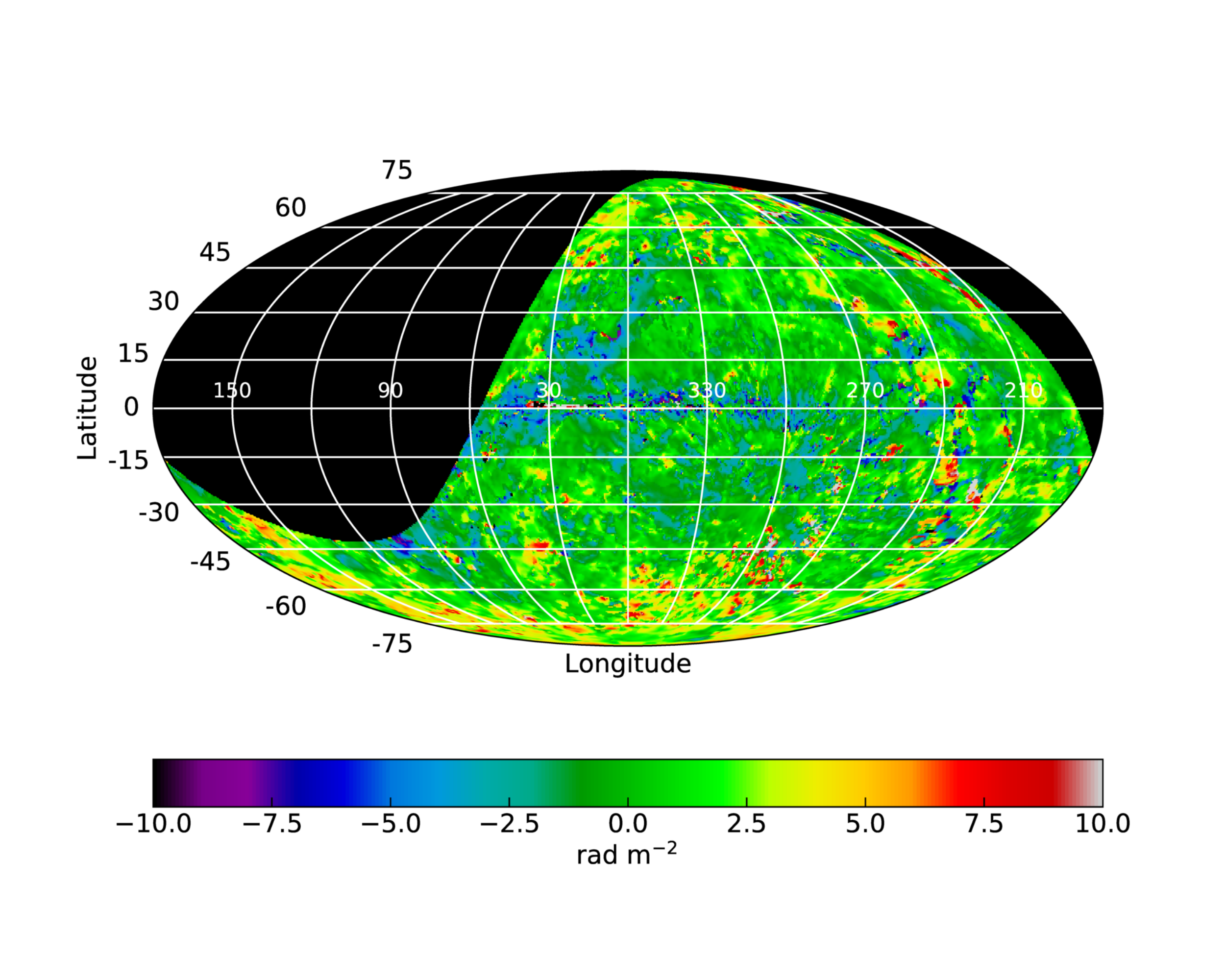}
\caption{The first moments ($M_1$) of the Faraday cubes of the DRAO (upper) and Parkes (lower) surveys.
The units are rad m$^{-2}$ from equation (\ref{eq:mom1}).  
%Note the slight residual striping in the DRAO data, that appears in 
%this projection as distorted circles around the North Celestial Pole 
%at $(\ell,b)=(122.9,27.1)$.  
The first moment shows the dominant $\phi$ value at each pixel.  Usually this
is $\phi$ at the center of the emission in the Faraday spectrum, $F(\phi)$.  The black areas are places where the emission is not strong 
enough to cross the threshold for computation of the moments,
or declinations not accessible to the telescopes.
\label{fig:mom1} }
\end{figure}

The first moment maps are shown on figure \ref{fig:mom1}.  These show for each pixel the
mean of $\phi$ weighted by the brightness temperature.
The bright areas around the NPS and the Fan Region show quite smooth first moment values
in the DRAO survey with values around +5 rad m$^{-2}$, whereas in the Parkes survey the smoothest region is in the fourth
quadrant at latitudes $+10 < b < +30$.  In the Parkes map, the Galactic plane is evident
in the first and fourth quadrants with significantly negative $\phi$ compared with most of
the rest of the sky.  
%The Parkes telescope suffers from leakage of Stokes $I$ into Stokes $Q$
%and U at low latitudes where the unpolarized brightness temperature is
%very high, hundreds of K in the first and fourth quadrants.  
%It is best not to analyse the Parkes data for latitudes below about $|b| = 3^o$ due
%to leakage of Stokes I into Stokes $Q$ and $U$.
The DRAO cube does not show the Galactic plane so clearly at all,
although there is some leakage of bright Stokes $I$ emission into Stokes $Q$ and $U$
in both surveys. Such leakage leads to unreliable values of the moments 
for both surveys for $\left|b\right| < 5^o$.

\begin{figure}[ht]
%\vspace*{2.5in}
\hspace{.5in}\includegraphics[width=5.5in]{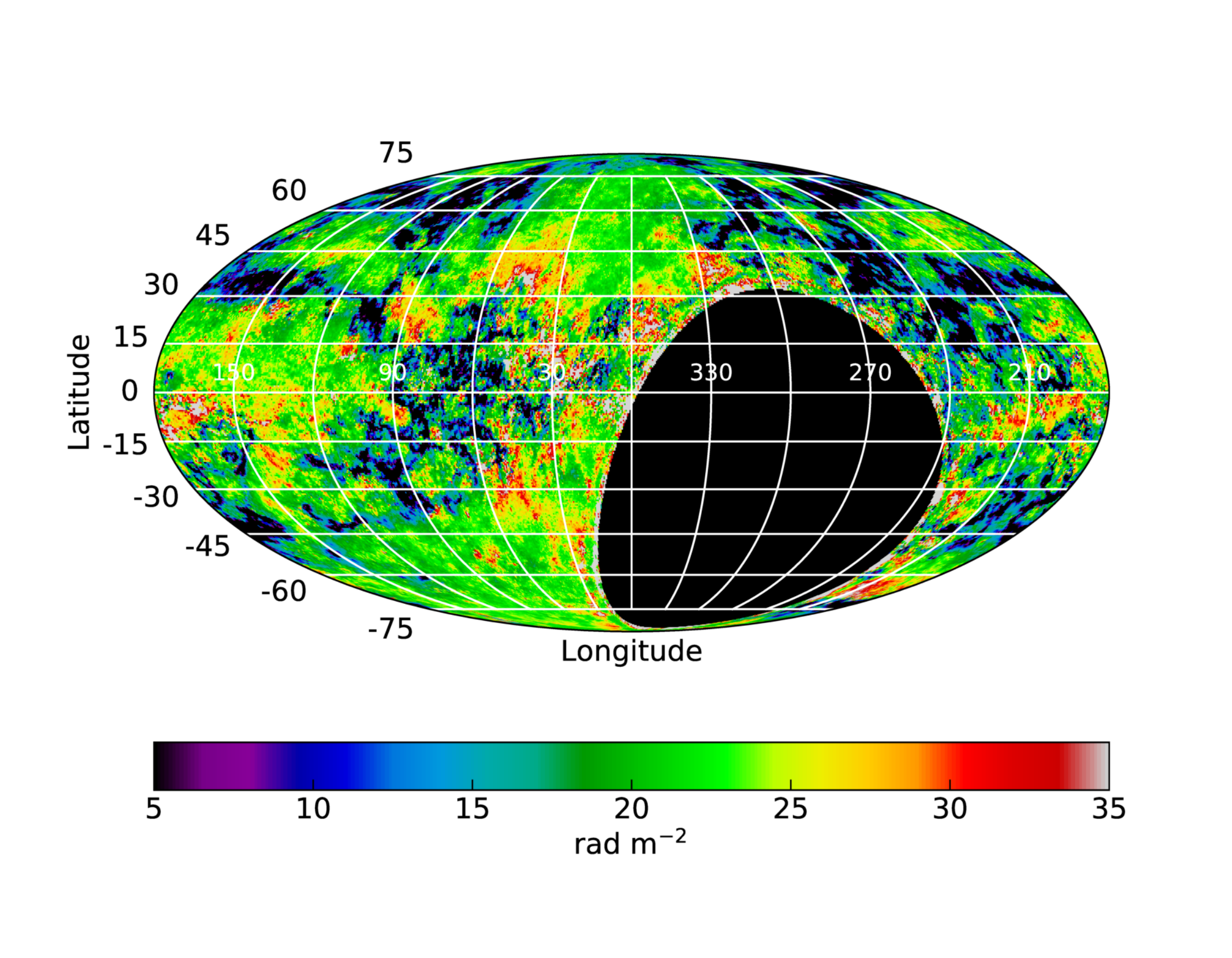}
\vspace{-.35in}

%\vspace*{2.5in}
\hspace{.5in}\includegraphics[width=5.5in]{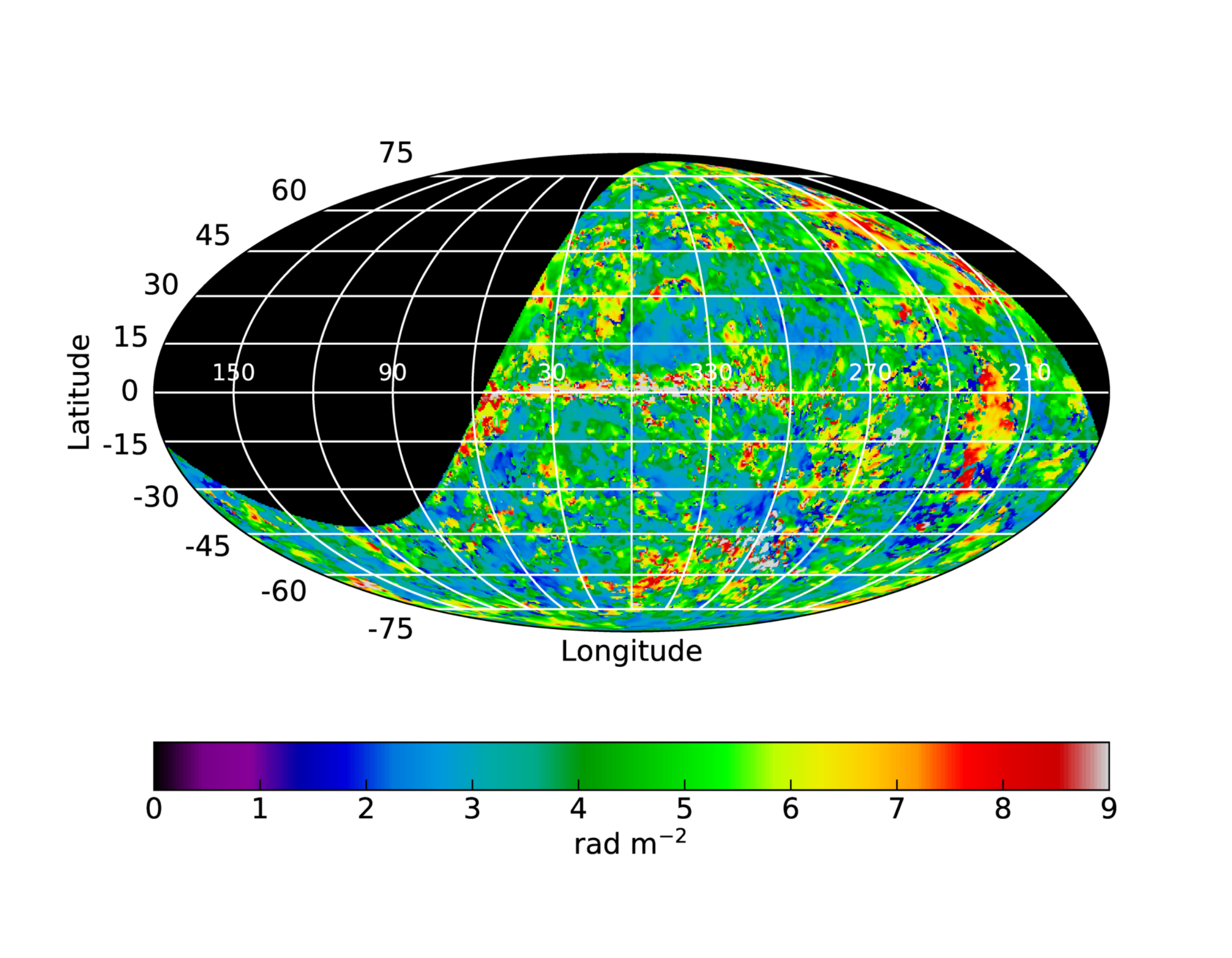}
\caption{The second moments ($m_2$) of the Faraday cubes of the DRAO (upper) and Parkes (lower) surveys.
The units are rad m$^{-2}$ from the square root of equation (\ref{eq:mom2}). 
The North Polar Spur and the Fan Region stand out in all three moments of the
DRAO survey.  The second moment can be thought of as the width of the emission in Faraday space,
similar to the velocity width of a spectral line.  For a Gaussian spectral feature, this is just $\sigma_{\phi}$,
but if there are several line components, it is the half-width of the range of $\phi$ that they cover.
\label{fig:mom2}}
\end{figure}

The second moment maps (Figure \ref{fig:mom2}) indicate the width of the brightness
distribution in $\phi$, similar to the widths of the Gaussians on figure \ref{fig:skyav},
but now shown for each pixel.  In both surveys, the second moment shows a mottled 
structure, but there is little correspondence between the two.  
%In the Parkes survey
%the second moment, like the zero moment, shows a slight residual hint of the scanning
%pattern around the south celestial pole, at $(\ell,b)=(303^o,-27^o)$

\begin{figure}[ht]
\hspace{.1in}\includegraphics[width=3.5in]{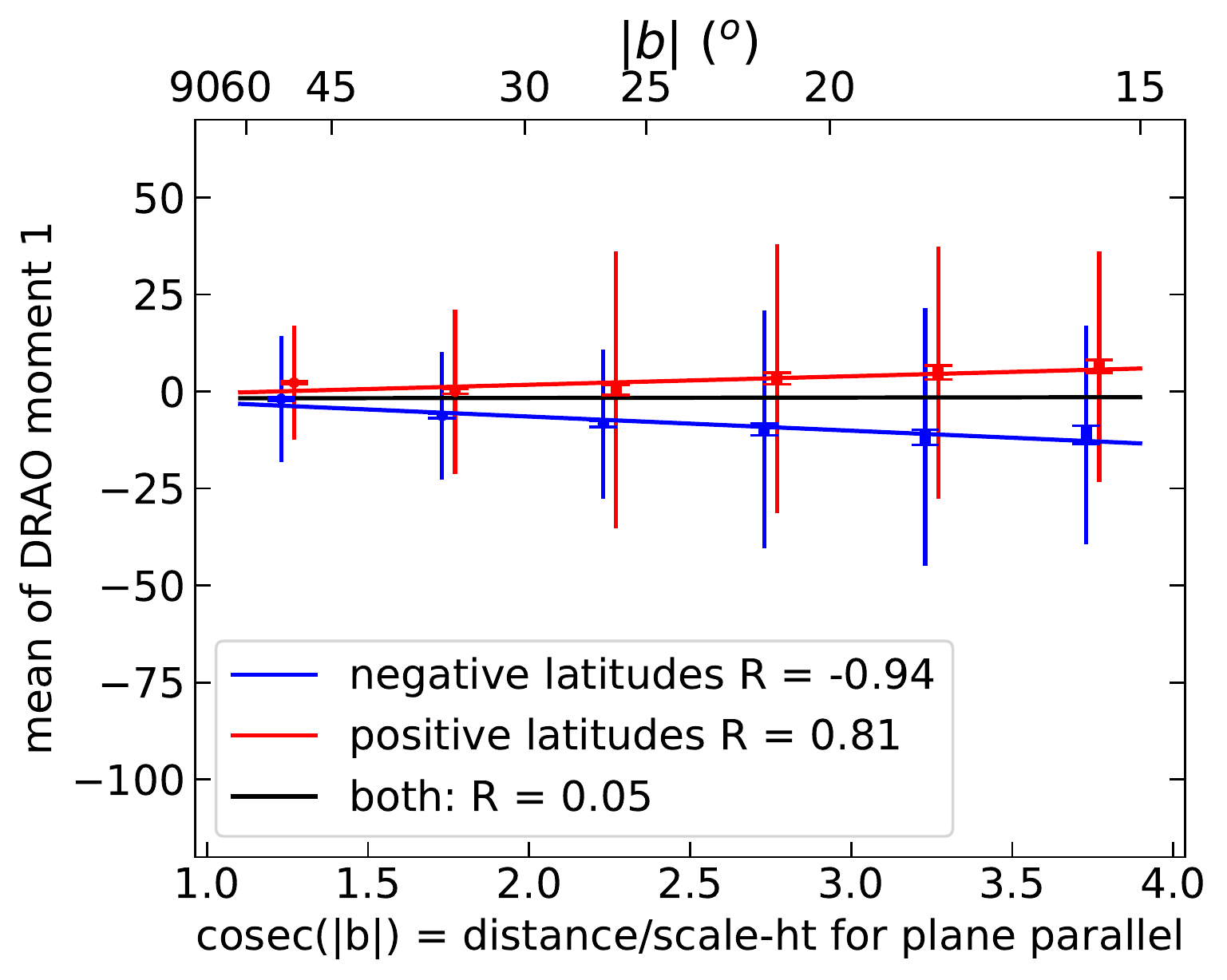}
\hspace{-.1in}\includegraphics[width=3.5in]{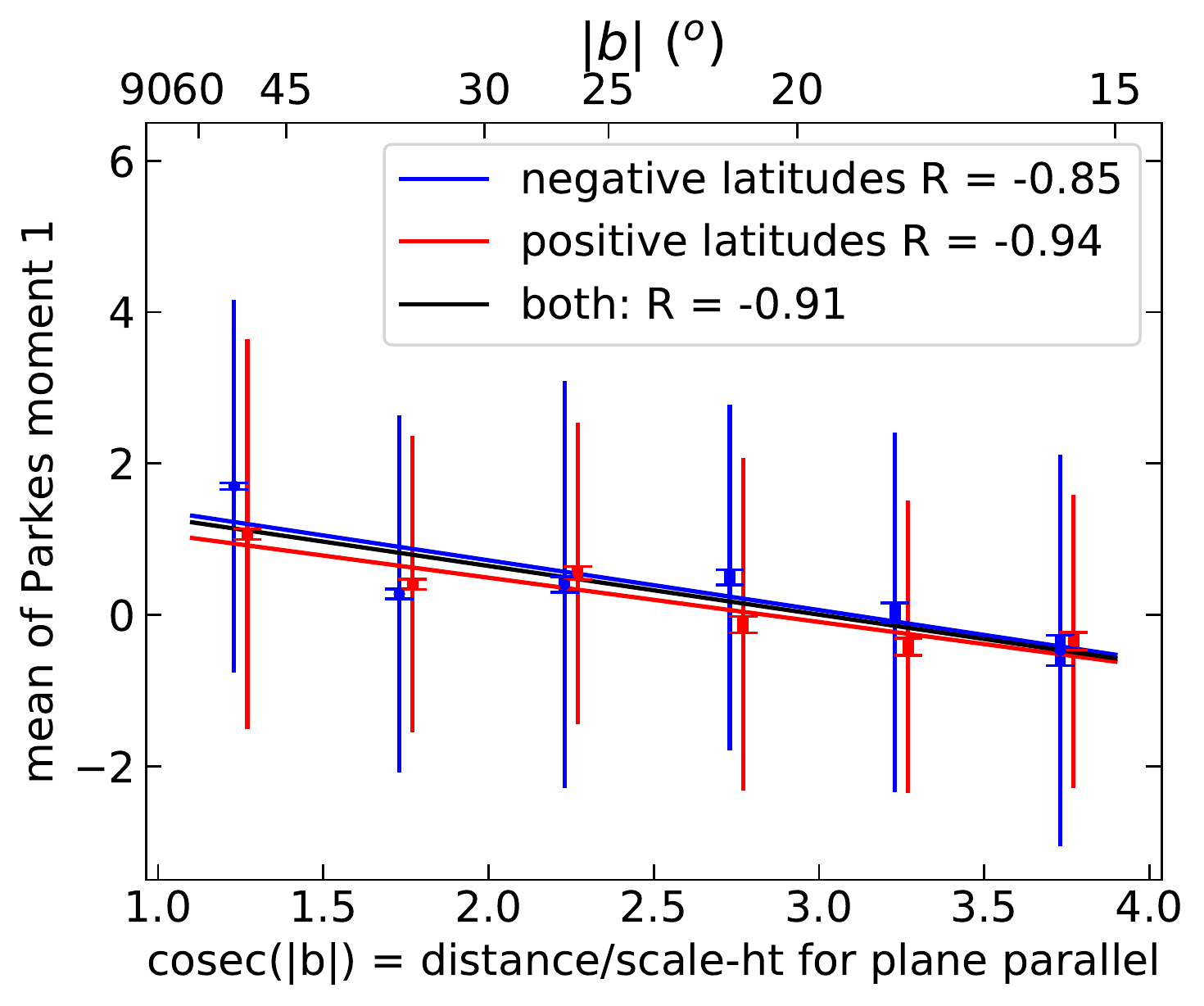}
\caption{Mean $\phi$ ($M_1$ in rad m$^{-2}$) vs. path length. The mean of moment 1 of the DRAO survey (left) and the Parkes
survey (right) are shown for different ranges of the path length through a plane
parallel layer (cosecant of latitude).  The points indicate the mean plus and minus the standard deviation
of the sample of directions in the range of cosec($|b|$)
as indicated on the x-axis.  The small error bars with thicker lines show the standard
error of the mean, i.e. the standard deviation divided by the square root of the
number of measurements.  
%Although the samples are taken at positions separated
%by more than a beamwidth, and so are not correlated due to the telescope resolution,
%they are correlated by the structure of the Galactic magnetic field, thus the
%standard error of the mean is an underestimate of the error bars on the points.
Values of the correlation coefficients (R) are shown for each linear fit.
 \label{fig:path-mom1} }
\end{figure}

%\subsection{The Overlap Region}
%The region where the two surveys overlap on the sky is shown on figure \ref{fig:overlap}.
%It is useful to compare the two surveys in this area, which makes a great circle that 
%crosses the Galactic plane in the first and third quadrants and reaches almost to both
%Galactic poles.  Thus almost all galactic latitudes are sampled in both hemispheres.
%In this area we take values of the moments spaced by 1.5 degrees in latitude and longitude,
%that is somewhat larger than the largest beam full-width in the two surveys.  Thus the samples are
%statistically independent. 

\subsection{Statistics of the First and Second Moments \label{sec:statistics}}

\begin{figure}[ht]
\hspace{.1in}\includegraphics[width=3.5in]{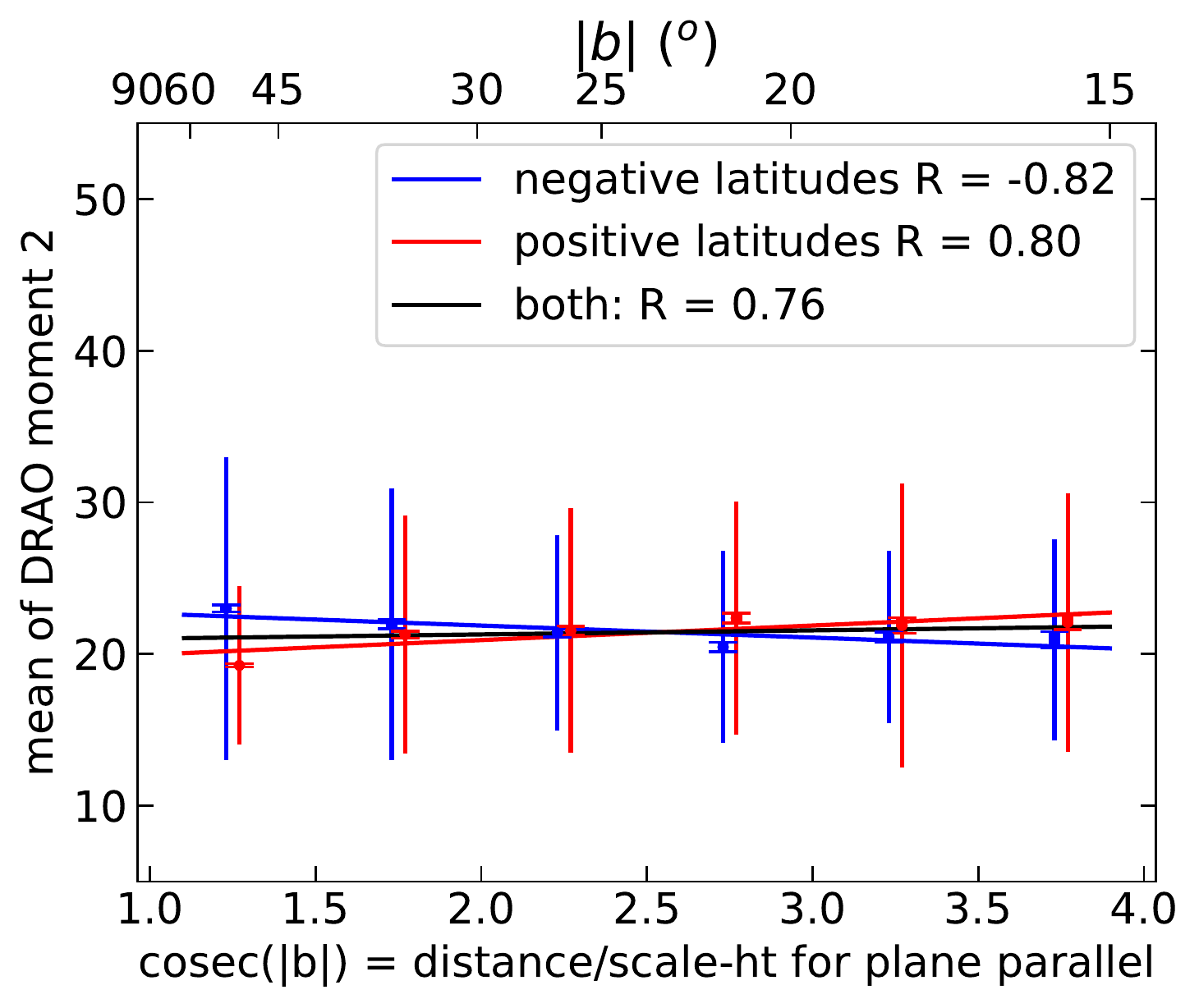}
\hspace{-.1in}\includegraphics[width=3.5in]{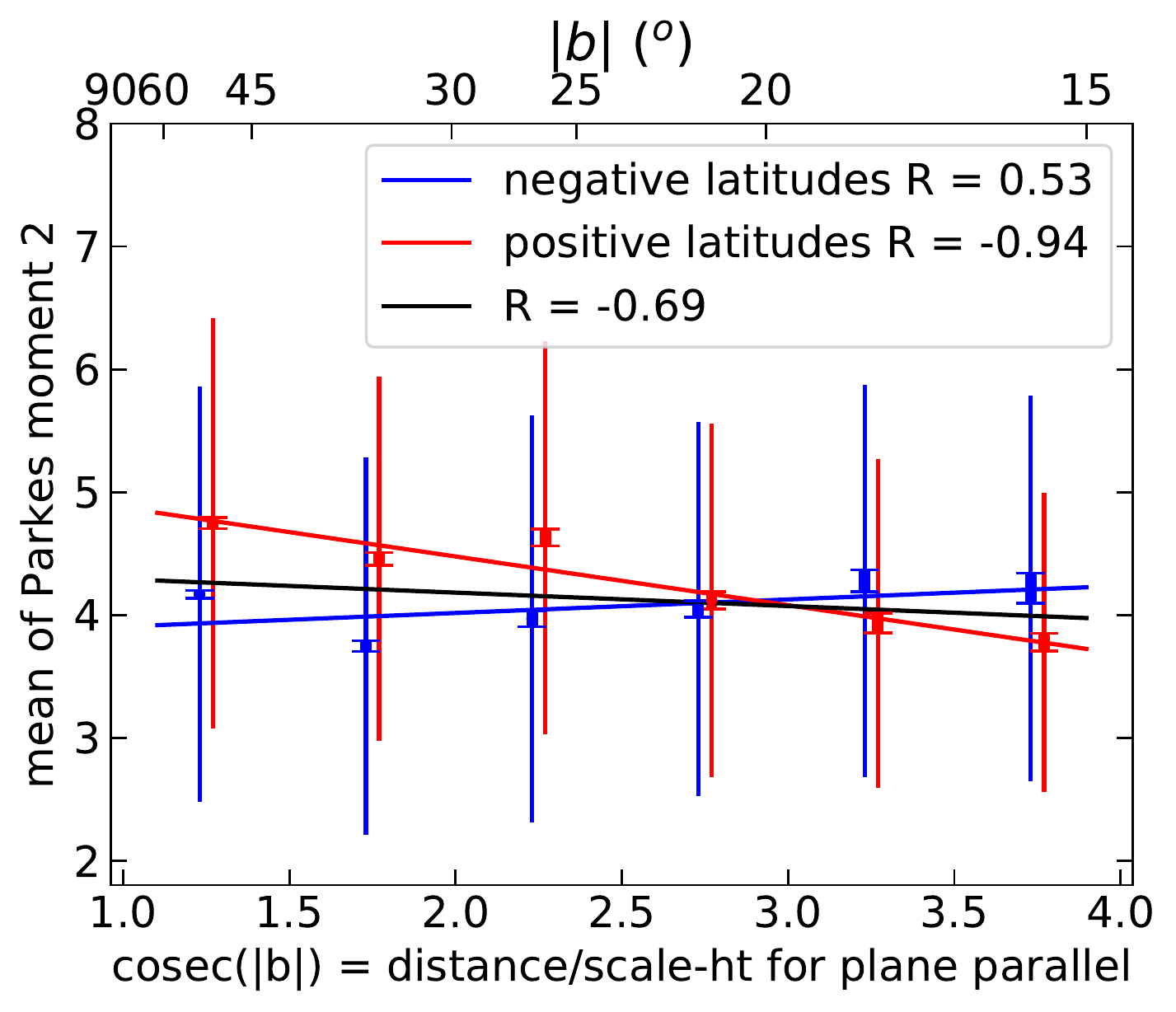}
\caption{$\phi$ width ($m_2$ in rad m$^{-2}$) vs. path length.  The mean of moment 2 of the DRAO survey (left) and the Parkes
survey (right) are shown for ranges of  cosecant($|b|$).  The points indicate the mean plus and
minus the standard deviation
with the thicker error bars indicating the standard error of the mean, as in figure
\ref{fig:path-mom1}.
\label{fig:path-mom2} }
\end{figure}

To study the statistics of the first and second moments we take a sample of points
separated by 90\arcmin \ in latitude and in longitude by 90\arcmin /cos($|b|$), i.e. by more than
the telescope beamwidths in both surveys.  We then separate the samples into 
sets for different ranges of latitude, $b$.
Figures \ref{fig:path-mom1} - \ref{fig:path-exgal} show the means and 
standard deviations of these samples, where the latitude boundaries are
set by steps of 0.5 in the cosecant of $|b|$.  For a plane-parallel geometry, this
is the ratio of the path length through the disk to the scale height of the disk, i.e.
\[ \mathrm{cosec}{(|b|)} \ = \ \frac{s_{eff}}{h} \]
where $h$ is the half-thickness of the plane, and $s_{eff}$ is the path length through
the disk at latitude $b$.
%The samples are taken at survey points spaced by 1.5 degrees in latitude and longitude,
%that is somewhat larger than the beam full-width for the Parkes survey.  Thus the sample values
%are not correlated due to the telescope resolution.  
Here we will not assume a value for $h$, but note that 
%in her comprehensive review
%Ferri\`{e}re (2001 eq. 5) suggests a commonly used function for the density of the
%warm ionized gas that is the sum of two exponentials with scale heights of 70 pc and 900pc, whereas 
\citet{Gaensler_etal_2008} find good evidence that $h\simeq 1.8$ kpc. 

On figures \ref{fig:path-mom1}, \ref{fig:path-mom2}, and \ref{fig:path-exgal} the points show
the means of distributions of several hundred independent measurements of the moments
in the latitude ranges set by the intervals of cosec$|b|$ on the x axis.  The number
of points in each sample ranges from $\sim$150 at the high latitudes to $\sim$1500 at
the lower latitudes.  The mean of each sample is plotted as the point, and the standard
deviation is plotted as positive and negative bars, without end caps, on each point.
The formal error of the mean, calculated simply as the standard deviation divided by the
square root of the number of samples, is plotted as the positive and negative error bars
with thicker lines and end caps.  Thus, although the correlations with cosec$|b|$
appear to be very weak relative to the longer bars, relative to the errors on the points 
they are statistically significant.   For example in the DRAO survey in the highest
latitude bin (plotted at 1.25 on the x-axis on the left panel of figure \ref{fig:path-mom1})
the positive latitude (red) point is 2.26$\pm$0.29 rad m$^{-2}$, while
the negative latitude point is -1.98$\pm$0.39 rad m$^{-2}$.  The difference is more than
ten times the standard errors.
%[xxx insert values : neg side -1.98 +- 0.39, pos side: 2.26 +- 0.29
%diff is 
The incomplete coverage of the sky in the two surveys
may be a factor in the trends of the moments with latitude.  Until the two hemispheres are fully
surveyed at both wavelengths it will be hard to be fully characterize the pattern of the local B field, but the averages shown on figure \ref{fig:path-mom1} strongly suggest that there is
a {\it z} component in the nearby Galactic magnetic field pointing from the northern toward the southern
hemisphere.

Looking at the distributions of the first moments vs. cosec$|b|$ on figure \ref{fig:path-mom1}, 
the Parkes points (right-hand panel) show
a smooth decrease in moment 1 from positive values at high latitudes (left side),
to negative values at intermediate latitudes (right side).  The highest value of cosec$|b|$
shown on the x-axis (4.0) corresponds to $|b| = \arcsin{0.25} \simeq 0.25$ rad = 14$^o$.
Note that {\bf both hemispheres show the same trend}, i.e. the values are very similar
% Copy Editor:  please leave the \bf in the above line
for positive and negative latitudes. Since positive rotation measure corresponds to
magnetic fields pointing toward the observer, the implication of the
right hand panel of figure \ref{fig:path-mom1} is that the $B$-field points
toward the solar neighborhood at high latitudes in both Galactic hemispheres, but
it points away at lower latitudes, in the longitude ranges covered by the 
Parkes survey (i.e. most of the Galactic southern hemisphere but only about half of
the northern hemisphere).  
If this or some other field geometry is the explanation
for the trend in the Parkes survey first moment points, it is indicated only for the
%region {\bf inside the polarization horizon} at the Parkes survey wavelength.
region visible in linear polarization at the Parkes survey wavelength, i.e. close enough
to be only weakly depolarized.

The DRAO first moment points (left panel of figure \ref{fig:path-mom1}), show
a weak but significant divergence between the two Galactic hemispheres 
as the path length increases.  The positive latitudes shift toward positive $\phi$, thus
$\vec{B}$ pointing toward the Sun, while the negative latitudes shift
the opposite way, with $\vec{B}$ pointing away from the Sun.  
The two strongest features at latitudes $b > +45^o$ in the first moment 
maps of the DRAO survey (figure \ref{fig:mom1}, upper panel) are the North
Polar Spur, at longitudes $-30^o < \ell < +60^o$, and another smooth
feature at longitudes $180^o < \ell < 240^o$.  Both of these show 
positive values of $M_1$, with $3 < \phi < 15$ rad m$^{-2}$.
There is very little emission at high positive latitudes that shows
negative $M_1$.
The black lines on figure \ref{fig:path-mom1} indicate the 
linear regression best fit to all the points in both hemispheres,
with the regression coefficients R=+0.04 and R=-0.91 indicated.  
Averaging the two hemispheres together, there is almost no correlation
of the path length (cosec$|b|$) with $M_1$ in the 
DRAO survey, but strong negative correlation between the path length and $M_1$ 
in the Parkes survey.  

The Parkes first moment correlated against path length
has an R of -0.91, which indicates a strong anti-correlation between the combined
data from the two hemispheres and the path length.  The lower panel of figure \ref{fig:mom1}
shows more yellow and red (positive first moments) at the highest latitudes in both hemispheres, 
and more dark green and blue color at lower latitudes.  This shift from positive $M_1$ at high latitudes
to negative $M_1$ at lower latitudes explains the behavior of the latitude averages
shown on figure \ref{fig:path-mom1}, right panel.  Since negative $\phi$ corresponds to line of sight $B$ field
component pointing away from the observer, these two figures suggest two distinct field geometries at
high latitudes.
The DRAO survey indicates a B field pointing toward the Sun in the North Galactic hemisphere, and away from us
in the Southern hemisphere.  On the other hand, in the Parkes survey we see the field pointing toward
us from {\bf both} the Galactic north and south poles, but away from us at intermediate latitudes.  
% Copy Editor:  please leave the \bf in the above line
%The two results are not necessarily inconsistent, as we will see below (section \ref{sec:depol}) that the two surveys sample quite different volumes.

%[Maybe add a table with cosecb, b-range, number of points, mean val +- std dev of mean,
%std dev for each hemisphere separately for Parkes mom1 data.]

The difference between the DRAO and Parkes second moments is evident on figure 
\ref{fig:path-mom2}.  The DRAO widths are much greater than for the Parkes features, 20 to 25 rad m$^{-2}$
compared with 3 to 5 rad m$^{-2}$ in the Parkes data.  The DRAO survey shows opposite trends in the two 
Galactic hemispheres; $m_2$ increases with path length at positive latitudes, but decreases slightly with
path length at negative latitudes.  A similar contrary effect is seen in the Parkes data, but it 
goes the other way around.  The strong negative correlation between $m_2$ and path length
in the Parkes data for positive latitudes suggests that the lower latitudes are not increasing
the scatter, as would be expected by increasing
the number of steps in a random walk process of field reversals.  This in turn suggests that the
polarization horizon is so nearby for the emission seen in the Parkes survey that the
local interstellar medium, including the local bubble 
\citep{Frisch_etal_2012, Alves_etal_2018}, is dominating the width of
features in the Faraday spectra.  In the Parkes data the negative latitudes show a weak positive
correlation between $m_2$ and path length.
A similar horizon effect in M51 might explain the difference in the width of $T(\phi)$ measured at 1-2 GHz
 compared with that measured at 5-8 GHz \citep{Mao_etal_2015b}.

\section{Comparison with Other RM surveys \label{sec:comparison}}

\begin{figure}[ht]
%\vspace*{2.5in}
\hspace{.5in}\includegraphics[width=5.5in]{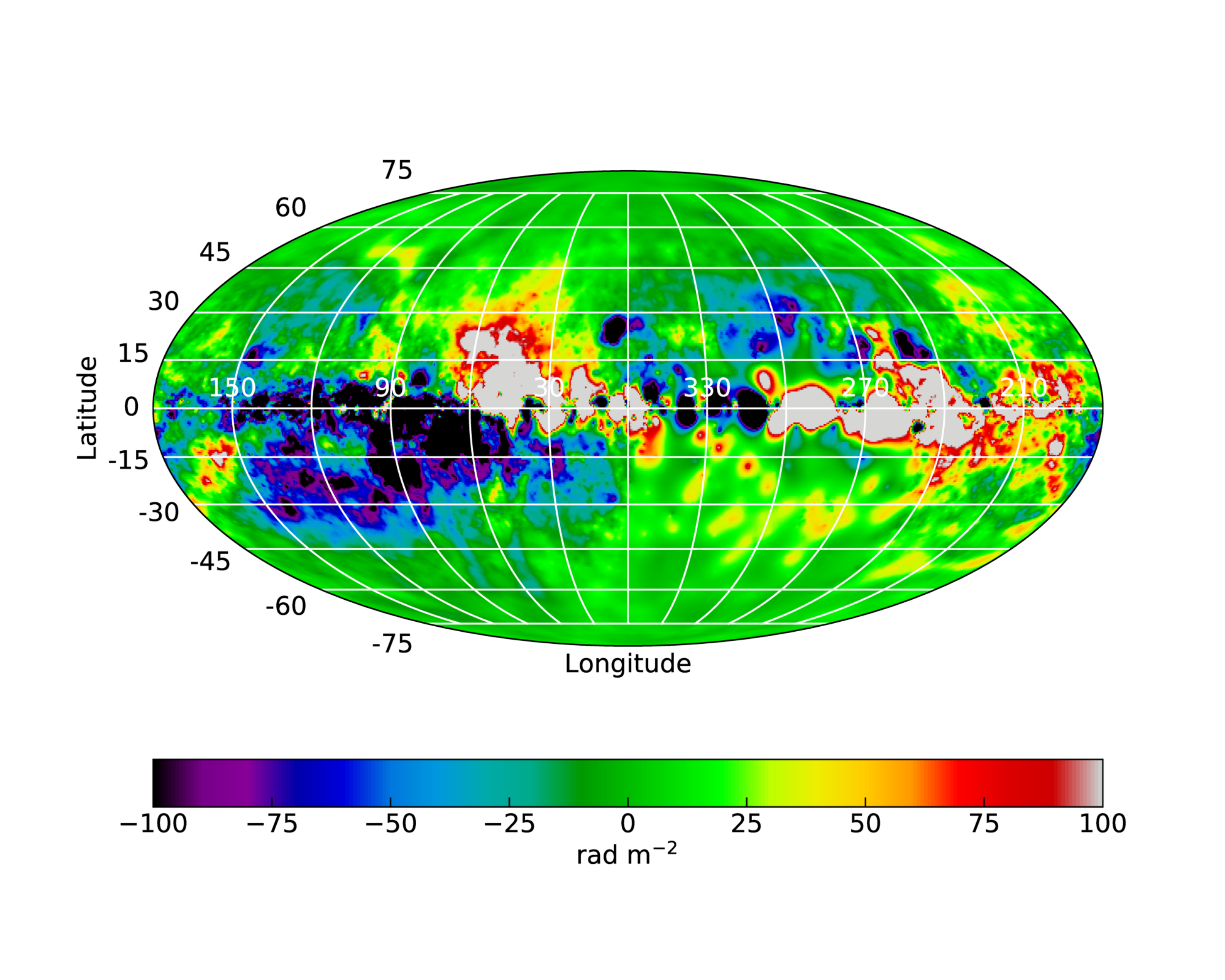}
\caption{The foreground Galactic RM grid from \citet{Oppermann_etal_2015} projected on the 
same coordinates as figures \ref{fig:mom0} - \ref{fig:mom2}.  At low latitudes ($|b|<5^o$)
the scale is saturated.  There the extremes are $-1124 < \phi < 1273$ rad m$^{-2}$.
\label{fig:Oppermann}}
\end{figure}

The Faraday spectra of the diffuse polarized emission illustrated
in the previous section can be compared with other tracers of the RM at high and
intermediate latitudes.  The most comprehensive
is a compendium of surveys of extragalactic radio source RMs compiled and gridded by 
\citet{Oppermann_etal_2012, Oppermann_etal_2015}.  We have made comparisons with both the 2012 and
2015 versions of the Galactic foreground rotation measure maps of Oppermann et al.,
as the former is more directly derived from the data, while the latter is based
on models that best reproduce the data.  In comparison with the GMIMS survey
results the two give similar information.
Below we use the 2015 map (``maps/phi'' available from
\url{https://wwwmpa.mpa-garching.mpg.de/ift/faraday/2014/index.html}
).  
%  RM cubes (using Astropy WCS, Calabretta et al 2011), 
The contribution of the Milky Way foreground 
derived from the Oppermann model is shown on figure \ref{fig:Oppermann}.
Although this is the estimate for the {\bf Galactic foreground}, we will refer to
% Copy Editor:  please leave the \bf in the above line
it as the ``extragalactic RM grid'' or just the ``extragalactic RMs''.

\begin{figure}[ht]
\hspace{1.5in}\includegraphics[width=3.5in]{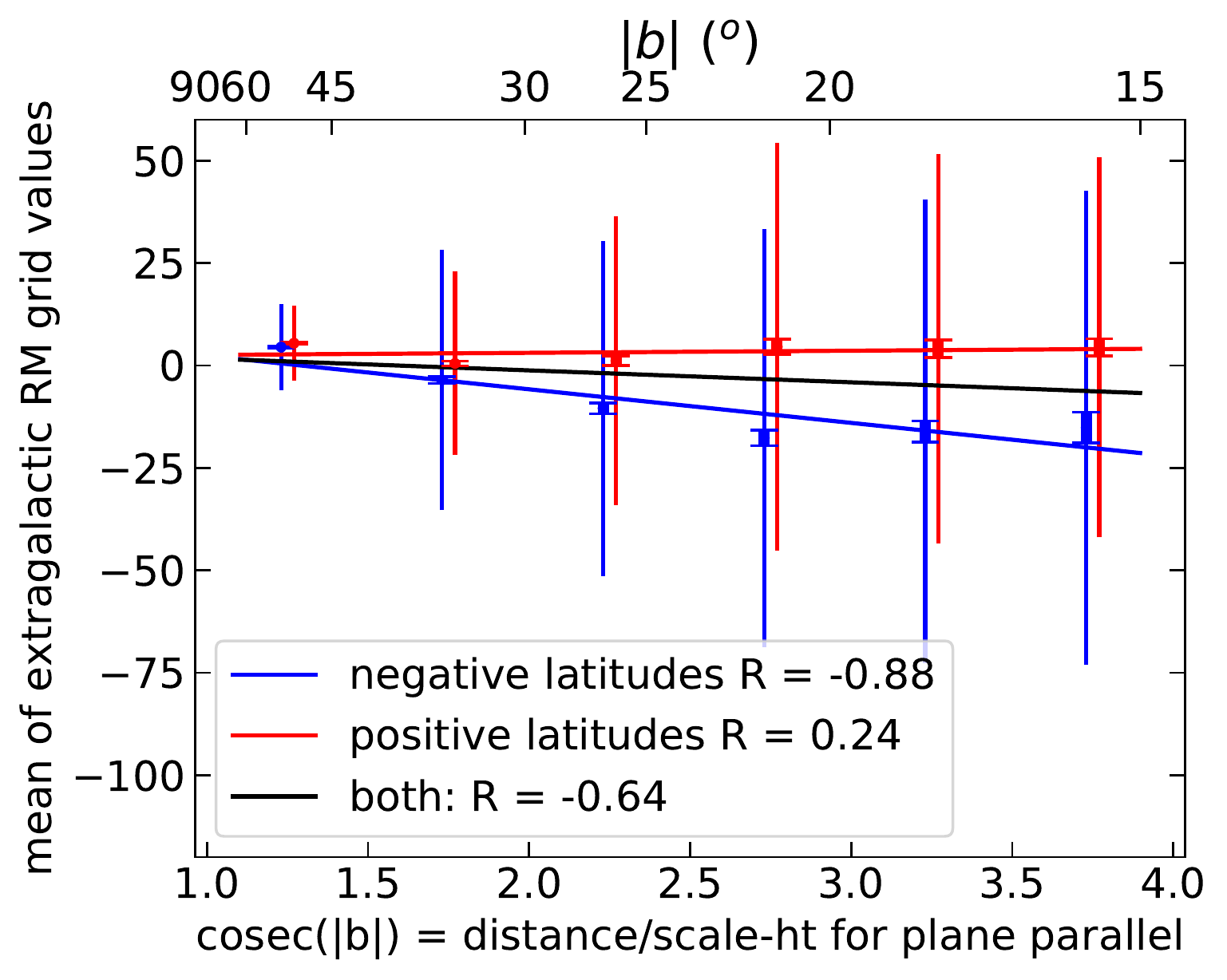}
\caption{The extragalactic RM grid, sampled at the same points 
that are used for figures \ref{fig:path-mom1} and \ref{fig:path-mom2},
data from \citet{Oppermann_etal_2015}.  Note that the standard deviation of the points
increases with path length, but there is no significant difference between
the means at high latitudes.
The gradual separation of the red and blue points at lower latitudes, similar
to that seen on the left panel of figure \ref{fig:path-mom1}, 
suggests that the North Galactic hemisphere has a net line of sight $\vec{B}$ field component
pointing toward the Sun, while the southern hemisphere has a net $\vec{B}$ field component
pointing away from the Sun.
%may come from the directions of the well known azimuthal or spiral field pattern seen in the Galactic plane
%at low latitudes, in the different longitude ranges sampled by the two surveys.
The leftmost points are at 4.52$\pm$0.19 rad m$^{-2}$ in blue (negative latitudes) and
5.45$\pm$0.16 rad m$^{-2}$ in red (positive latitudes), a five sigma difference.
\label{fig:path-exgal}}
\end{figure}

Sampling the extragalactic RMs 
at the same points as for figures \ref{fig:path-mom1}
and \ref{fig:path-mom2} gives figure \ref{fig:path-exgal}.  The RM values on the y-axis
of figure \ref{fig:path-exgal} and on the scale
of figure \ref{fig:Oppermann}, are {\bf much} larger than the range of $\phi$ with
% Copy Editor:  please leave the \bf in the above line
bright emission in the Parkes survey.  The width
of the distribution of RMs (vertical bars) increases rapidly with cosec$|b|$, as expected for a random walk process
where the line of sight passes through many uncorrelated regions where
the $\vec{B}$ field component is sometimes toward
the observer (positive $\phi$), sometimes away (negative)
coupled with the higher average density of the ionized medium at low $|z|$, and higher $B$ field intensity at
low $|z|$.  This is similar to the increase in the DRAO first 
moments, $M_1$, with increasing cosec$|b|$ shown on figure \ref{fig:path-mom1}, left panel.
%Computing the mean of the rotation measures of the extragalactic foreground, using the same set of
%directions spaced by 90\arcmin as used for figures \ref{fig:path
In addition, the extragalactic rotation measure means (red and blue points on figure \ref{fig:path-exgal}) separate to positive and negative values 
for the northern and southern hemisphere samples, in the same way
that the DRAO first moment points do on figure \ref{fig:path-mom1}, left panel.
The consistency of this effect suggests that it is caused by the large scale {\bf ordered}
% Copy Editor:  please leave the \bf in the above line
$\vec{B}$ field similar to that seen at low latitudes in surveys of rotation measures
toward compact sources 
%(e.g. Ordog et al. 2017, Mao et al. 2012, Han et al. 2017).
\citep[e.g.][]{Ordog_etal_2017, Mao_etal_2012, Han_2017}.
In both figure \ref{fig:Oppermann} and the upper panel of figure \ref{fig:mom1}, the 
overall picture for the inner Galaxy is positive RMs at positive latitudes ($0^o < b < 30^o$) in the first
quadrant, negative RMs at negative latitudes in the first quadrant, and the opposite
in the fourth quadrant. 
The larger absolute numbers in the extragalactic sample are expected based on the factor of two
between the peak $\phi$ measured for the emission from a slab, and the RM seen toward a source behind the slab.
The implication is that the DRAO first moments are tracing roughly the same
ordered field component as traced by the extragalactic and pulsar RMs, whereas
the Parkes first moments are tracing something quite different
\citep[see][figure 6]{Han_2017}. 

\begin{figure}[ht]
%\hspace{.1in}\includegraphics[width=3.5in]{exgal_Parkes_vs_DRAO_a.pdf}
\hspace{-.2in}\includegraphics[width=3.5in]{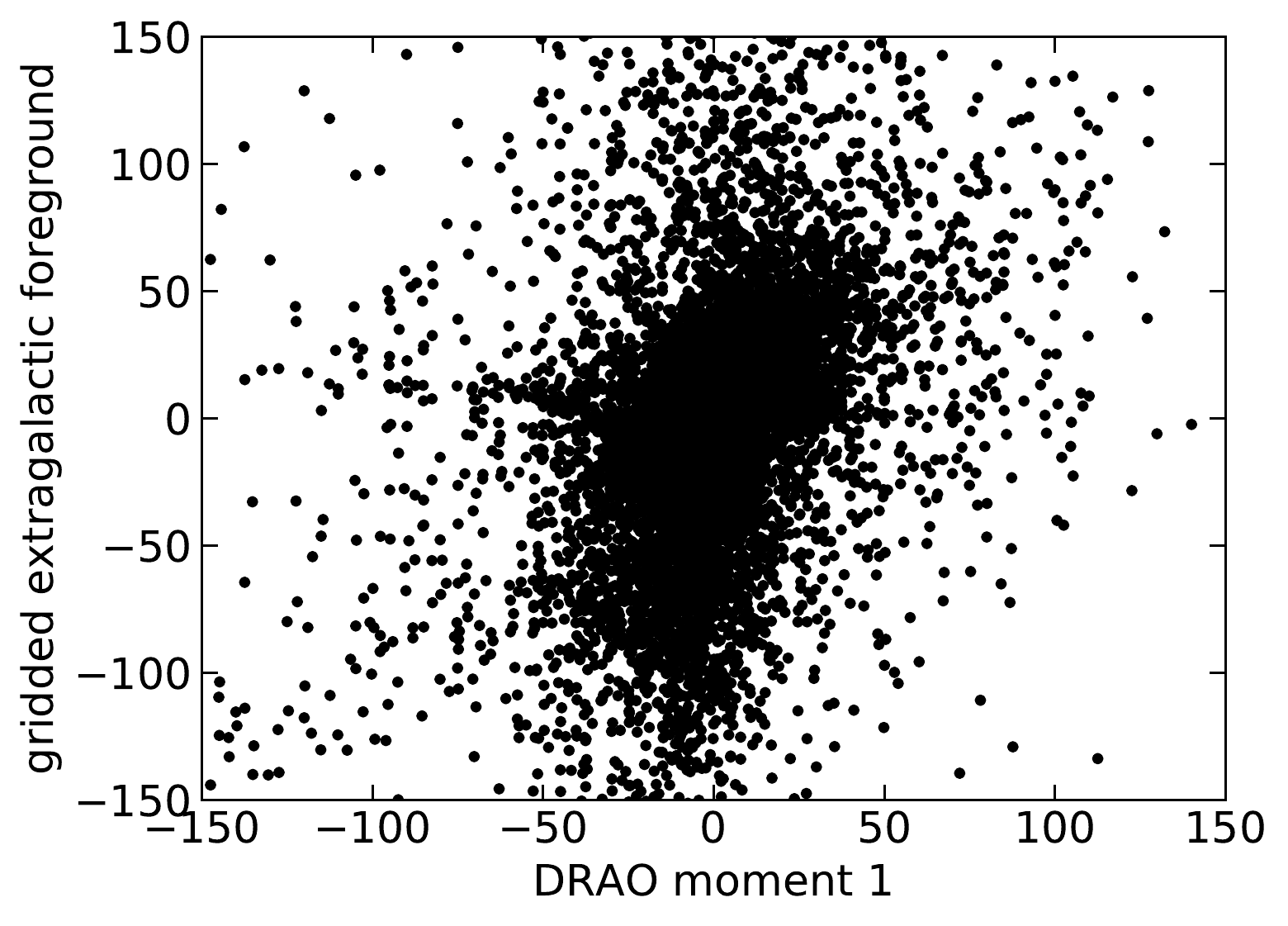}
\hspace{.1in}\includegraphics[width=3.5in]{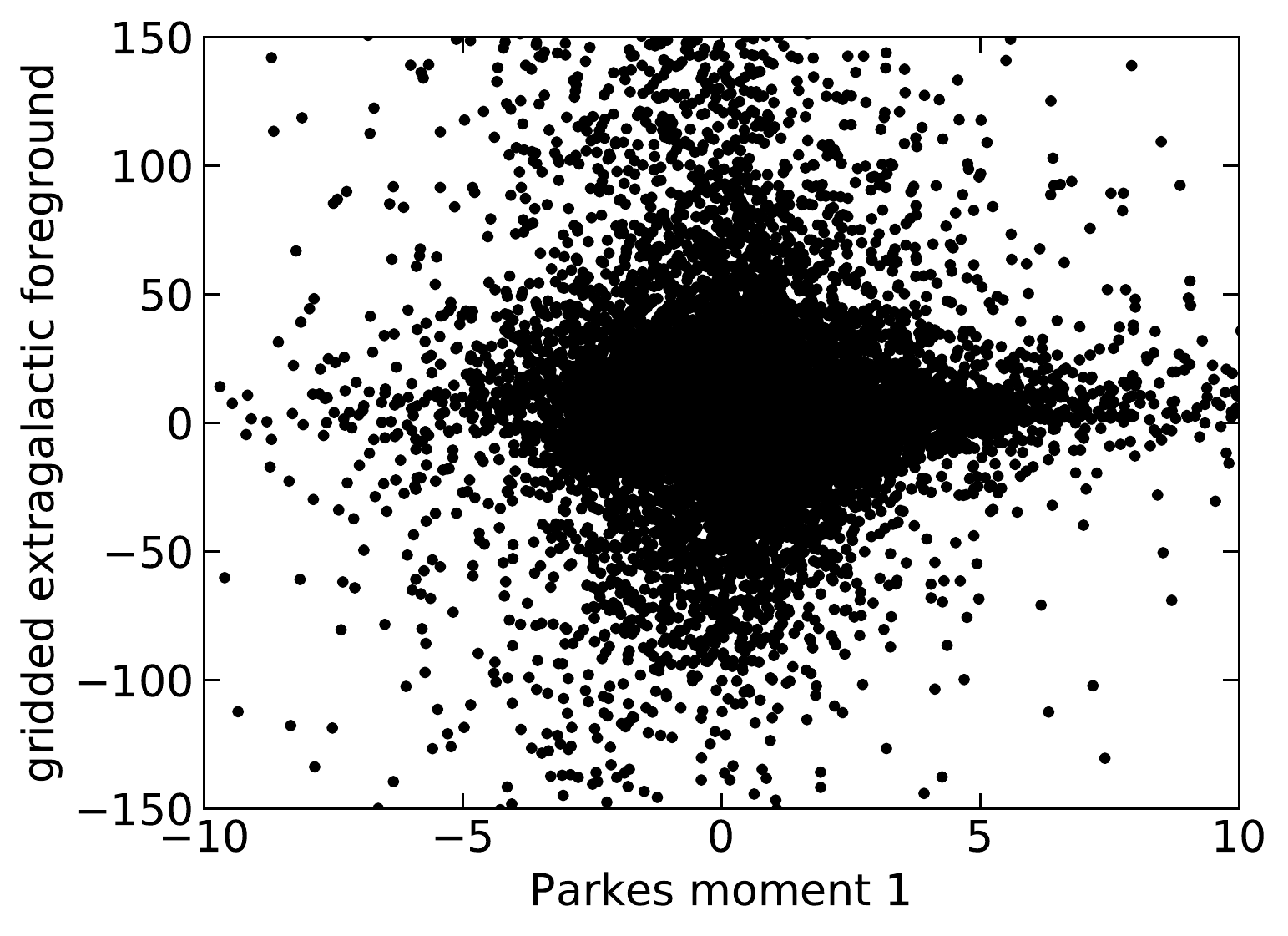}

\caption{Left Panel: Comparison between the extragalactic rotation measures and the 
DRAO survey first moment rotation measures for independent points spaced by 1.5$^o$.
The units on the axes are rad m$^{-2}$.
Outlier points extend to several hundred rad m$^{-2}$ on the y axis, to about $\pm100$ rad m$^{-2}$
on the x axis.  There is weak correlation between these two quantities (correlation
coefficient R=+0.25).
Right panel: Comparison between the Parkes survey first moments and the
extragalactic rotation measures, with points spaced by 1.5$^o$ as in the left panel.
Note the change of scale on the x axis.  Outlier points extend in x to $\pm 15$ rad m$^{-2}$.
There is no significant correlation between these two quantities (correlation
coefficient R=-0.02).}

\label{fig:exgal_Parkes_vs_DRAO}
\end{figure}

The correlations with distance that appear on figures \ref{fig:path-mom1} and \ref{fig:path-mom2} 
are all the more interesting considering that {\bf the moments are not correlated with each
% Copy Editor:  please leave the \bf in the above line
other}, especially at low latitudes.  The first moments of the two surveys are compared 
with the extragalactic foreground sample 
%for the overlap region (-25$^o$ $<\ \delta\ <$ +15$^o$) shown 
on figure \ref{fig:exgal_Parkes_vs_DRAO}. 

\subsection{Pulsar Rotation Measures \label{sec:pulsar}}

Rotation measures have been determined for 1001 pulsars (\citealt{Manchester_etal_2005} version 1.56 supplemented
by \citealt{Han_etal_2018b}, see \citealt{Han_etal_2018a}),
most of these also have distance determinations, either from dispersion, parallax, or other means.  
Although the RM and dispersion measure (DM) are not physically
independent, since both involve the line of sight 
integral of the electron density, observationally they constitute 
entirely separate measurements.  Most of the pulsar distances
are based on combining DM values with
an electron density model of the Milky Way.  Although this does not give a 
very precise distance, it is in some ways just what we want for comparison
of the pulsar and diffuse RMs, since we might expect more RM, or more 
fluctuation in RM, on a path with higher DM, i.e. a higher path integral of
electron density.  So we will make use of the pulsar distances and RMs
as milestones to compare with the $\phi$ distribution
of the diffuse polarized emission,
keeping in mind that most individual pulsar distances are not reliable to
better than about 30\% at latitudes $|b|>20^o$. 

\begin{figure}[ht]
\includegraphics[width=3.5in]{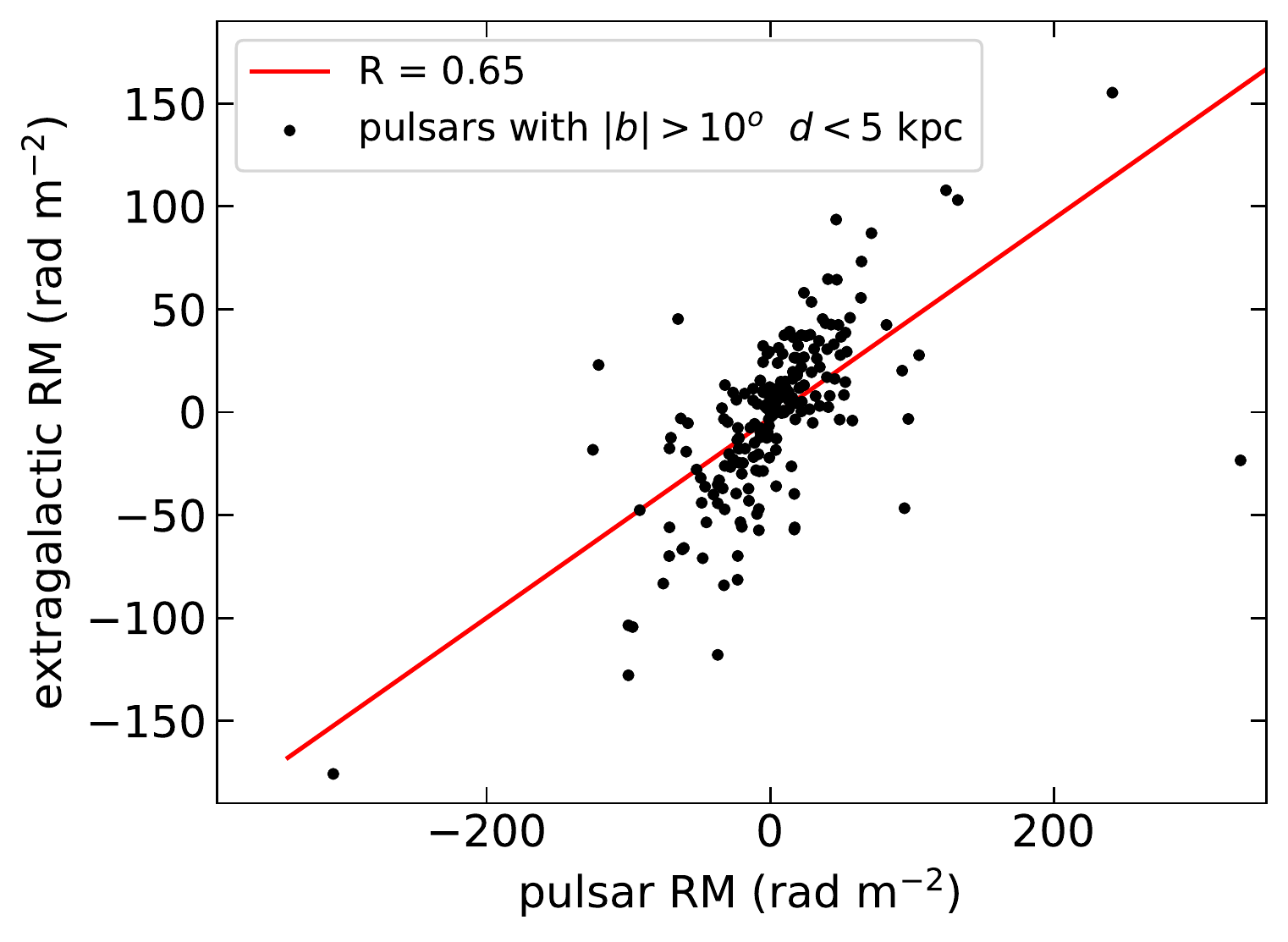}
\hspace{-.2in}\includegraphics[width=3.5in]{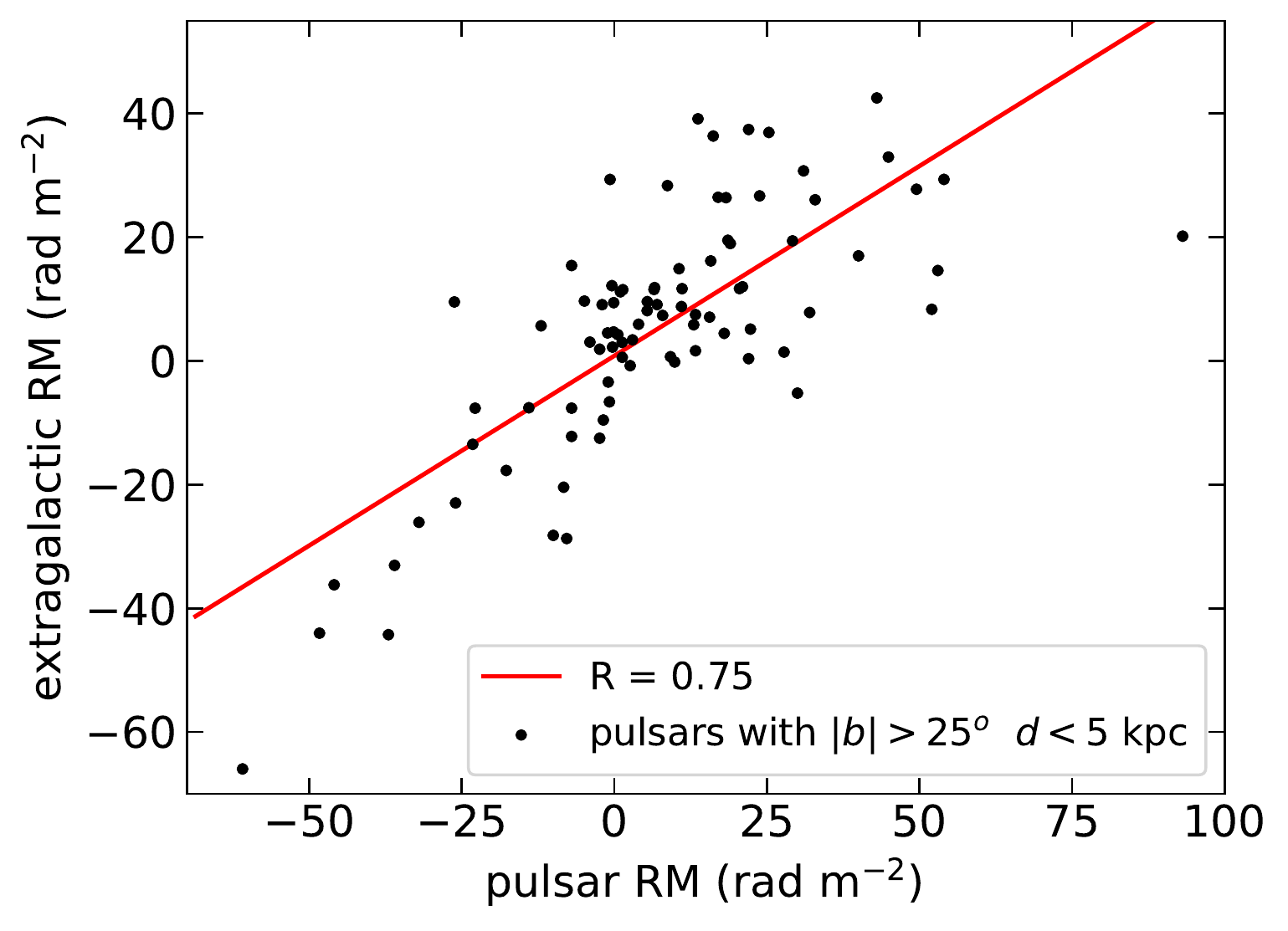}
%\includegraphics[width=3.5in]{Han_pulsar_vs_exgal_rms_a.pdf}
%\hspace{-.2in}\includegraphics[width=3.5in]{Han_pulsar_vs_exgal_rms_b.pdf}

\caption{Pulsar rotation measures compared with the extragalactic foreground RM at the same
positions.  Including all pulsars with $|b|>10^o$ and distance $d < 5$ kpc there is
significant correlation (R=0.65, left panel), but
there is stronger correlation for pulsars above $|b|=25^o$ with distances less than
5 kpc (R=0.75, right panel).  Including pulsars with latitudes
below $|b| = 5^o$ washes out the correlation with the extragalactic foreground.
The red lines are least squares linear fits to the data points, with correlation
coefficients R as indicated.
}
\label{fig:pulsar_vs_exgal_rms}
\end{figure}

Note that pulsars, like the extragalactic sources that have been used to form the Oppermann extragalactic
RM grid, are all compact sources, so they do not suffer depolarization due to the
Galactic magneto-ionic medium.  But the pulsar distances are often much less than the path length entirely
through the ionized interstellar medium (Reynolds Layer) that can cause Faraday rotation
and depolarization of the diffuse emission.  Thus pulsar RMs are not perfectly
correlated with the extragalactic RMs.  This is particularly true when the large
number of pulsars at low latitudes are included in the sample, as shown on the left panel of
figure \ref{fig:pulsar_vs_exgal_rms}.  On the other hand, for pulsars at latitudes
above $|b| = 25^o$ and distances less than
a few kpc, the correlation is better (R = 0.75, right panel of figure \ref{fig:pulsar_vs_exgal_rms}).  

\section{Distances \label{sec:distances}}

\begin{figure}[ht]
%\hspace{.1in}\includegraphics[width=3.5in]{pulsar_drao_1_a.pdf}
%\hspace{.1in}\includegraphics[width=3.5in]{pulsar_drao_1_b.pdf}
\hspace{.1in}\includegraphics[width=3.5in]{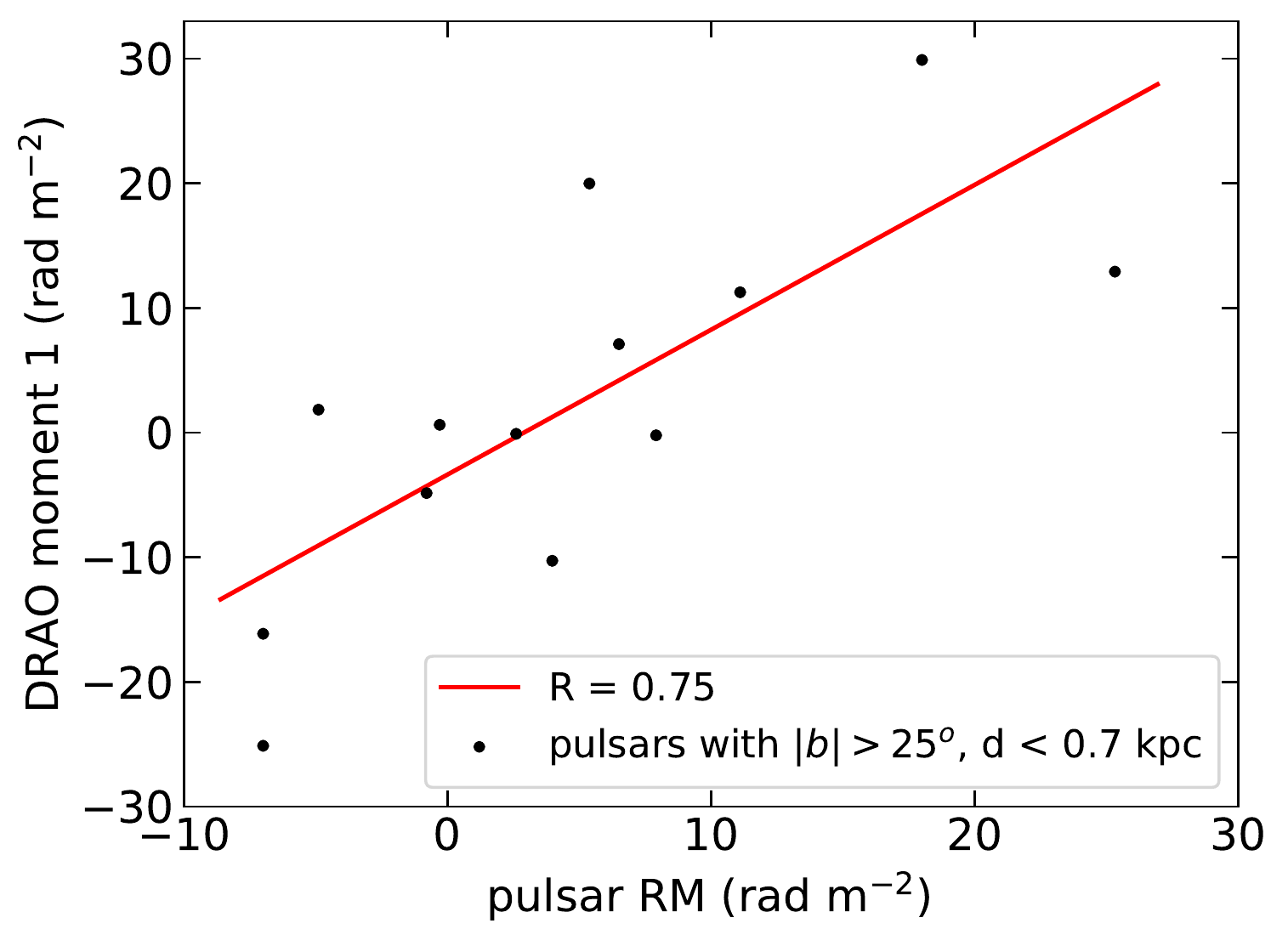}
\hspace{.1in}\includegraphics[width=3.5in]{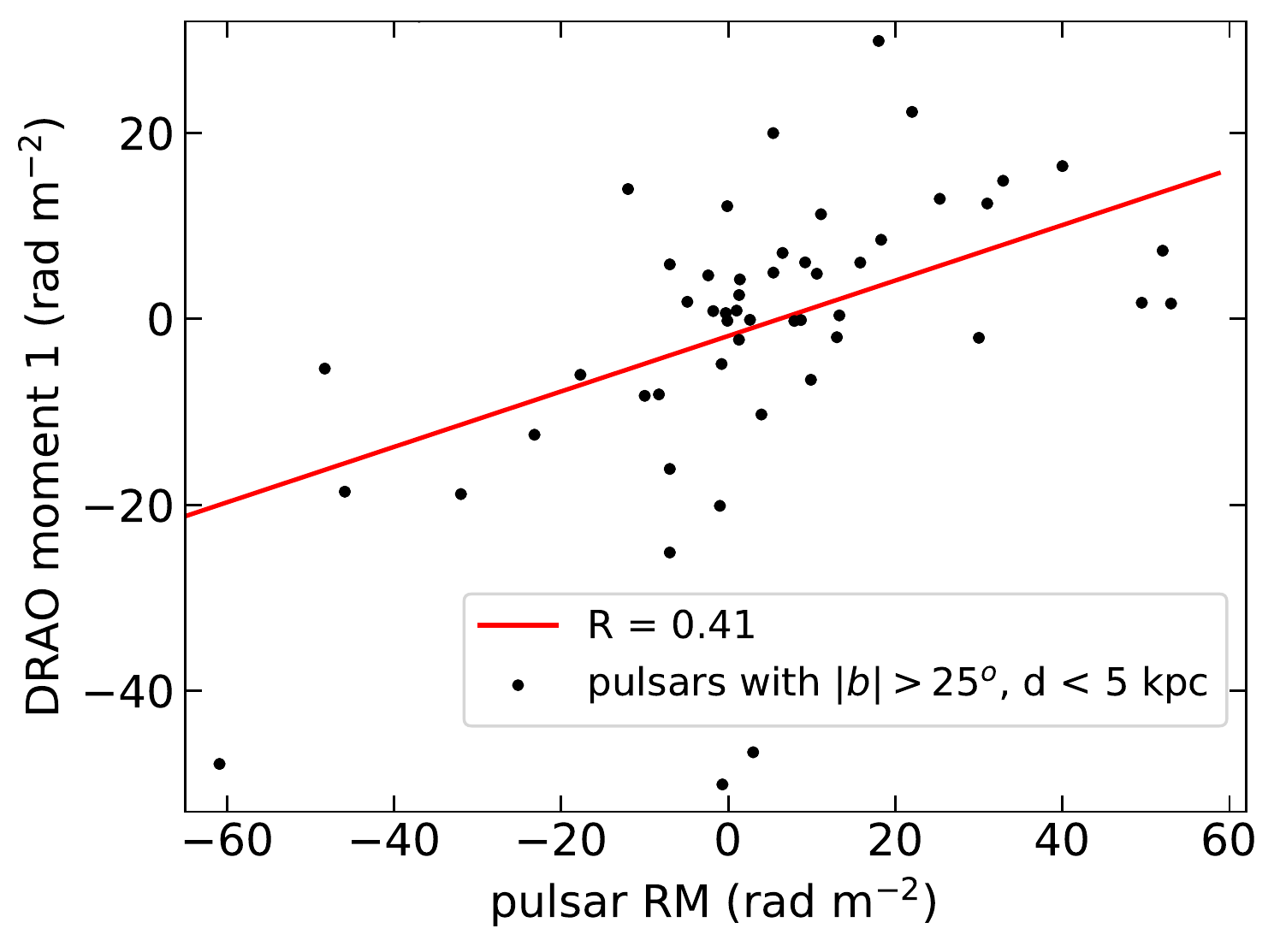}

\caption{
DRAO survey first moments in the directions of pulsars with known rotation measures.
Left panel:  13 pulsars above $|b|=25^o$ with distance $d < 0.7$ kpc.
Right panel: 53 pulsars above $|b|=25^o$ with distance $ d < 5$ kpc.  
The smaller sample of the more nearby pulsars shows
considerably better correlation than the larger sample (R=0.75 vs. R=0.41).
They both have quite low P values (0.0032 and 0.0023 see table \ref{tab:R_andP})
indicating that the probability of null hypothesis is well below 1\%.}
\label{fig:pulsar_drao_1}
\end{figure}

Comparison of the pulsar RMs to the first moments of the polarization surveys in the directions
of the pulsars is useful to see roughly the range of distances from which the bulk of the
polarized emission must come.  Figure \ref{fig:pulsar_drao_1} shows moderate correlation
between pulsar RMs and first moments in the DRAO survey, with the pulsars selected to
be at latitudes above $|b|=25^o$ and with distances,
$d < 5$ kpc (right panel).  The correlation is better if we restrict
the distances of the pulsars to $d < 700$ pc, as shown on the left 
hand panel of figure \ref{fig:pulsar_drao_1}.  Although the number of points
is less ($n$=13) the correlation coefficient is higher, R = 0.75 vs. R = 0.41 for 
the larger sample.  The probability of null-hypothesis, i.e. the chance
that the sample is taken from a population with R = 0, formally the
``two-sided P value'', is
0.0032 for the left panel ($n$=13 points) and 0.0023 for the right panel
($n$=52 points) of figure \ref{fig:pulsar_drao_1}.

On table \ref{tab:R_andP} are shown the R and P values for samples of
pulsars selected by distance ($d < d_{max}$).  The first moments of the 
DRAO and Parkes surveys and the Galactic foreground computed from the extragalactic grid of RMs are correlated
against the RMs of the pulsars in the same directions.  The effects illustrated on figures 
\ref{fig:pulsar_vs_exgal_rms} and \ref{fig:pulsar_drao_1}
are similar for many of the samples on the table. However, although the extragalactic
RMs are correlated with the pulsar RMs with R between 0.69 and 0.88 for the full
range of distances, the DRAO first moments show stronger correlation with
pulsar RMs for samples with distances less than about 1.5 kpc, and the strongest correlation
is for $d < $700 pc.  Note that the numbers of pulsars in each sample, and their
values of $\sigma_{RM}$, shown in the second and third columns
on table \ref{tab:R_andP}, are computed over the whole sky.  The numbers of pulsars
in the areas of the DRAO and Parkes surveys are smaller, indicated by n in columns 5 and 8 on
the table.

%  old version \begin{deluxetable}{|lrr|rr|rr|rr|}[h]
\begin{deluxetable}{|lrr|rrr|rrr|rr|}[h]
\tablecolumns{9}
\tablecaption{Correlation Coefficients \label{tab:R_andP} -- pulsars with $|b| > 25^o$}
\tablehead{\multicolumn{3}{c}{pulsar sample}&
\multicolumn{3}{c}{DRAO moment 1} &
\multicolumn{3}{c}{Parkes moment 1}  &
\multicolumn{2}{c}{Extragalactic}\\ \colhead{$d$ (kpc)} & \colhead{$n$} &
\colhead{$\sigma_{RM}$}
&\colhead{R}&\colhead{n}&\colhead{P}&\colhead{R}&\colhead{n}&\colhead{P}&\colhead{R}&\colhead{P} 
%  pre_28_sept_18: &\colhead{R}&\colhead{P}&\colhead{R}&\colhead{P}&\colhead{R}&\colhead{P} 
}
\startdata
%  old version (pre 9 aug 2018)  not much change...
%$d < 0.3$ & 8 & 3.3 & 0.69 & 0.20 & -0.08 & 0.85 & 0.80 & 0.16 \\
%$d < 0.5$ & 14 & 3.1 & 0.69 & 0.039 & -0.08 & 0.85 & 0.86 & 9$\cdot 10^{-5}$ \\
%$d < 0.7$ & 19 & 3.0 & {\bf 0.75} & {\bf 0.0033} & -0.34 & 0.19 & 0.88 & $< 10^{-5}$ \\
%$d < 1.0$ & 33 & 3.9 & 0.57 & 0.0044 & -0.33 & 0.10 & 0.70 & $< 10^{-5}$ \\
%$d < 1.5$ & 49 & 4.0 & 0.48 & 0.0050 & -0.19 & 0.24 & 0.64 & $< 10^{-5}$ \\
%$d < 2.0$ & 59 & 4.2 & 0.26 & 0.11 & -0.27 & 0.06 & 0.74 & $< 10^{-5}$ \\
%$d < 3.0$ & 72 & 4.6 & 0.28 & 0.85 & -0.19 & 0.15 & 0.75 & $< 10^{-5}$ \\
%$d < 5.0$ & 76 & 4.6 & 0.03 & 0.83 & -0.19 & 0.15 & 0.74 & $< 10^{-5}$ \\
%\enddata
%  Han's file only (no Manchester data)
%  now with DRAO threshold 0.03 instead of 0.04
$d < 0.3$ & 8 & 11.6 & 0.74 &5& 0.15  & -0.17 &7& 0.70 & 0.75 & 0.033 \\
$d < 0.5$ & 14 & 11.0 & 0.70 &9& 0.03 & -0.16 &10& 0.66 & 0.80 & $6\times10^{-4}$  \\
$d < 0.7$ & 19 & 13.2 & {\bf 0.75} &13& 0.0032 & -0.38 &15& 0.16 & 0.88 & $1\times 10^{-6}$ \\
% Copy Editor:  please leave the \bf in the above line
$d < 1.0$ & 34 & 15.6 & 0.45 &22& 0.036 & -0.36 &26& 0.08 & 0.73 & $ 1\times10^{-6}$ \\
$d < 1.5$ & 54 & 16.8 & 0.42 &33& 0.014 & -0.21 &44& 0.18 & 0.69 & $< 10^{-6}$ \\
$d < 2.0$ & 65 & 19.8 & 0.27 &40& 0.09 & -0.26 &53& 0.07 & 0.74 & $< 10^{-6}$ \\
$d < 3.0$ & 78 & 21.4 & 0.28 &48& 0.06 & -0.18 &62& 0.17 & 0.74 & $< 10^{-6}$ \\
$d < 5.0$ & 86 & 23.9 & 0.41 &52& 0.0023 & -0.06 &69& 0.62 & 0.75 & $< 10^{-6}$ \\
%\enddata
% Manchester data with Han values added for new pulsars only
%$d < 0.3$ & 8 & 10.7 & 0.69 & 0.20 & -0.08 & 0.85 & 0.77 & 0.026 \\
%$d < 0.5$ & 14 & 10.7 & 0.69 & 0.04 & -0.08 & 0.81 & 0.82 & 0.0004  \\
%$d < 0.7$ & 19 & 9.5 & {\bf 0.75} & {\bf 0.0033} & -0.35 & 0.19 & 0.89 & $< 10^{-6}$ \\
%$d < 1.0$ & 33 & 8.9 & 0.44 & 0.038 & -0.33 & 0.10 & 0.74 & $ 1\cdot10^{-6}$ \\
%$d < 1.5$ & 49 & 15.3 & 0.36 & 0.049 & -0.19 & 0.24 & 0.67 & $< 10^{-6}$ \\
%$d < 2.0$ & 59 & 17.4 & 0.09 & 0.58 & -0.27 & 0.06 & 0.76 & $< 10^{-6}$ \\
%$d < 3.0$ & 72 & 20.5 & 0.14 & 0.34 & -0.19 & 0.15 & 0.74 & $< 10^{-6}$ \\
%$d < 5.0$ & 76 & 20.3 & 0.15 & 0.31 & -0.19 & 0.15 & 0.74 & $< 10^{-6}$ \\
\enddata
\end{deluxetable}
%Since the pulsar RMs correlate with both the DRAO first moments and
%the extragalactic foreground grid for these samples of 
%nearby, high latitude pulsars, then the DRAO first moments should correlate with
%the extragalactic RMs also. 
%In fact they do, as shown on figure \ref{fig:exgal_drao}.  Here we use
%the same sample of directions determined by the pulsar samples shown on
%the left panel of figure \ref{fig:pulsar_drao_1}, i.e. distances less than 0.7 kpc
%and latitudes $|b| > 25^o$.

%\begin{figure}[ht]
%\hspace{2.1in}\includegraphics[width=3.5in]{exgal_drao.pdf}
%
%\caption{
%DRAO survey first moments vs. extragalactic foreground RMs in the same directions as the pulsars
%shown on  figure \ref{fig:pulsar_drao_1} (right panel).
%The correlation has R = 0.61. This is similar to the correlation
%between the pulsar RMs and the extragalactic RMs, which gives R = 0.89
%for this sample of directions.
%} \label{fig:exgal_drao}
%\end{figure}
%
For the Parkes survey first moments, there is {\bf negative correlation} with the
% Copy Editor:  please leave the \bf in the above line
pulsar RMs for similar samples of nearby, high latitude pulsars (e.g. R = -0.38 for
a sample of 15 pulsars in the Parkes declination range with distance less than 0.7 kpc).  This anti-correlation
is not statistically significant (P = 0.16).  Similarly, there is no significant correlation
between the Parkes first moments and the extragalactic RMs in the directions of
nearby pulsars.  The absence of correlations between the Parkes first moments and
other RM tracers suggests that the high latitude polarized emission seen at the
low frequencies of the Parkes survey is mostly quite nearby, probably within a few hundred
parsecs.  There may be more distant emission in some areas; discrete 
structures at greater distances would be missed
by these small samples of pulsar-selected directions.

\begin{figure}[ht]
\hspace{2.1in}\includegraphics[width=3.5in]{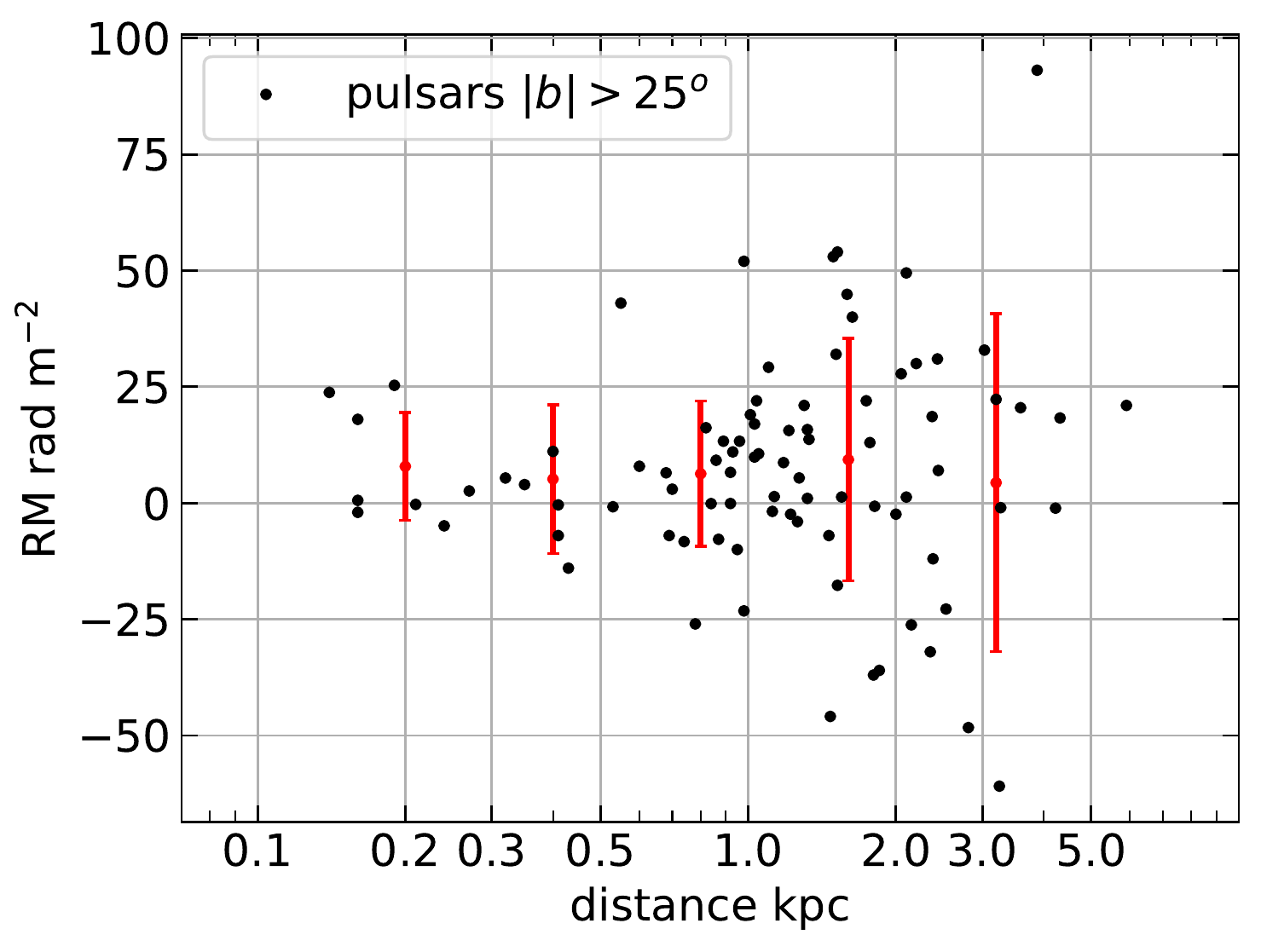}

\caption{The distribution of pulsar RMs vs. distance on a semi-log scale.
The dispersion of the RM distribution increases with distance, 
as indicated on column 3, table \ref{tab:R_andP} and shown by the red error bars.
} \label{fig:pulsar_rm_log_dist}
\end{figure}

A rough idea of the distance to the bulk of the Parkes survey polarized emission
is indicated by figure \ref{fig:pulsar_rm_log_dist}. 
Selecting pulsars with $|b| > 25^o$ as in the analysis above, the
standard deviation of the pulsar RMs increases with
the distance of the sample, starting from about $\sigma_{RM}$ = 12
rad m$^{-2}$ for distance of about 200 pc, and increasing smoothly to 
26 rad m$^{-2}$ for distance 1.6 kpc (in bins of width a factor of two in distance).  The dispersion
of the all-sky average Faraday spectrum of the Parkes survey data is just 4.5
rad m$^{-2}$ (table \ref{tab:Gaussians} and figure \ref{fig:skyav}).
From this we draw the conclusion that the bulk of the polarized emission at 300 to 500 MHz
is coming from distance less than 0.3 to 0.5 kpc if we assume that the
scatter of the pulsar RMs is generated by the same process as the 
width of the Faraday spectrum of the polarized emission, and
bearing in mind that a background source should show twice the mean RM of a slab that
has mixed emission and Faraday rotating material.
At high latitudes, most of the scatter in both quantities probably comes from a random walk
through the line of sight distribution of magnetic
field directions and interstellar electron densities.
This process leads to the depolarization of the emission from distances greater than
about 1 kpc. 

\section{Faraday Depolarization \label{sec:depol}}

%The RM is fundamental to studies of the diffuse polarized Galactic emission
%like this one.  A very similar but distinct concept is Faraday depth, $\phi$.
%The line of sight integral of the electron density ($n_e$) times the 
%parallel component of the magnetic field ($B_{||}$) gives $\phi$
%\begin{equation} \label{eq:FD_def}
%\phi \ = \ -0.81 \ \int_0^d \ n_e \ B_{||} \ ds 
%\end{equation}
%where the integral is taken from the observer to the source.  If 
%$n_e$ is in units of cm$^{-3}$, $B_{||}$ is in $\mu$G, and the distance to the
%source $d$ is in parsecs, then
%$\phi$ is given by equation (\ref{eq:FD_def}) in  rad m$^{-2}$.  Then the
%rotation angle of the plane of linear polarization is 
%\[ \chi \ = \ \phi \ \lambda^2 \]
%with $\lambda^2$ in m$^2$ and $\theta$ in radians.  
%This is sometimes thought of as the definition of RM, but it involves a hypothetical
%smooth, uniform density medium with a $\vec{B}$ field that is
%constant in magnitude and direction that effects the background polarized emission without adding
%any synchrotron emission internally, so it is obviously based on an idealized case.
%But this equation is useful for estimating the distance to the polarization horizon, since
%such a hypothetical rotating slab will begin to depolarize the background emission when
%$\chi \simeq \frac{\pi}{2}$, the condition for changing $Q$ into $U$, $U$ into $-Q$.

%Just as the extinction and obscuration due to dust grains in interstellar
%clouds causes an optical horizon that is a strong function of Galactic latitude,
Although the rotation measure does not increase monotonically with distance along the line
of sight, the Faraday depolarization does.  Thus
the polarization horizon at any given wavelength may
recede or approach the observer by factors of three or even ten from one direction
to another.  Several different physical processes contribute to depolarization, falling
into four groups:  {\bf depth} depolarization, {\bf beam} depolarization,
% Copy Editor:  please leave the \bf in the above line
{\bf bandwidth} depolarization, and {\bf geometric} depolarization 
% Copy Editor:  please leave the \bf in the above line
%(Burn 1966, Tribble 1991, Sokoloff et al. 1998).
\citep{Burn_1966, Tribble_1991, Sokoloff_etal_1998}.
Bandwidth
depolarization depends on the resolution
of the spectrometer, as given on table \ref{tab:surveys}
$\delta \phi, \ \phi_{max}$, and $\phi_{max-scale}$.
These depend on the survey parameters and on the Faraday depth.  For
high values of $\phi$, close to $\phi_{max}$, the finite
channel width attenuates the strength of the polarized signal.
Depth depolarization is a radiative transfer effect in a medium with mixed
thermal and cosmic ray electrons and magnetic field, where Faraday rotation changes
the plane of polarization of the radiation
as it moves toward the observer along the line of sight.  After propagating through
a medium for a distance such that $\chi \simeq \pi$ radians, the polarization from the
near side destructively interferes with that from the far side.  The distance required
is inversely proportional to $\lambda^2$, by equation (\ref{eq:FD_def}).  This occurs
even in an entirely uniform medium, but also in a medium with irregularities in the 
electron density and/or the strength or direction of the $\vec{B}$ field.  Geometric
depolarization occurs when two emission regions along the same line of sight have different
projections of the $\vec{B}$ field on the plane of the sky, so that their polarization 
adds in a random way, and Stokes $Q$ or $U$ or both can sum to zero.  Geometric depolarization
is not a Faraday effect, it is independent of $\lambda$, but its effect can be mixed with Faraday rotation
to give a $\lambda$ dependence.
Finally, beam depolarization comes from variation of the position angle of 
the linear polarization on different lines of sight within the area of the telescope beam, caused either by the
geometry of the emission or by changes in the Faraday depth along nearby lines of
sight, that are not resolved by the telescope.  For the single dish observations described here,
with beam widths of 30\arcmin \ to 80\arcmin, beam depolarization and depth depolarization are the most
significant effects that limit the distance that these surveys can see.
%of depolarization, particularly at lower latitudes where the line of sight distance 
%through the magneto-ionic medium is long.

For the simplified case of beam depolarization arising from
varying Faraday rotation mixed inhomogeneously with polarized emission along
different lines of sight within the telescope
beam,
\citet{Sokoloff_etal_1998}
derive a result (their equation 34) for
the combined effects of depth and beam depolarization, $DP$, 
based on a single complex parameter, $S$.  If $T(\lambda^2)$ is the observed
polarized intensity (equation \ref{eq:Fourier}), and $T_0(\lambda^2)$ is the 
intrinsic polarized brightness of the source, then 
\begin{equation} \label{eq:DP}
DP \ \equiv \ \frac{T}{T_0} \ \approx \ \left| \frac{1\ - \ e^{-S}}{S} \right| 
\end{equation}
where the parameter $S$ is defined as
\begin{equation} \label{eq:depol}
S \ = \ 2 \ \sigma_{RM}^2 \ \lambda^4 \ - \ 2 \ i \ \lambda^2 \ \mathbb{R} \ = \ A \ + \ i C
\end{equation}
where $\mathbb{R}$ is the (maximum) Faraday depth of the emission region,
that we take equal to the absolute value of the extragalactic foreground RM, and $\sigma_{RM}$
is the rms fluctuation of the RM measured on the scale of the beam width
%Note that both a $\lambda^2$ dependence and a $\lambda^4$ dependence 
%appear in equation \ref{eq:DP} and \ref{eq:depol} 
%(see also Burn 1966 eq. 18).
\citep[see also][eq. 18]{Burn_1966}.
Taking $A$ and $C$ as the real and imaginary parts of $S$ in equation \ref{eq:DP} gives
\begin{equation}\label{eq:DP2}
DP \ \approx \left| \frac{1}{A^2 + C^2} \ \left\{ A \ - \ e^{-A} \  \left(A \cos{C} - C \sin{C}\right) 
\ -\ i \left[ C \ - \ e^{-A} \  \left( A \sin{C} \ + \ C \cos{C} \right) \right] \right\} \right|
\end{equation}
and finally
\begin{equation}
DP \  \approx \ \frac{1}{A^2 + C^2} \ \sqrt{\left[A - e^{-A} \left(A \cos{C} - C \sin{C}\right) \right]^2 \ + \ 
\left[ C - e^{-A} \left( A \sin{C} + C \cos{C}\right) \right]^2 }
\end{equation}
where $A$ increases as $\lambda^4$ and $C$ increases as $\lambda^2$ as we go to longer wavelengths.  Thus equation \ref{eq:depol} is consistent with the conclusion
of \citet{Tribble_1991} that $DP \propto \lambda^4$ at short wavelengths and 
$DP \propto \lambda^2$ at long wavelengths.
%(table \ref{tab:surveys}).

\begin{figure}[ht]
\hspace{1in}\includegraphics[width=5in]{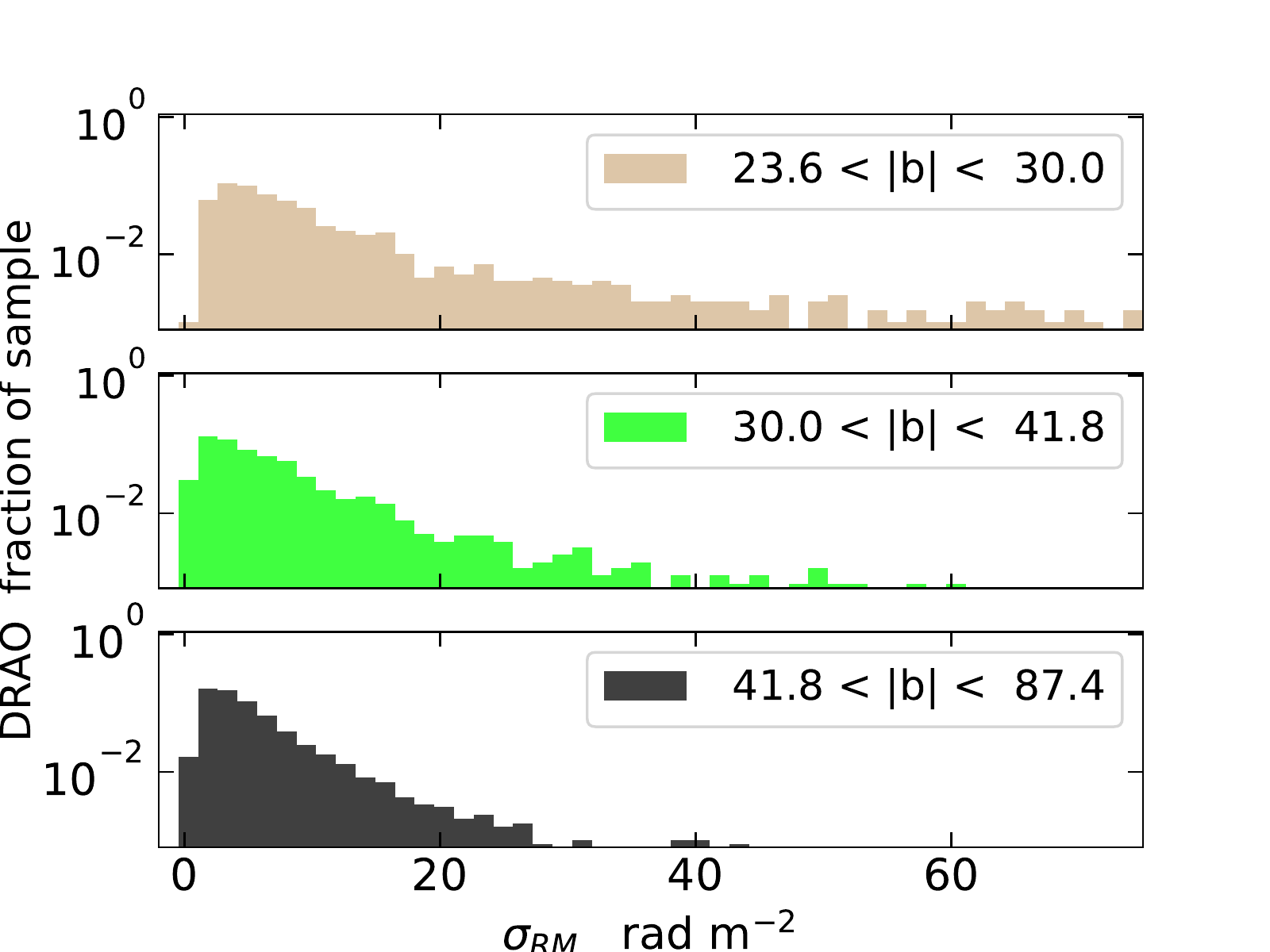}

\caption{The distribution of standard deviations of the first moments, $\sigma (M_1)$ 
for the DRAO survey.  The values of $\sigma$ are computed for each point in a grid of centers
separated by 90\arcmin, with $\sigma$ the standard deviation of the values of the first moment
for each pointing center,
in an annulus of points between 58\arcmin and 68\arcmin (equation \ref{eq:depol}) from the center.
Samples of $\sigma$ values measured in different latitude ranges, corresponding to
steps of 0.5 in cosec$|b|$, are shown separately.  
Note that the y-axis is logarithmic,
and the distributions decrease roughly exponentially above their peaks (linear
on the semi-logarithmic axes of these figures).
\label{fig:sdistrib_DRAO} 
}
\end{figure}
\begin{figure}[ht]

\hspace{1in}\includegraphics[width=5in]{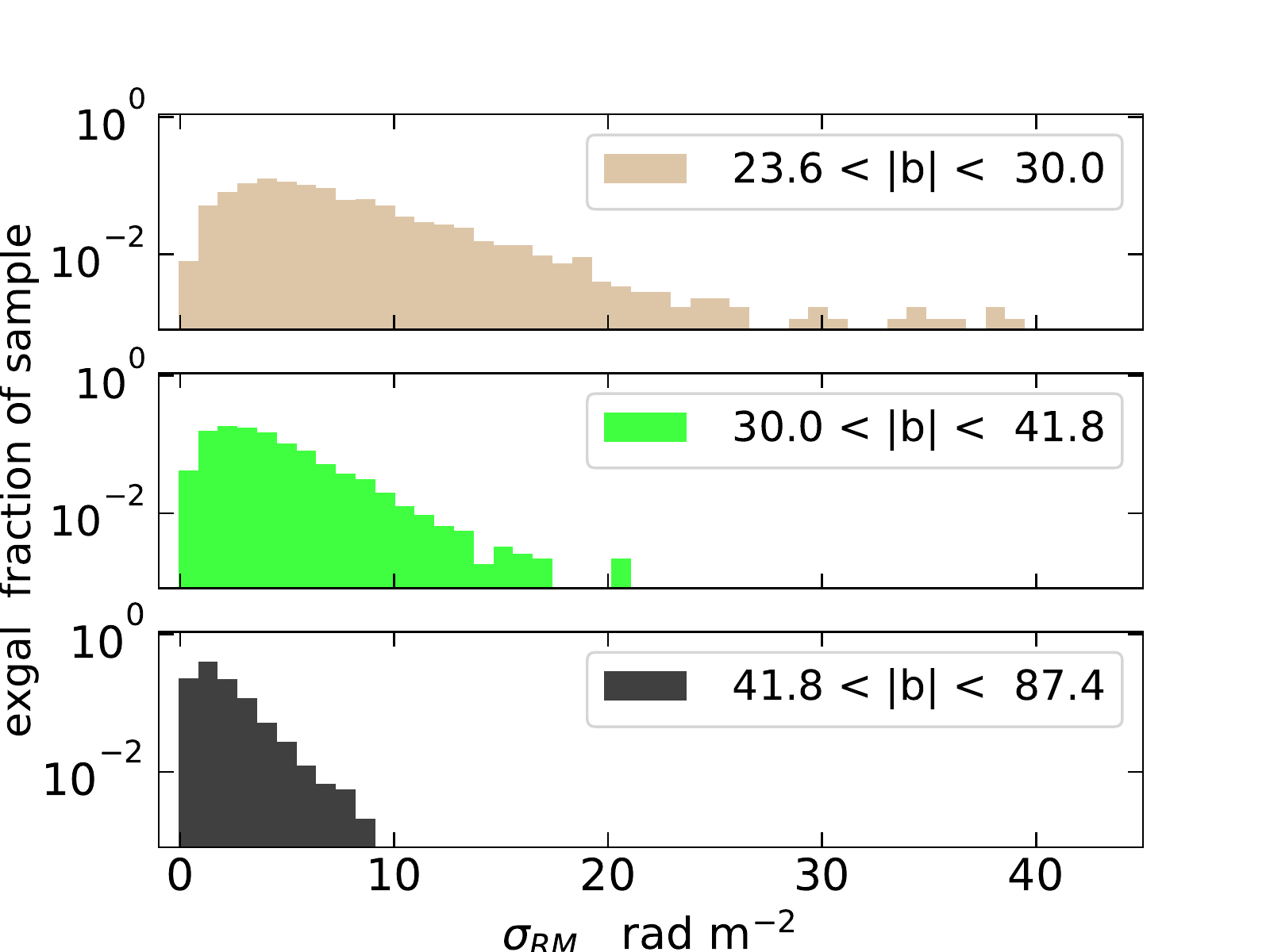}

\caption{The distribution of standard deviations of rotation measures, $\sigma (RM)$ computed
for the extragalactic foreground rotation measures.  The center points and annular areas are the same
as used for figure \ref{fig:sdistrib_DRAO} and \ref{fig:sdistrib_Parkes}
, as are the axes on the figure, but note the different scale on the x axis.
\label{fig:sdistrib_exgal}
}
\end{figure}
\begin{figure}[h]

\hspace{1in}\includegraphics[width=5in]{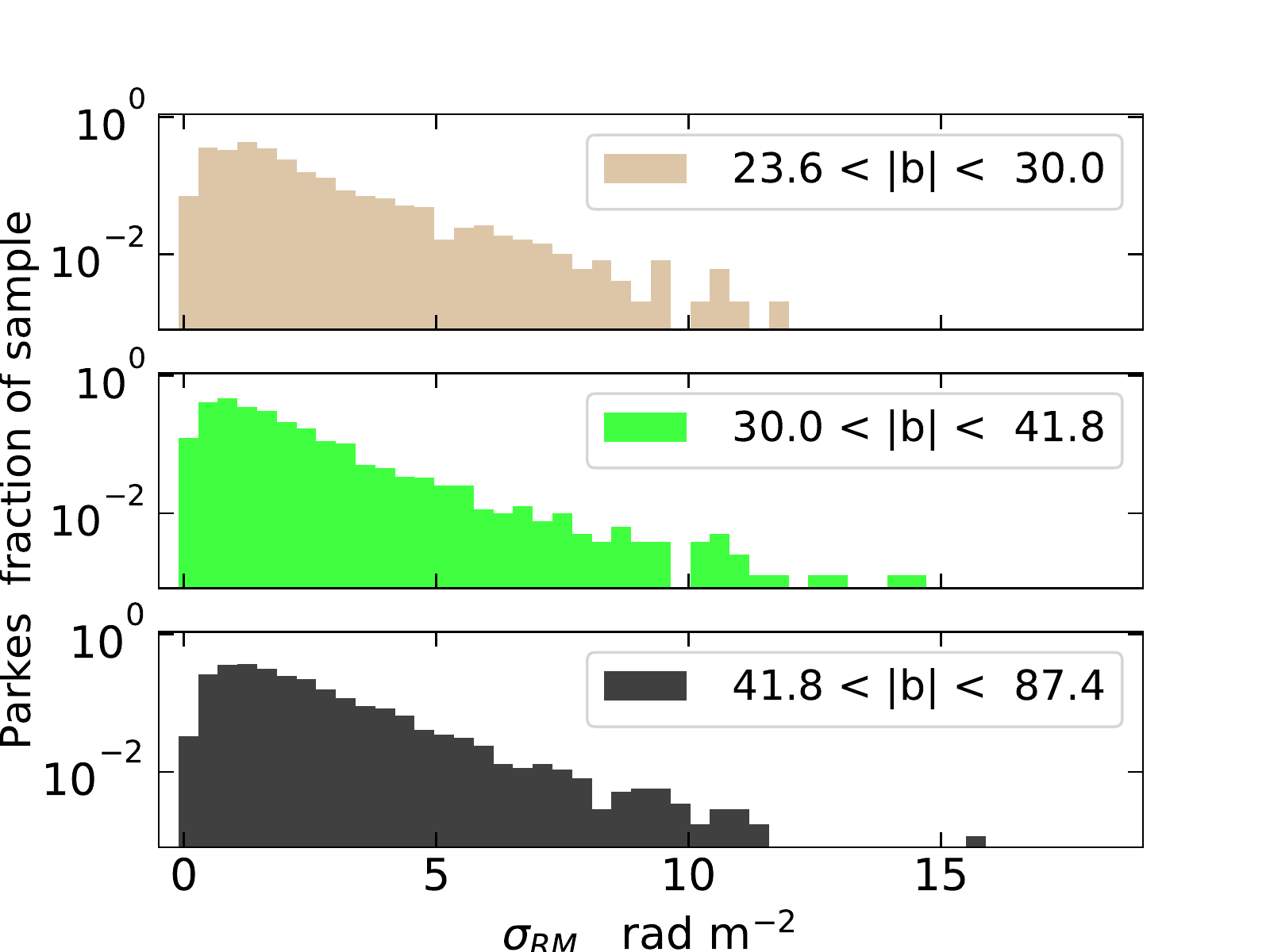}

\caption{The distribution of standard deviations of rotation measures, $\sigma (M_1)$ computed
for the Parkes survey first moments.  The center points and annular areas are the same
as used for figure \ref{fig:sdistrib_DRAO}, as are the axes on the figure, but note the different scale on the x axis.
\label{fig:sdistrib_Parkes}
}
\end{figure}

To evaluate $\sigma_{RM}$ for the two GMIMS surveys, we compute the standard deviation of
the observed first moments over an annulus just outside the 40\arcmin \ 
beam radius of the Parkes survey. 
%  xxx get the number of pointint centers...
 For each independent pointing center, i.e. pixels spaced
by 90\arcmin \ in latitude and in longitude/cos($|b|$), we take the standard deviation 
over all pixels $i = 1 \dots N$ that are in  an annulus with inner radius 58\arcmin \ and outer radius 68\arcmin:
\[ \sigma_{M1} \ = \ \sqrt{\frac{1}{N}\sum_{i=1}^N \left(M_{1,i} - \bar{M_1}\right)^2} \]
(note that the {\it NumPy} `nanstd' function used here gives the population standard deviation rather than the sample standard
deviation, which has $N-1$ in the denominator instead of $N$).
The number of pixels $N$ contributing to these samples depends on the latitude,
but it is typically twelve or more.  The distribution of values of $\sigma$ determined 
for these annuli for the two surveys is shown on figures \ref{fig:sdistrib_DRAO} - \ref{fig:sdistrib_Parkes}.  Also shown are
sigmas computed for the extragalactic sample of RMs, over the same areas with
the same centers.  The progressively narrower distribution of RMs at higher latitudes was
noted by \citet{Schnitzeler_2010}, using a similar cosec$|b|$ approach to
separate the Galactic and extragalactic contributions to the RMs of NVSS
sources.  Here the averaging associated with the \citet{Oppermann_etal_2015}
model separates the Galactic
foreground from the extragalactic RM contribution, at least nominally.

\begin{figure}[h]
\hspace{.1in}\includegraphics[width=3.7in]{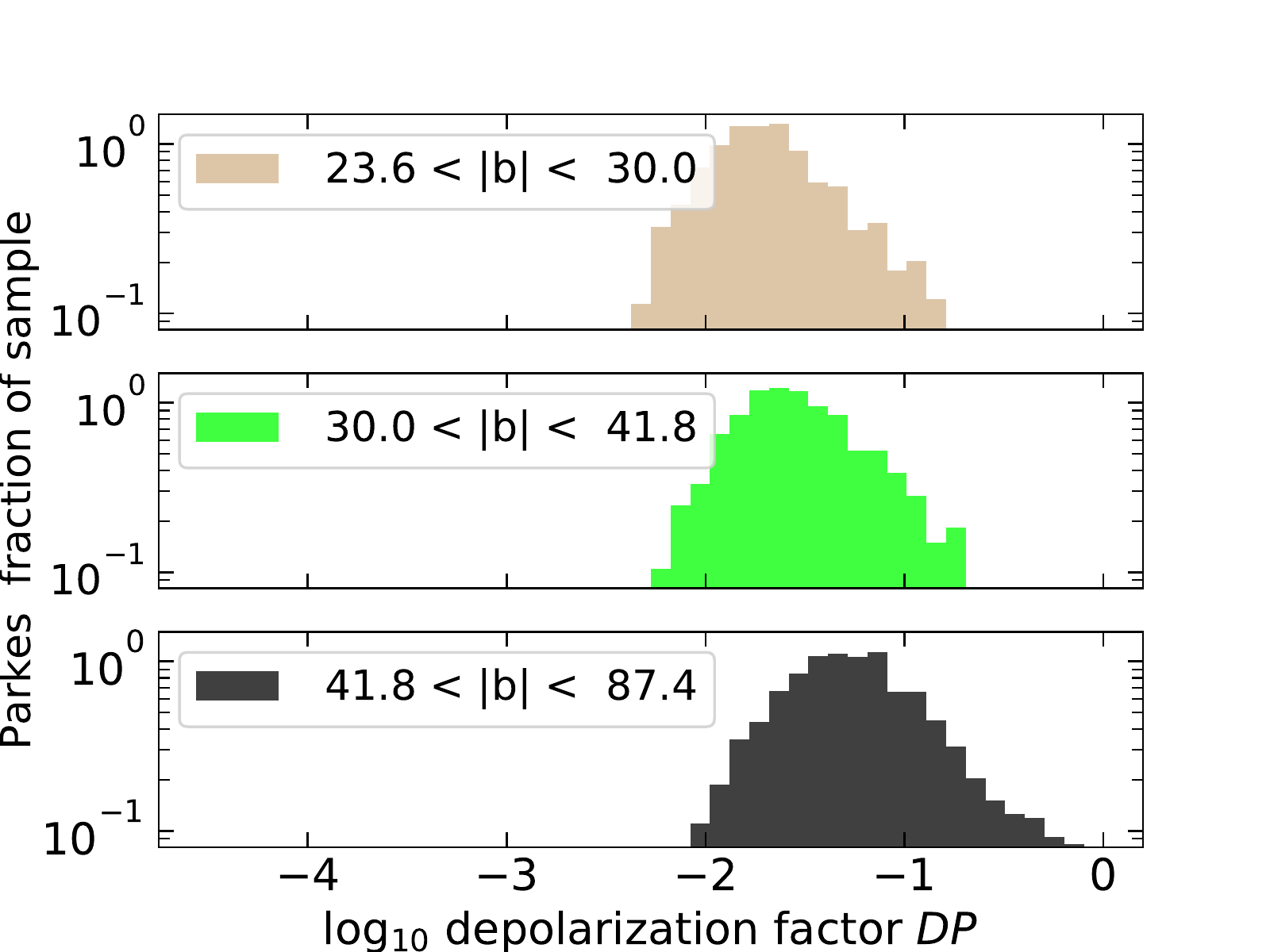}
\hspace{-.2in}\includegraphics[width=3.7in]{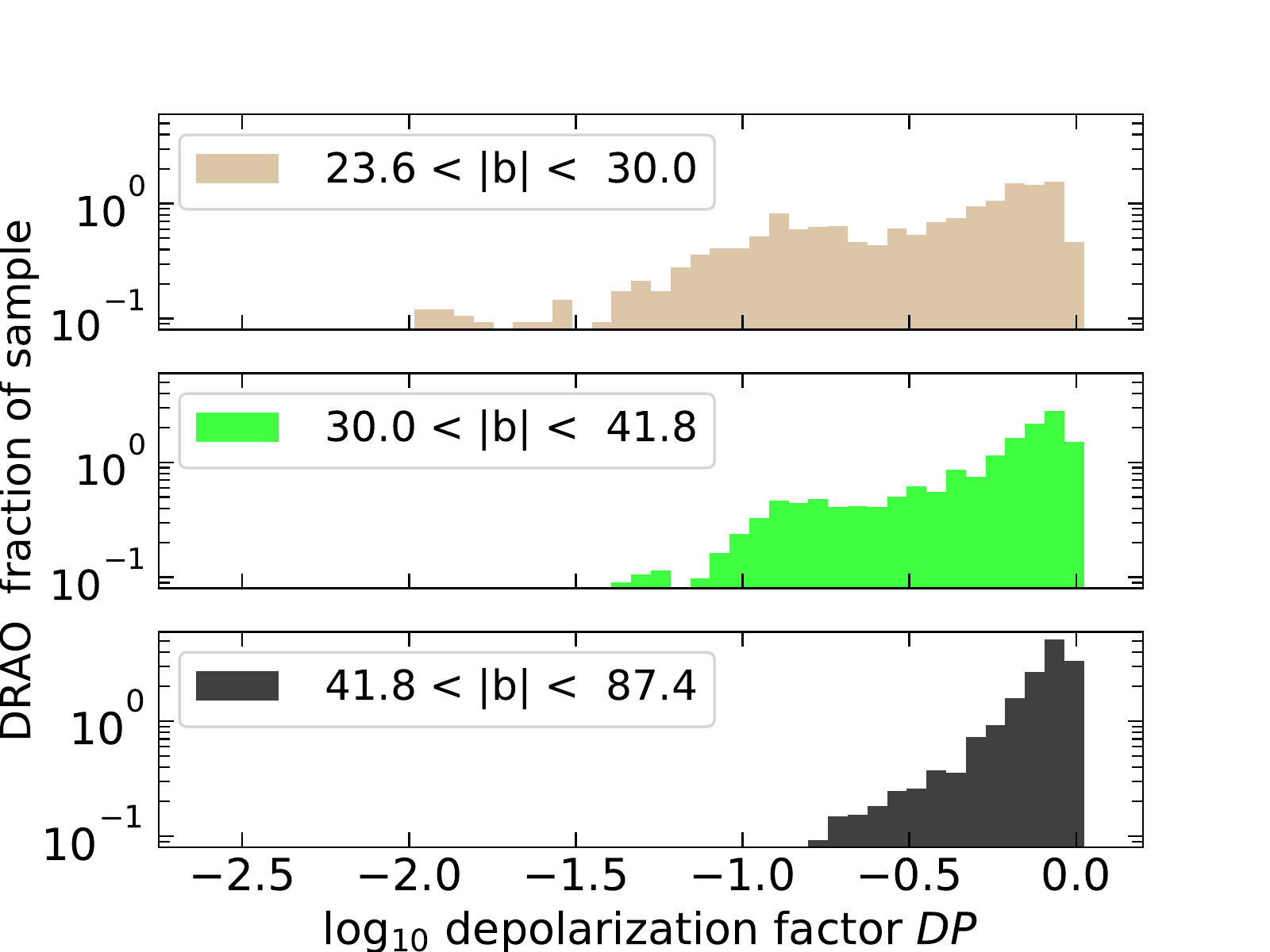}
\caption{The depolarization factors predicted by equations \ref{eq:DP} - \ref{eq:DP2}, 
for the distribution of pointing centers described in the text.  The left panel shows 
that the Parkes survey should be highly depolarized, with a median value of $DP \sim 0.05$ even at
the highest latitudes, decreasing to $DP \sim 0.02$ at the intermediate latitudes plotted on the
top panel of the left figure.  For the DRAO survey, depolarization is not so strong, particularly
at the higher latitudes, as shown in the right figure bottom panel, for which the median value
of $DP \sim 0.77$  decreasing to $DP \sim 0.29$ at the intermediate latitudes shown in the top
panel.
}
\label{fig:depols}
\end{figure}

Although $\sigma_{M1}$ is computed over areas on the sky about two to four times larger than the beamwidth of
the telescope, we will assume that it gives an estimate, probably an overestimate, of $\sigma_{RM}$, which
is the rms fluctuation of the $RM$ in a single beam area.  We cannot measure $\sigma_{RM}$ {\bf inside} the DRAO and
% Copy Editor:  please leave the \bf in the above line
Parkes beams without going to
higher resolution, either with a larger single dish or an aperture synthesis telescope.  For the
simplified analysis in this section, we will assume the two are roughly equal.
Using these values of $\sigma_{RM} \simeq \sigma_{M1}$ measured over
the annular areas around each of the grid of pointing
centers we can determine the expected depolarization using equations
(\ref{eq:DP}) and (\ref{eq:depol}). These are shown on figure
\ref{fig:depols} for three ranges
of Galactic latitude, with cosec$|b|$ in the ranges 1 to 1.5, 1.5 to 2, and 2 to 2.5.
For the Parkes data (left panel), the values of $DP$ are mostly less than $10^{-1}$, with some below
10$^{-2}$ at the lower latitudes.  On the other hand, for the DRAO survey, the median value of DP
predicted for latitudes above $|b|=42^o$ is 0.77.  Thus depolarization should not be very significant
for this survey at high latitudes.  At lower latitudes the medians decrease to 0.53 and 0.29 in
the middle and upper right hand panels of figure \ref{fig:depols}.  So depolarization is becoming
significant at intermediate latitudes.  This result explains why the correlation between $M_1$ from the
DRAO survey and the pulsar RMs weakens for pulsar distances greater than 700 pc to 1 kpc.  For example,
if the median $DP = 0.5$ at cosec$|b|$=2 ($b = 30^o$) and this corresponds to a distance of 800 to 1000 pc, then
the scale height of the magneto-ionic layer causing the depolarization at this wavelength
should be about 400 to 500 pc.
Although the depolarization estimates derived from equations 7 and 8 appear to be conclusive in
explaining the difference between the Parkes and DRAO survey volumes, these equations were derived
for an idealized situation more relevant to supernova remnants or other
galaxies than to the all-sky surveys
discussed here.  More analysis and simulations will give a better understanding of
the wavelength dependence of the Faraday depolarization.

\section{Conclusion}

   Our Galaxy presents many faces; various tracers of the interstellar medium
show the effects of
the many different physical processes at work.  The magnetic field shapes the features
of these faces, even for the spectral line tracers of the cool neutral medium and
the molecular medium 
\citep[e.g.][]{Clark_2018, ZamoraAviles_etal_2018}.
For the ionized medium and the cosmic ray electrons,
the magnetic field is an important and often dominant factor in their
dynamics and evolution.   The polarization of the diffuse synchrotron emission 
observed at high frequencies 
%(Page et al. 2007, Miville-Desch\^{e}nes et al. 2008)
\citep{Page_etal_2007, Miville-D_etal_2008}
shows the structure of the magnetic fields and the cosmic ray electrons that fill
the disk and extend into the halo. 
At lower frequencies, the Faraday spectrum of the Galactic diffuse synchrotron emission
shows the juxtaposition of the emission regions with the diffuse ionized medium that causes the
Faraday rotation.  The Faraday rotating medium is thermal plasma, again
with a magnetic field, although this time it is the line of sight
component of the field that matters, in contrast to the component in the plane
of the sky that determines the position angle of the polarized emission.
Thus the Faraday spectrum holds the promise of providing distance information;
someday it may be one of several observational techniques that will allow an
accurate  three-dimensional model of the Galaxy
to be constructed including the magnetic field, the cosmic ray electrons,
and the diffuse ionized medium \citep{Su_etal_2018}.
This goal overlaps that of much recent work by low frequency arrays such as LOFAR and the MWA
\citep{Iacobelli_etal_2013, Jelic_etal_2014, Jelic_etal_2015, Lenc_etal_2016, vanEck_etal_2017}.

Two other pieces of this puzzle are pulsar rotation measures and the large samples
of extragalactic rotation measures that will be available soon, e.g. from the
POSSUM survey with the Australian Square Kilometre Array Pathfinder \citep{Gaensler_2009}.
In this paper we make an attempt to compare and contrast these three sets of
data on Faraday rotation, starting with latitudes above $\sim 25^o$ where the
path length through the magneto-ionic medium is short.  There are not quite enough
pulsars to determine distances to specific features in the Faraday spectra, but
there are fairly strong correlations between the pulsar and extragalactic foreground
RMs, and between the nearby pulsars and the first moments of the Faraday spectra from
the DRAO survey.

The spectra from the Parkes survey have much better resolution in $\phi$, and
they show compelling structure that will someday be traceable to structures in the
nearby interstellar medium, most likely at distances of a few hundred parsecs or less.
Some of these can be associated with known structures, including HII regions 
(\citealt{Thomson_etal_2018a, Thomson_etal_2018b} and see \citealt{HarveySmith_etal_2011, Gaensler_etal_2001} and
\citealt{Madsen_etal_2006} for other examples) and 
neutral interstellar clouds \citep{vanEck_etal_2017}.  But at the low
frequencies of the Parkes
survey, magnetised plasma that has a significant effect on the Faraday spectrum can
be so diffuse as to be completely undetectable in H$\alpha$ or any other  
spectral line tracer at any wavelength.  Thus as low frequency polarization
surveys like the Parkes survey improve, they will reveal more and more of the
structure of the local interstellar diffuse ionized medium.

Based on the pulsar correlation with the first moments of the DRAO survey on figure
\ref{fig:pulsar_drao_1} and table \ref{tab:R_andP} we find the best correlation for 
a sample with maximum pulsar distance 700 pc.  The correlation for a sample with maximum
distance of 1 kpc is significantly worse. 
The absence of correlation between the first moment of the Parkes survey data and the RMs of pulsars
in any distance sample suggests that 
%the horizon distance, $d_h$, in the Parkes
%survey is $d_h(Parkes) <
the polarized emission seen in that survey is mostly within about 300 pc, but the number of pulsars
closer than this (five in the DRAO survey area and seven in the Parkes survey
area, table \ref{tab:R_andP}) is too small to search for correlations effectively.
The comparison of the Parkes second moment with the dispersion of the pulsar RMs on figure 
\ref{fig:pulsar_rm_log_dist} also suggests a distance less than 500 pc.

A more sophisticated approach to determine, or at least to set limits on, the polarization horizon
in the Parkes survey is to simulate
the random magnetic field based on its turbulent spectrum, apply an electron density model,
and compute typical values of $\phi$ \citep{Hill_2018}.  That is beyond the scope of this paper, but 
several recent studies have reported impressive results that could be applied to the GMIMS 
survey, including \citet{Herron_etal_2016, Herron_etal_2018a, Herron_etal_2018b}, \citet{Hill_2018}, and \citet{Beck_etal_2016}, and see also 
%survey, including Herron et al (2016, 2018a, 2018b), Jones et al. (2018), and Beck et al. (2016), and see also 
the statistical approach used by \citet{Iacobelli_etal_2014}.

%   Many spectral line tracers of the interstellar medium show morphological
%correlations or 
%anti-correlations with each other, e.g. H$\alpha$ emission and 21-cm emission that trace 
%ionized vs. neutral hydrogen.  The Faraday medium is surprising in how little it seems to
%match any other ISM tracer, either in correlation or contrast.  Although we have various
%tracers of the diffuse ionized medium (Hill et al. xxx ), most of them depend on the local density 
%squared, because they are collisional processes such as free-free absorption or recombination
%line emission.  Pulsar dispersion measures are an exception.  Like the Faraday rotation, 
%dispersion depends on the line of sight integral of the electron density, but without the
%magnetic field.  This paper has compared pulsar distances, that mostly come from
%dispersion measures plus various electron density models of the nearby Milky Way,  
%with the rotation measure of the diffuse synchrotron emission.

The behavior of the different moments of the Faraday spectra vs. path length (cosec$|b|$)
is consistent with a paradigm where
the long wavelength polarization (Parkes data) is coming from a relatively small volume 
around the Sun, considerably smaller than the scale height of the magneto-ionic medium. 
In this region there is evidence for a vertical component of the B field
at high latitudes, with the field pointing toward the Sun from both the north and south
Galactic poles, (figure \ref{fig:path-mom1} right panel).  A similar trend is
not seen in the DRAO survey first moments, so this is
apparently a local phenomenon.  On the longer lines of sight sampled by
the DRAO survey
% added 24 nov
the first moments indicate a $\vec{B}$ field component in the $-\hat{z}$
direction, i.e. from the northern to the southern Galactic hemispheres
(figure \ref{fig:path-mom1} left panel).
This is a small effect, at high and intermediate latitudes
random variations of the field lead to a dispersion in the
measured first moments that is generally on the same order as 
the systematic effect (figure \ref{fig:sdistrib_DRAO}).  But the
trend of the first moments with latitude is confirmed
by similarly placed samples the RM-foreground map from \citet{Oppermann_etal_2015} 
(figure \ref{fig:path-exgal}).
The second moments of the DRAO survey increase
as the square root of the number of B-field structures, as in a random
walk process, hence the correlation of $m_2$ with cosec$|b|$ (left panel of figure 
\ref{fig:path-mom2}), although the less well sampled southern hemisphere points
show a weak opposite trend.  

The ultimate significance of the GMIMS survey data will depend on how much it influences the
development of comprehensive 
models of the Galactic magnetic field and the related physics of cosmic ray propagation, such
as GALPROP \citep{Strong_etal_2010, Grenier_etal_2015}.
An approach with a robust statistical basis is the IMAGINE 
Consortium Bayesian platform \citep{Boulanger_etal_2018}, that has the 
goal to unify observations of many different kinds.  Simulations of the
magneto-ionic medium to predict and study the results of rotation measure surveys 
are showing which analysis techniques are most robust and revealing
\citep{Haverkorn_etal_2008, Beck_etal_2016, Herron_etal_2018b, Reissl_etal_2018}.
As rapid progress is made in the field of Faraday spectroscopy, we can hope for 
improved models of the nearby magnetic field and its interaction with structures in
the ionized interstellar medium.

\acknowledgments
%\section{Acknowledgments}

The Parkes Radio Telescope is part of the
Australia Telescope National Facility which is funded by the Commonwealth of
Australia for operation as a national facility managed by CSIRO. 
The DRAO 26-m Telescope is operated as a national facility by the National Research Council Canada.
The Dunlap Institute is funded through
an endowment established by the David Dunlap family and the University of
Toronto.
We are grateful to JinLin Han for providing his complete table of 1001 pulsars with
RM and distance measurements.
We are grateful to Rainier Beck for a critical reading of the manuscript and helpful suggestions.
JD is grateful for the hospitality of the Boston University Center for Astrophysical 
Research where some of this work was done.
AT acknowledges the support of an Australian Government Research Training Program (RTP) Scholarship.
This research made use of Astropy,\footnote{http://www.astropy.org} a community-developed core Python package
for Astronomy \citep{astropy_2018}.

%\section{References}

\appendix

\section{Advantages of Spectral Moments vs. Two Alternatives \label{app1}}

The spectral moment analysis in this paper is one approach to simplify and convey the
information contained in the Faraday cube in the form of a small number of two-dimensional
images or maps.  In ordinary spectroscopic imaging, e.g. with an aperture synthesis telescope
observing a spectral line in emission, the moments are useful to characterise the kinematics
of the source.
The first moment may be used to trace the radial velocity field, the second moment the
turbulence, and the zero moment often gives the column density of the atoms or molecules
emitting the line.  Rotation curves of galaxies are usually fitted to the first moment map.
An alternative approach that is sometimes simpler is to fit a Gaussian line profile to
the spectrum at each pixel, and use the resulting maps of the peak, center, and width
($T_o$, $\phi_o$, and $\sigma_{\phi}$ in the notation of equation \ref{eq:Gaussian_eq})
to characterize the velocity field and to study the variation
in linewidth from point to point.
An even simpler approach is simply to find the highest point on the spectrum,
$T_{peak}$ and the corresponding velocity, or in our case Faraday depth, $\phi_{peak}$.

Most previous surveys of Galactic synchrotron polarization at frequencies above 250 MHz
have suffered from observing too narrow a range of wavelengths,
$\Delta \lambda^2$, so that the width of the RMSF, $\delta \phi$, is very broad
(table \ref{tab:surveys}).  
The result is that the Faraday spectrum is convolved with
a very broad Gaussian that smooths away the detailed structure of $T(\phi)$.
This can be seen in the black traces on figures \ref{fig:spectra_1} - \ref{fig:spectra_3}
corresponding to the DRAO survey data; since $\delta \phi = 140$ rad m$^{-2}$ for
the DRAO survey, the spectrum is effectively convolved with a broad smoothing function.
There is structure in the spectra in some directions that is
broader than this width, but mostly the observed
spectra in the DRAO survey could be approximately fitted
by Gaussians without losing much information.
Simply measuring the peak of the spectrum and its rotation measure gives a quick 
characterisation of the strength of the polarization and a single value for the RM.
This is the way that polarization surveys were done in the last century,
where a single value of the polarized brightness temperature and a single RM were calculated
over a narrow bandwidth at a given center frequency.

\begin{figure} 
\hspace{-.2in}\includegraphics[width=7.5in]{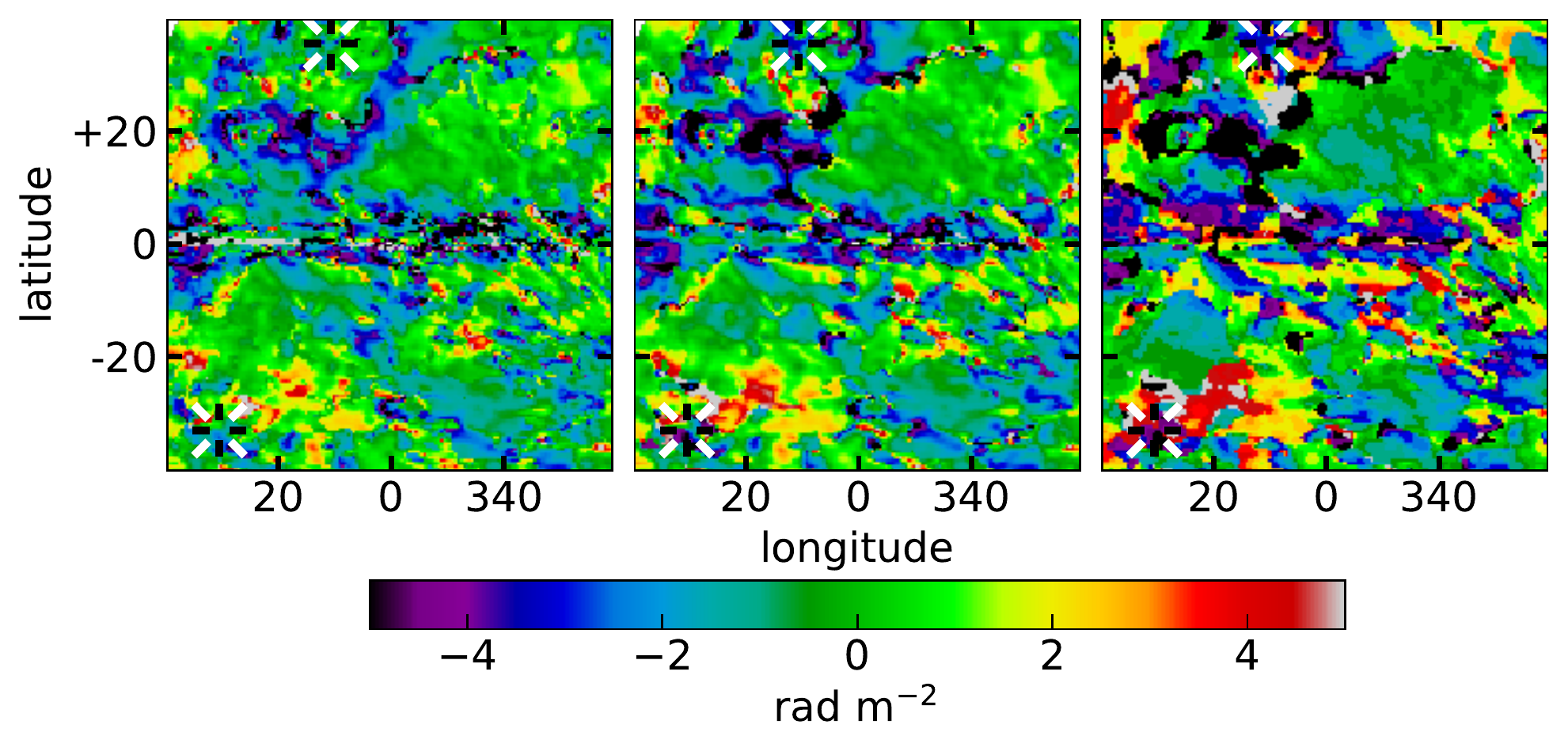}

\caption{Comparison of three different methods of calculating the central RM
of a Faraday spectrum of polarized emission.  On the left is the first moment map of a section of
the Parkes survey (taken from figure \ref{fig:mom1}, lower panel).
In the center is the same area, with the center value, $\phi_o$ of a Gaussian fit to
each pixel.  On the right is the value of $\phi_{peak}$ for the center of the channel with
the peak or highest value of $T(\phi)$.  Markers in the lower left corners and near the 
top, left of center, show the positions of the spectra illustrated in figure \ref{fig:spectra_1}.
 \label{fig:appA_mom1_comp} }
\end{figure}

For polarization surveys like GMIMS, that are attempting to measure the Faraday
spectrum with $\delta \phi$ small enough to show detailed line shapes
like those seen in the red profiles on figures \ref{fig:spectra_1} - \ref{fig:spectra_3} from
the Parkes survey, a more subtle approach is needed to characterise the distribution
of the emission over $\phi$ at each position.  The Faraday moments are a good tool
for this if there is more than one feature present in
the spectrum.  Figure \ref{fig:appA_mom1_comp}
shows three maps of the Parkes survey data, illustrating the
effect of taking the first moment,
fitting Gaussians and making a map from the fitted center ($\phi_o$), and simply finding the
peak and plotting the resulting value of $\phi_{peak}$.  The left panel is an expanded
version of the lower panel of figure \ref{fig:mom1}, showing the first moment of the Parkes survey on
an area of the inner Galaxy with $-40^o < (\ell,b) < +40^o$.  The middle panel shows
the center value, $\phi_o$, of a Gaussian fitted to the channels of the Faraday
spectrum above the threshold used in the moment calculation, and the right panel
shows the value of $\phi_{peak}$ of the highest channel of the spectrum.  

The two positions shown in the spectra on figure \ref{fig:spectra_1} are indicated
by black and white markers 
on each panel of figure \ref{fig:appA_mom1_comp}, at (longitude, latitude) = (10.72,+35.47)
and (30.51,-33.08).  The values at the centers of the markers are (-1.2, -3.1, -3.5)
rad m$^{-2}$ for the former position, and (-1.0, -1.9, -4.5) for the latter
for the left, center, and right panels.
Over the entire area shown in figure \ref{fig:appA_mom1_comp}, the mean and standard deviation
of the difference between the mean $\phi$ calculated using the moment formula and the  
Gaussian fitted $\phi_o$ are 0.09 and 1.80 rad m$^{-2}$.  The mean difference between the 
first moment and the peak $\phi$ over this area is 0.13 rad m$^{-2}$  with standard
deviation 2.6 rad m$^{-2}$.  
These differences are small, but the first moment calculation takes account of the structure
of the Faraday spectrum more carefully than taking $\phi_{peak}$.  This makes a difference
as long as the spectrum has not been heavily smoothed by a broad RMSF.  
The GMIMS survey is designed to minimize $\delta \phi$ by using wideband receivers to
cover a large fractional range of $\lambda$.  The moment calculations make the most
of this narrow Faraday spectral resolution. 
%but much larger than the error bars on the points on the right
%hand panels of figures \ref{fig:path-mom1} and \ref{fig:path-mom2}.
As \citet{vanEck_etal_2017} show, Galactic Faraday spectra measured with low
frequency telescopes commonly exhibit two or more distinct components that can
be identified with separate structures on the line of sight.
Whenever the Faraday spectrum shows multiple features, the moment calculation
gives a much better estimate of the center $\phi$ and $\phi$-width
than a single Gaussian fit or simply the peak value.  

\begin{figure} 
\hspace{-.2in}\includegraphics[width=7.5in]{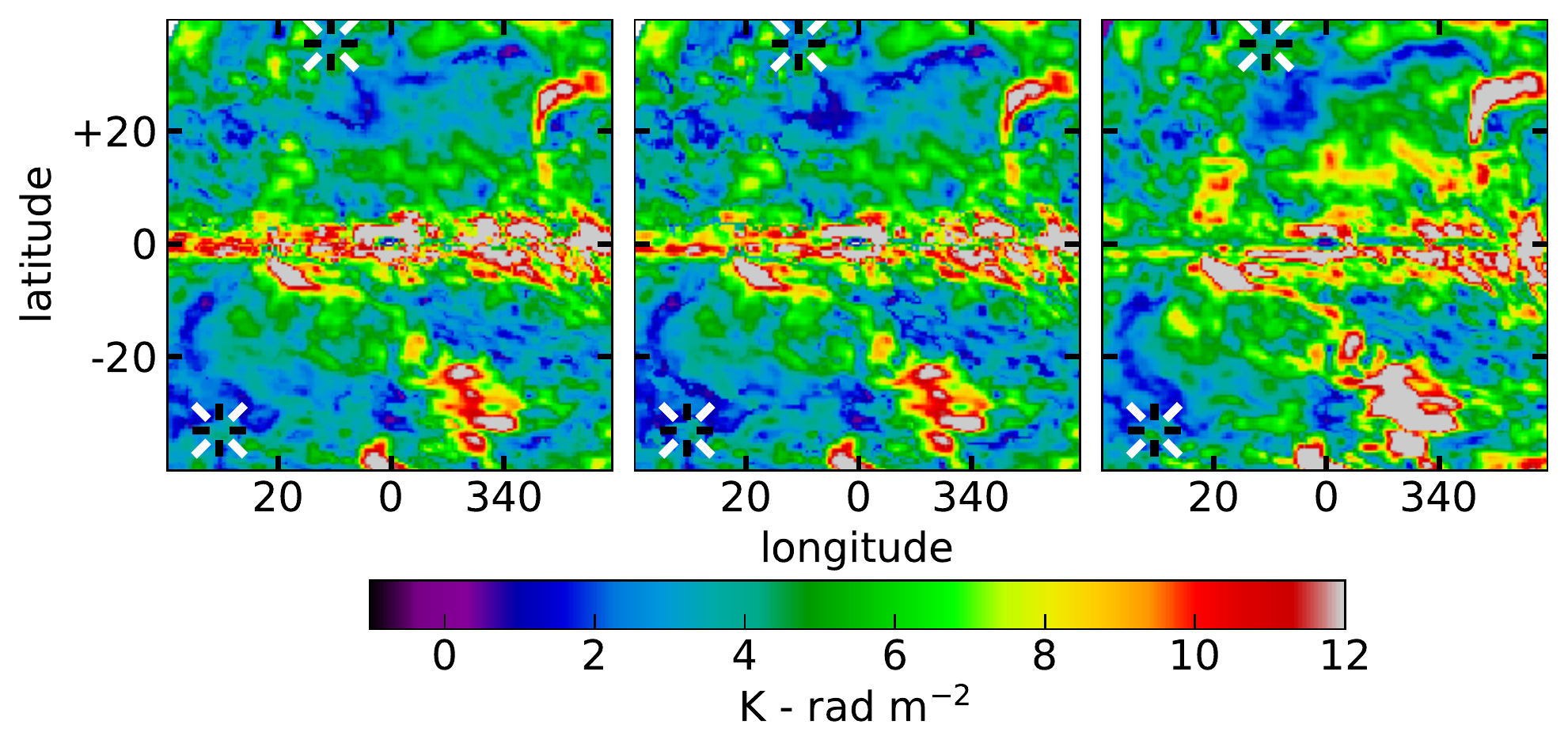}

\caption{Comparison of three different methods of calculating the integral of the
Faraday spectrum, the zero moment.  On the left is the zero moment map of a the same area
shown on figure \ref{fig:appA_mom1_comp}, the complete map is shown on figure \ref{fig:mom0},
lower panel.  In the center is the line integral computed from the best fit Gaussian,
and on the right is the peak value of the spectrum, multiplied by 
$\sqrt{2 \pi} \ \overline{\sigma_o}$
where $\overline{\sigma_o} = 4.9$ is the mean value over this area.
 \label{fig:appA_mom0_comp} }
\end{figure}

Figure \ref{fig:appA_mom0_comp} shows a similar comparison of the moment zero maps
for the region shown in figure \ref{fig:appA_mom1_comp}.  Here the 
fitted Gaussian parameters are combined to give the Gaussian integral
$G0 = \sqrt{2 \pi } \  T_o \ \sigma_{\phi}$.  The peak values are shown on the 
right hand panel scaled to match the zero moments as $10.4\ T_{peak} = 4.9 \cdot \sqrt{2 \pi } \ T_{peak}$,
where 4.9 is the mean value of the half-width, $\sigma_{\phi}$ of the Gaussian fits in this
area.  There is a very good match between the integrals of the Gaussian fits (center panel)
and the zero moment values (left panel); the highest point values (right panel) match pretty well
with the other two.  For the two pulsar positions in figure \ref{fig:spectra_1}
the numbers are (2.8, 2.3, 3.7) K rad m$^{-2}$ for the northern point,
and (3.5, 3.6, 3.8) K rad m$^{-2}$ for the southern point. 
For the entire area, the difference between the calculated 
zero moment value (left panel, figure \ref{fig:appA_mom0_comp}) and the peak value (right panel)
has standard deviation 2.7 K - rad m$^{-2}$. The Gaussian integral (center panel)
matches the zero moment much better, their difference has standard deviation 0.76
K - rad m$^{-2}$.  Since the peak value does not take into account the width of the
Faraday spectral feature(s), it is not surprising that it gives a rougher estimate of 
the total linearly polarized brightness temperature.  The peak value formally 
equals the brightness temperature of the polarized emission at
just the single rotation measure ($\phi$) corresponding to $\phi_{peak}$.
Faraday spectroscopy allows the separation of many different contributions
to the observed spectrum of linear polarized brightness, each with
a different rotation measure.  The spectral moment calculation is designed to
capture the richness of the resulting Faraday cube.

\vspace{.5in}

\section{The Effect of Missing Short Wavelengths on the 
Faraday Spectral Moments \label{app2}}

The shortest wavelengths in a polarization survey limit the sensitivity to
broad features in the Faraday spectrum, as discussed in section 1.3.  Given
the values of $\lambda_1^2$ for the Parkes and DRAO surveys (table 
\ref{tab:surveys}) leads to the values of $\phi_{max-scale}$ of
8.0 and 110 rad m$^{-2}$.  This is much less than the maximum $RM$ detectable,
that is set by the width of the spectrometer channels, $\Delta \lambda^2$.
The two surveys are sensitive to features in the Faraday spectrum up to very
high values ($> 10^3$ rad m$^{-2}$), much more than needed for a survey of
high and intermediate latitudes in the Milky Way.  Thus there is no bias 
against detecting features at high values of $\phi$ in our spectra.  But
there is a strong bias in the Parkes survey against detecting {\bf broad}
% Copy Editor:  please leave the \bf in the above line
features centered at any value of $\phi$.  This effect has been analysed
in several papers, starting with \citet{Brentjens_deBruyn_2005}, other illustrations 
are given by \citet{Frick_etal_2011} and \citet{Beck_etal_2012}.  Here we
consider the effect of the missing short wavelength data on the moments
calculated as described in section \ref{sec:moments}.  

\begin{figure} 
\hspace{-.2in}\includegraphics[width=7.5in]{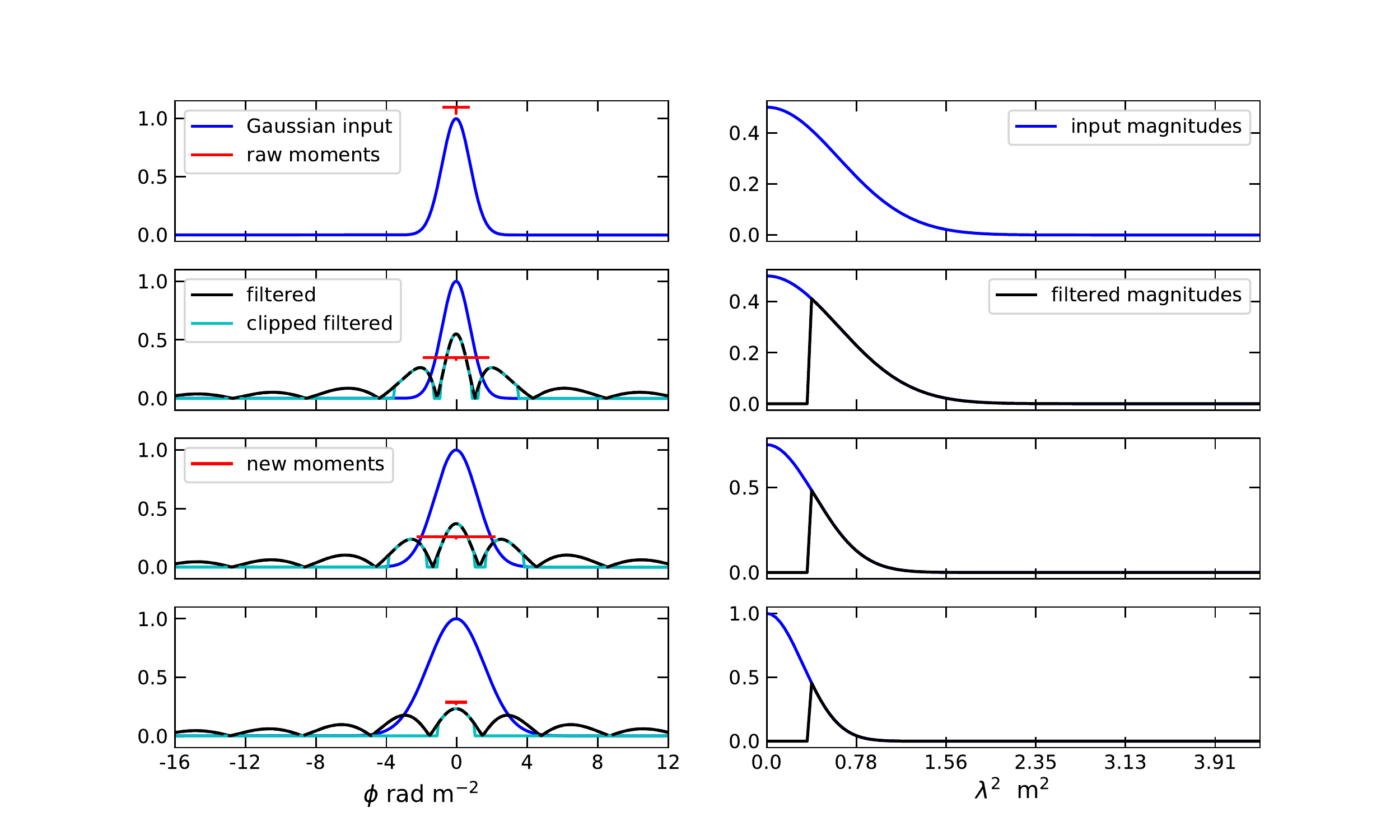}

\caption{The effect of the missing short wavelengths on a Gaussian line profile in 
Faraday space, and the resulting moments calculated after clipping the filtered
profile.  
The blue curves in the panels on the left side are Gaussian-shaped Faraday line profiles, $F(\phi)$,
and the corresponding blue Gaussians on the right hand panels are their transforms to $\lambda^2$ space
by equation \ref{eq:Fourier}.  In the second, third, and fourth rows are shown in
black the profiles after filtering out the short wavelengths, $\lambda^2 < 0.391$ m$^2$,
corresponding to the Parkes survey maximum frequency of 480 MHz.  Thresholding as described in 
section \ref{sec:thresh} changes the black curves on the left panels to the green curves,
and the moments calculated after thresholding are shown by the red bars, as on figures
\ref{fig:spectra_1} - \ref{fig:spectra_3}.
\label{fig:Gauss_hollow} 
}
\end{figure}

The simplest line profile function to consider is a Gaussian.  If a Faraday
spectral feature has a Gaussian shape in $\phi$ space, then it will
have a Gaussian shape in $\lambda^2$ space as well.  Figure
\ref{fig:Gauss_hollow} shows the effect of the missing short wavelengths
on progressively broader Gaussian features.  On the left side are Faraday
spectra, on the right the corresponding spectra in the $\lambda^2$ space.
The calculations are made with 10$^3$ equally spaced channels, but the
figure expands the ranges of significance on both sides for clarity.
The y scales are in arbitrary units, with zero points indicated.  
%For
%simplicity the analysis uses symmetric functions (positive and negative
%values of $\phi$ and $\lambda^2$), but only positive values have
%physical significance.

On figure \ref{fig:Gauss_hollow}
the top row shows a complete Gaussian on both sides, the ideal case with
no missing short wavelengths.  The second through fourth rows show the effect
of a gap in the measured values of $P(\lambda^2)$ for progressively 
broader features in the Faraday spectrum, $F(\phi)$ (equation
\ref{eq:Fourier}).  The left hand panels for each row show the effect
of this filtering on the Faraday profile function, before and after the
clipping applied in the moments calculation, and the red bars above the lines
show the resulting moments.  The zeroth moment is translated into an
equivalent line peak by the Gaussian formula
$T_0 = M_0 / (\sqrt{2 \pi} \sigma_{\phi})$, as on figures 
\ref{fig:spectra_1} - \ref{fig:spectra_3}.
As the feature width grows from $\sigma_{\phi}$ = 8 to 12 and then 16 rad m$^{-2}$,
on the second, third, and fourth rows of figure \ref{fig:Gauss_hollow},
the width of the feature in $\lambda^2$ conjugate space narrows.
The black curves on the right hand panels show the effect of the missing
short wavelengths on the line profile in $\lambda^2$, and the black and green 
curves on the corresponding panels on the left side show the effect of this
filtering on the line profile in Faraday space ($\phi$).  The green curve,
that partially covers the black curve, shows the result after the clipping at
15\% of the peak value, applied in the calculation of the moments.  The red bars
above the profiles show the values of the first and second moments that would 
then be calculated from the filtered, clipped line profile in $\phi$.
%The height of the bar is the zero moment divided by $\sqrt{2 \pi \sigma_{\phi}}$

\begin{figure} 
\hspace{-.2in}\includegraphics[width=7.5in]{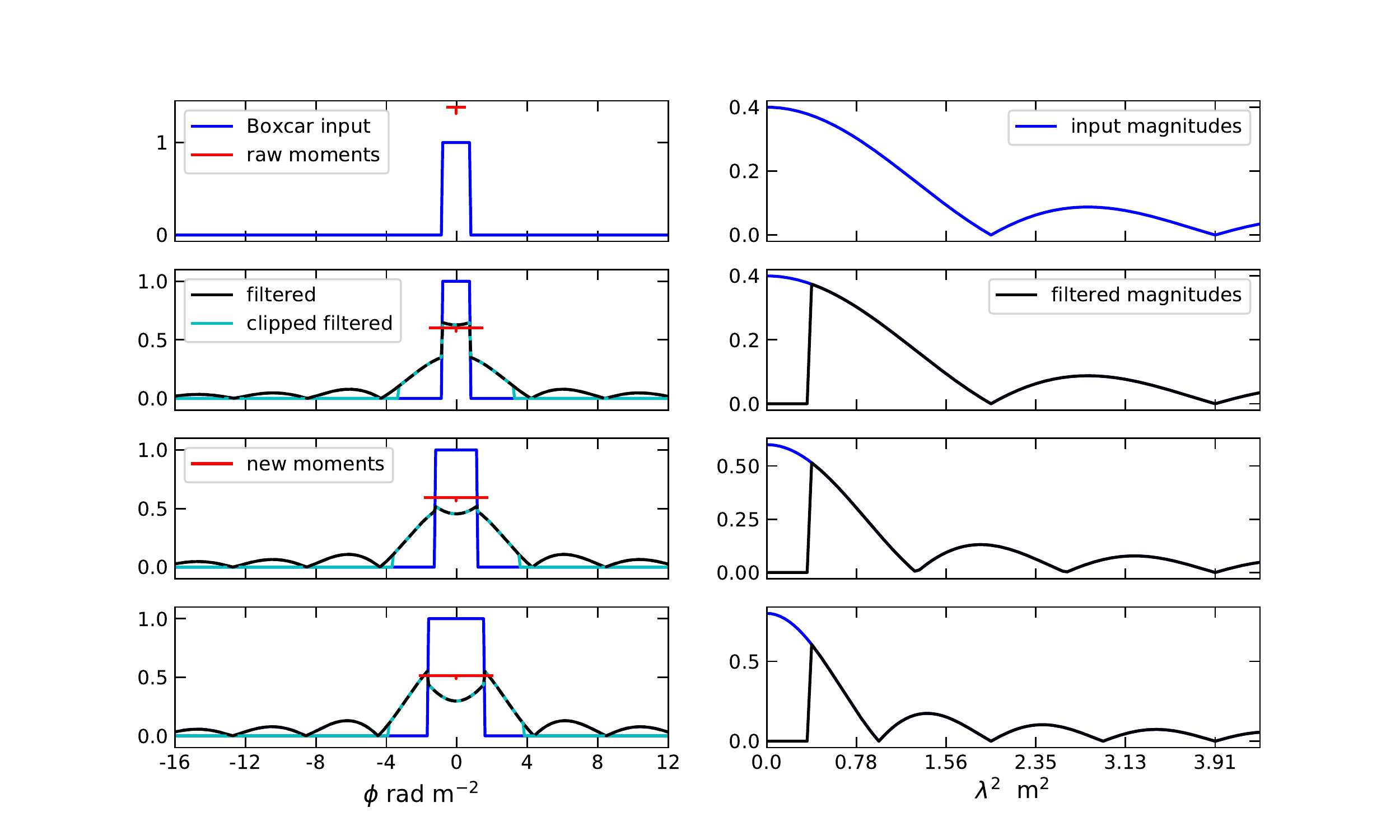}

\caption{The effect of the missing short wavelengths on a boxcar line profile in 
Faraday space, and the resulting moments calculated after clipping the filtered
profile.  The colors and layout are similar to those on figure \ref{fig:Gauss_hollow}.
Because of its discontinuous edges, the boxcar function turns into a ``two-horned''
profile after heavy filtering.  This is probably unrealistic; the Faraday profile
of a slab of mixed synchrotron emission and magnetised plasma would have smooth,
continuous edges due to irregularities in the density and $\vec{B}$ fields.
This will lead to profiles more like the Gaussians in figure \ref{fig:Gauss_hollow}.
 \label{fig:Boxcar_hollow} }
\end{figure}

An alternate profile shape that has been considered by several authors
is a boxcar or top-hat function, shown on figure \ref{fig:Boxcar_hollow}.
Here the effect of the missing short wavelengths is dramatic, because the discontinuous 
edges of the boxcar become two spikes after filtering.  The second moment, $m2$, is not
very sensitive to the filtering in this case, because the spacing between the two 
spikes or horns on the filtered profile does not change much.  The zero moment for
the functions shown on figure \ref{fig:Boxcar_hollow} is not much affected either,
because this is the integral of the {\bf magnitude} of $T_{pol}(\phi)$.
% Copy Editor:  please leave the \bf in the above line
%For narrower input boxcar functions, the zeroth moment actually increases due to the 
%filtering, because the main effect is to raise wings on the sides of the boxcar.

\begin{figure} 
\hspace{.9in}\includegraphics[width=5in]{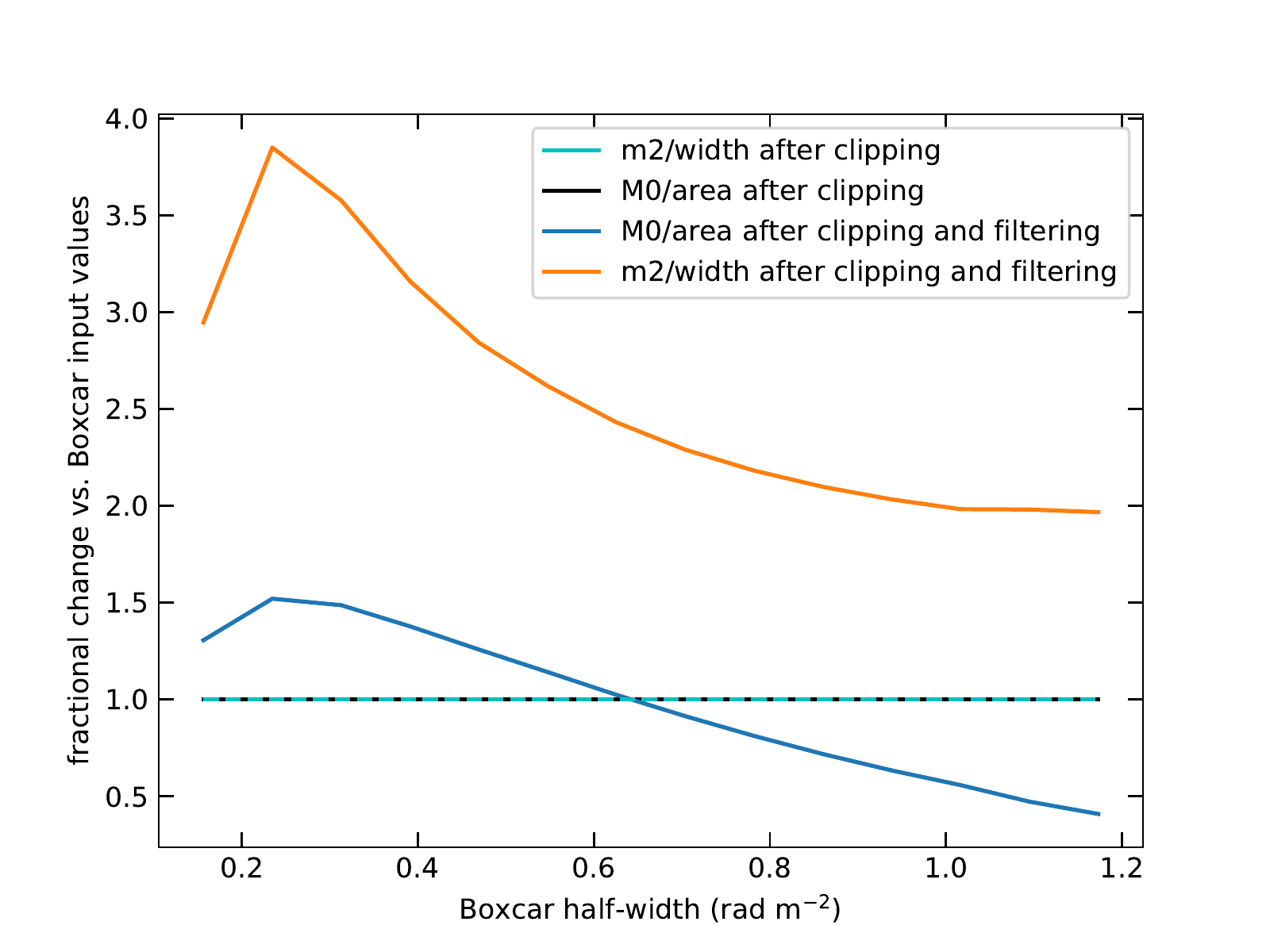}

\caption{The effect of the filtering caused by the missing short wavelengths
on the measured moments of synthetic spectra like those on figure \ref{fig:Boxcar_hollow}.
The width of the boxcar increases from left to right on the x-axis, and the
ratio of the values of $M_0$ and $m2$ measured for the filtered, clipped spectrum
to their corresponding values for the original boxcar is shown on the y-axis.
As discussed in the text, the result of the filtering due to missing short
wavelengths for narrow input lines is to more than triple the apparent width of the line.
As the input linewidth increases, both moments decrease due to the filtering.
The curves show the ratio of the moments ($M_0$ and $m2$) measured on 
the clipped and filtered Faraday spectra to the input values of the
area under the boxcar, i.e. the  area or true value of the moment $M_0$,
and the half-width of the boxcar, i.e. ``width'' which is the true value of $m2$.
 \label{fig:Boxcar_hollow_results} }
\end{figure}

The effects of the filtering and clipping on the computed values of moment zero and moment 2
are shown for a wider range of widths of boxcar functions in figure \ref{fig:Boxcar_hollow_results}.
The curves show the ratios of the computed values of $M_0$ and $m2$ to their values for a simple
boxcar of the same width, as a function of width.  The clipping alone (blue and green lines)
makes almost no difference at all, since a clipped boxcar is the same as a boxcar, but the
filtering increases and then decreases $M_0$ as the line width increases.  The filtering
greatly increases the second moment, even for the broadest lines the effect is a factor of two, and for narrower lines it is as high as 3.5.

\begin{figure} 
\hspace{.9in}\includegraphics[width=5in]{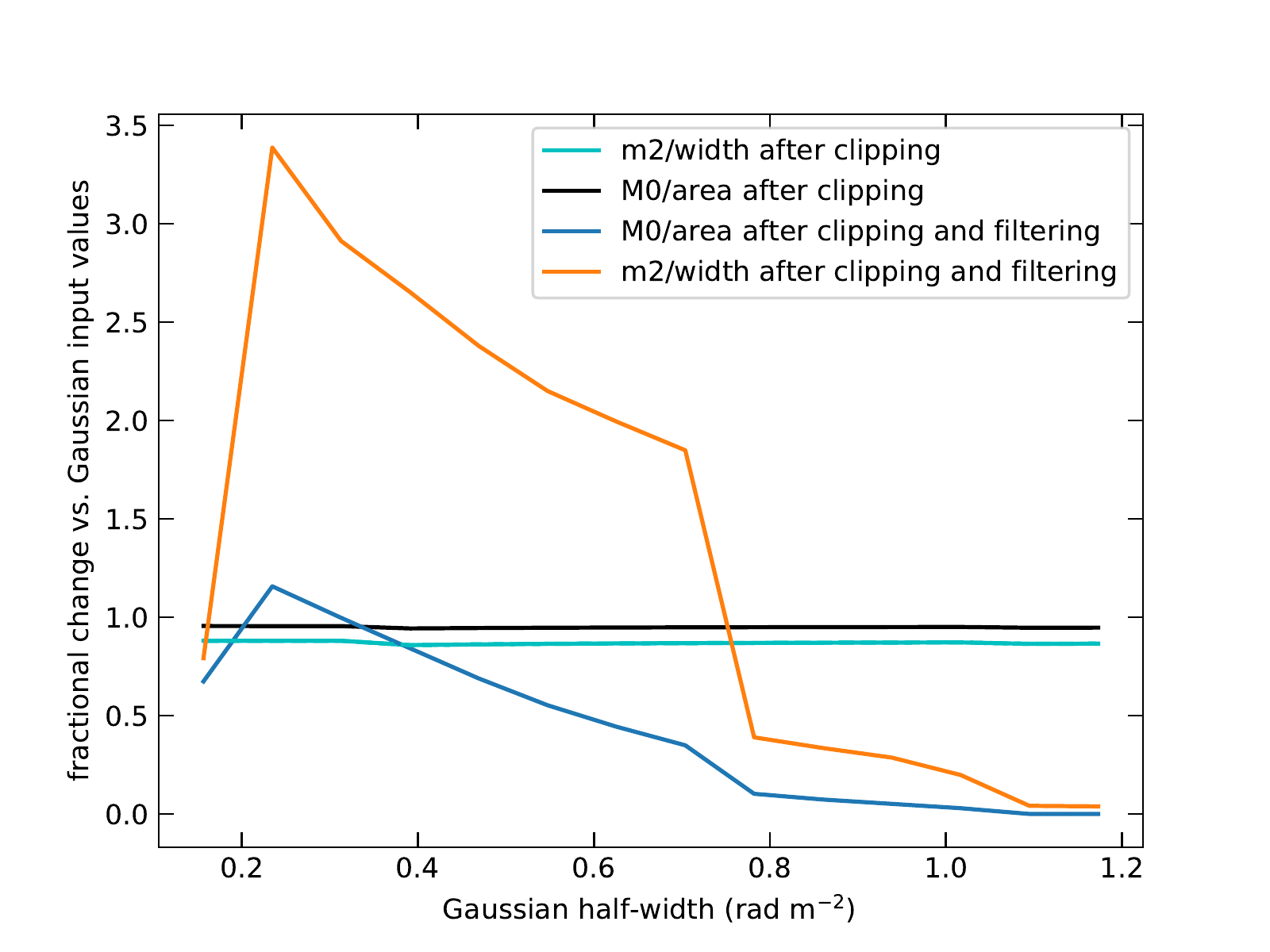}

\caption{The effect of the filtering caused by the missing short wavelengths
on the measured moments of synthetic Gaussian spectra like those on figure
 \ref{fig:Gauss_hollow}. 
The width of the Gaussian increases from left to right on the x-axis, and the
ratio of the values of $M_0$ and $m2$ measured for the filtered, clipped spectrum
to their corresponding values for the original Gaussian is shown on the y-axis.
As discussed in the text, the result of the filtering due to missing short
wavelengths for narrow input lines is to more than triple the apparent width of the line.
As the input linewidth increases, both moments decrease due to the filtering.
The curves show the ratio of the measured moments ($M_0$ and $m2$) to the
corresponding  values for the input function, i.e. the
area under the Gaussian, $M_0$,
and the width of the Gaussian, $\sigma = m_2$. 
\label{fig:Gaussian_hollow_results} }
\end{figure}

Similar to figure \ref{fig:Boxcar_hollow_results} is figure \ref{fig:Gaussian_hollow_results}, but
for Gaussian line profiles similar to those on figure \ref{fig:Gauss_hollow}.  Here the
clipping has a weak effect in reducing both $M_0$ and $m2$, but the filtering effect is much more severe.
On the right (widest input Gaussians) the filtering has attenuated the line below the threshold
at all values of $\phi$, so that both moments are zero.  For narrower input Gaussians,
the filtering increases $m2$, because the main line is surrounded by sidebands or spurious
secondary features on either side.  In a more realistic case of an asymmetric line profile,
it is likely that only one of the sidebands would be above the threshold, leading to a smaller
increase in the measured value of $m2$.  It is common in the Parkes spectra to see features
with two peaks, that may be the result of the missing short wavelength data.

The ultimate goal of the GMIMS collaboration is to combine surveys with different
telescopes that will cover the full wavelength range from $\lambda \sim 1$ m to 
$\lambda \sim $ 10 cm, so as to be sensitive to the full range of feature widths
in the Faraday spectrum.  The Parkes and DRAO surveys are the first big steps
toward that important objective.  When it is achieved it will provide an 
excellent view of the Galactic magneto-ionic medium
that cannot be traced in any other way.

\end{document}